\documentclass[11pt,letterpaper]{article}
\usepackage[letterpaper,margin=0.8in]{geometry}
\usepackage[leqno]{amsmath}
\numberwithin{equation}{section}
\usepackage{graphicx}
\usepackage{amssymb, amsthm}
\usepackage{eepic}
\usepackage{epic}
\usepackage{url} 
\usepackage{float} 
\usepackage{placeins} 
\usepackage{booktabs} 
\usepackage{makecell} 
\usepackage{multirow} 
\usepackage{adjustbox} 
\usepackage{forest} 
\usepackage{tikz}
\usetikzlibrary{positioning,arrows.meta}
\usepackage{comment}

\theoremstyle{definition}

\theoremstyle{remark}

\newcommand{\TB}{\mathop{\mathrm{TB}}} 

\newcommand{\TBn}{\ensuremath{\TB(n)}}

\newcommand{\vPhOne}{\mathit{ph1}}

\newcommand{\vRow}{\mathit{row}}

\newcommand{\PP}{{\textsc{P}}}
\newcommand{\AML}{\textsc{AMLs}}
\newcommand{\AMLfull}{Algebraic Modeling Languages}
\newcommand{\AMLlong}{\AMLfull{} (\AML)}
\newcommand{\CIP}{\textsc{CIP}}
\newcommand{\IP}{\textsc{IP}}
\newcommand{\PO}{\textsc{PO}}
\newcommand{\POfull}{Phase Optimization}
\newcommand{\POlong}{\POfull{} (\PO)}
\newcommand{\CS}{\textsc{CS}}
\newcommand{\CSfull}{Constraint Sharing}
\newcommand{\CSlong}{\CSfull{} (\CS)}
\newcommand{\UO}{\textsc{UO}}
\newcommand{\UOfull}{Un-optimized Compiler}
\newcommand{\UOlong}{\UOfull{} (\UO)}
\newcommand{\HSB}{\textsc{HSB}}
\newcommand{\HSBfull}{Hierarchical Synchronization Barriers}
\newcommand{\HSBlong}{\HSBfull{} (\HSB)}
\newcommand{\HSBCS}{\textsc{EDA}}
\newcommand{\HSBCSfull}{Execution Domain Aware}
\newcommand{\HSBCSlong}{\HSBCSfull{} (\HSBCS)}
\newcommand{\OPT}{\textsc{OPT}}

\newcommand{\lpnum}{37}
\newcommand{\ETI}{\textsc{ETI}}
\newcommand{\ETIfull}{Execution Time Interval}

\newcommand{\ETIlongpl}{\ETIfull{}s (\ETI{}s)}

\newcommand{\vMi}{\mathit{mi}}

\newcommand{\MM}{\textsc{MM}}

\newcommand{\HMM}{\textsc{HMM}}
\newcommand{\MS}{\textsc{MS}}
\newcommand{\MST}{\textsc{MST}}
\newcommand{\BS}{\textsc{BS}}
\newcommand{\SPDJ}{\textsc{SP}}
\newcommand{\MMfull}{Maximum Matching}

\newcommand{\HMMfull}{Hoisted Maximum Matching}
\newcommand{\MSfull}{Makespan Scheduling}
\newcommand{\MSTfull}{Minimum Spanning Tree}
\newcommand{\BSfull}{Binary Search}
\newcommand{\KS}{\textsc{KS}}
\newcommand{\KSfull}{Knapsack}
\newcommand{\KSlong}{\KSfull{} (\KS)}
\newcommand{\SPDJfull}{Shortest Path}
\newcommand{\LISproblem}{Longest Increasing Subsequence}
\newcommand{\LIS}{\textsc{LIS}}
\newcommand{\LIStimecomp}{\ensuremath{\mathcal{O}(n^2)}}
\newcommand{\LISfull}{\LISproblem{} (\LIStimecomp)}
\newcommand{\LISlong}{\LISfull{} (\LIS)}
\newcommand{\LISL}{\textsc{LISL}}
\newcommand{\LISLtimecomp}{\ensuremath{\mathcal{O}(n \log n)}}
\newcommand{\LISLfull}{\LISproblem{} (\LISLtimecomp)}
\newcommand{\LISLlong}{\LISLfull{} (\LISL)}
\newcommand{\SAM}{\textsc{SAM}}
\newcommand{\SAMfull}{Sorted Arrays Merging}
\newcommand{\SAMlong}{\SAMfull{} (\SAM)}
\newcommand{\MMlong}{\MMfull{} (\MM)}

\newcommand{\HMMlong}{\HMMfull{} (\HMM)}
\newcommand{\MSlong}{\MSfull{} (\MS)}
\newcommand{\MSTlong}{\MSTfull{} (\MST)}
\newcommand{\BSlong}{\BSfull{} (\BS)}
\newcommand{\SPDJlong}{\SPDJfull{} (\SPDJ)}
\newcommand{\TS}{\textsc{TS}}
\newcommand{\TSfull}{Topological Sort}
\newcommand{\TSlong}{\TSfull{} (\TS)}
\newcommand{\gurobi}{\textsc{gurobi}}

\newcommand{\cplex}{\textsc{cplex}}
\newcommand{\scip}{\textsc{scip}}
\newcommand{\soplex}{\textsc{SoPlex}}
\newcommand{\OOM}{\textsc{OOM}}
\newcommand{\OOT}{\textsc{OOT}}
\newcommand{\CPU}{\textsc{CPU}}
\newcommand{\GB}{\textsc{GB}}
\newcommand{\MB}{\textsc{MB}}
\newcommand{\FAIL}{\textemdash}
\newcommand{\IQR}{\textsc{IQR}}
\newcommand{\IQRlong}{interquartile range (\IQR)}
\newcommand{\MAE}{\textsc{MAE}}
\newcommand{\MAElong}{Mean Absolute Error (\MAE)}
\newcommand{\RMSE}{\textsc{RMSE}}
\newcommand{\RMSElong}{Root Mean Square Error (\RMSE{})}
\newcommand{\Sec}{\text{sec}}
\newcommand{\Min}{\text{min}}
\newcommand{\Hour}{\text{hr}}
\newcommand{\Med}{\text{Med.}}
\newcommand{\Diff}{\text{Diff.}}
\newcommand{\RSS}{\textsc{RSS}}
\newcommand{\RSSlong}{\text{resident set size (\RSS)}}
\newcommand{\BlockName}[1]{\textsf{#1}}
\newcommand{\blLone}{\BlockName{$L_1$}}
\newcommand{\blLtwo}{\BlockName{$L_2$}}
\newcommand{\blLthree}{\BlockName{$L_3$}}
\newcommand{\blLfour}{\BlockName{$L_4$}}
\newcommand{\blCone}{\BlockName{$C_1$}} 
\newcommand{\blCtwo}{\BlockName{$C_2$}} 
\newcommand{\blCthree}{\BlockName{$C_3$}} 
\newcommand{\blCfour}{\BlockName{$C_4$}} 
\newcommand{\blCfive}{\BlockName{$C_5$}} 
\newcommand{\blCsix}{\BlockName{$C_6$}} 
\newcommand{\blIone}{\BlockName{$I_1$}}
\newcommand{\blItwo}{\BlockName{$I_2$}}
\newcommand{\lineOne}{\BlockName{$l_1$}}
\newcommand{\lineTwo}{\BlockName{$l_2$}}
\newcommand{\lineThree}{\BlockName{$l_3$}}
\newcommand{\lineFour}{\BlockName{$l_4$}}
\newcommand{\lineFive}{\BlockName{$l_5$}}
\newcommand{\enquote}[1]{``#1''}

\newcommand{\Sparktope}{\textsc{sparktope}}
\newcommand{\Asm}{\texttt{Asm}}

\title{Efficient Conversion of Algorithms to Linear Programs%
\thanks{This work provides an extended evaluation of techniques introduced in conference papers~\cite{khosravi_efficient_2025,khosravi_smaller_2026} and addresses a limitation of the technique introduced in~\cite{khosravi_efficient_2025}.}}

\author{Shermin \textsc{Khosravi}\thanks{Faculty of Computer Science, University of New Brunswick, Fredericton, Canada. Email: \texttt{shermin.khosravi@unb.ca}}
  \and David \textsc{Bremner}\thanks{Faculty of Computer Science, University of New Brunswick, Fredericton, Canada. Email: \texttt{bremner@unb.ca}}}    
\date{}

\begin{document}
\maketitle

\begin{abstract}      
The expressiveness of Linear Programming (LP) has motivated research on an LP compiler that generates polynomial-size LPs for polynomial-time solvable problems, including those having exponential extension complexity. 
By automating modeling from high-level algorithmic descriptions, the compiler offers a promising alternative to \AMLlong{}. 
However, the resulting LPs are extremely large, posing challenges for existing solvers. 
Recently, we have introduced two LP compiler optimizations, namely \HSBlong{} and \CSlong{}, that address this challenge by exploiting structural and similarity-based patterns derived from the source code and the design of compiler-generated LPs, beyond what conventional compiler and presolve techniques can achieve in isolation.
In this paper, we provide a comprehensive evaluation of these optimization methods on fundamental combinatorial algorithms and introduce the \HSBCSlong{} optimization framework to address a limitation revealed by this analysis. 
The framework reinterprets structural synchronizations as a time abstraction layer within which other optimizations can target localized reductions while remaining agnostic to time-step transitions. 
Across a broad suite of benchmarks, \HSB{} and \CS{} consistently generate smaller LPs, achieving LP file-size reductions of up to 96\% and 53\%, respectively, while also improving numerical conditioning, solver performance, and solution quality. \HSB{} provides the strongest and most robust overall gains. The \HSBCS{} framework further broadens the range of benchmarks that can benefit from \HSB{} when its underlying structural assumptions are less effective.
We view this work as a step toward systematic generation of Compact Integer Programming (\CIP{}) formulations for problems with exponential-size Integer Programs having polynomial-time separation oracles.
\end{abstract}

\noindent\textbf{Keywords:} Linear Programming, Algebraic Modeling Languages, Extension Complexity, Polytopes, Matching, Makespan

\medskip

\section{Introduction}
\label{sec:introduction}
Extensive research has focused on the expressive power of Linear Programming (LP) as a modeling tool. 
An active area of research studies compact reformulations for LP models having exponential-size, particularly for polynomial-time problems whose natural\footnote{Formulations derived from the problem description, such as a polytope defined as the convex hull of characteristic vectors of \enquote{yes} instances of a decision problem.} LP formulations are exponential in size.
Although compact formulations exist for some of these problems, such as spanning tree polytope~\cite{martin_using_1991}, research continues to investigate whether even smaller formulations can be obtained for these problems~\cite{fiorini_smaller_2017,aprile_smaller_2021}. 
These research efforts can result in tighter LP relaxations for Integer Programs (\IP{}s), enhance solver performance, and potentially lead to deeper theoretical insights.

However, some of the polynomial-time problems have natural LP formulations that have exponential extension complexity, including Edmonds' matching polytope~\cite{rothvoss_matching_2017}, the makespan polytope~\cite{tiwary_extension_2020}, and polytopes associated with certain matroid families~\cite{rothvoss_01_2013}.
Although the Ellipsoid method~\cite{grotschel_ellipsoid_1981} can solve exponential-size LPs having polynomial-time separation oracles in polynomial time, it is rarely used in practice because of its slow convergence. Smaller LP formulations enable the use of practical LP algorithms that require an explicit description of the model~\cite{carr_compact_2002}, thereby eliminating the need for external separation oracles.

In contrast to LPs, where larger formulations are typically harder to solve~\cite{ted_integer_2025}, some small \IP{}s remain unsolved even today, whereas many large instances can be solved fairly quickly.
Nevertheless, model size remains important for \IP{}s solved by iteratively computing LP relaxations, since it affects the computational cost of each iteration.

Moreover, separation oracles sometimes fail to generate quality cuts. For problems such as the contact map overlap problem~\cite{carr_compact_2002}, the cuts generated at each iteration may improve the objective value only slightly, resulting in a large number of iterations and degraded solver performance.

Weak extended formulations (WEF)~\cite{avis_polynomial_2019}, implemented within an LP compiler~\cite{avis_sparktope_2022}, enable the conversion of polynomial-time algorithms into LP models of polynomial size, thereby providing a systematic way to obtain polynomial-size LPs for $\PP$ problems, particularly for those with natural polytopes that have exponential extension complexity.

The resulting LPs are guaranteed to have integral solutions whenever the input variables take 0/1 values. This is referred to as the $x$-0/1 property of the polytope~\cite{avis_polynomial_2019}. 
Other LPs in the literature are also known to have integral solutions without explicit integrality constraints. Examples include polytopes for combinatorial problems such as Edmonds' matching polytope~\cite{edmonds_maximum_1965}, the order polytope~\cite{stanley_two_1986}, and transportation polytopes with integer marginals~\cite{loera_combinatorics_2013}, which also appear in neural representation metrics~\cite{khosla_soft_2024}, tropical geometry~\cite{comaneci_tropical_2024}, and statistical testing~\cite{berrett_optimal_2023}. 

In the context of model construction, modeling languages serve a variety of purposes.
These range from Rosette, which compiles programs into logical constraints for solver-aided verification and synthesis and has been applied in practice in systems such as Topaz for DNS policy verification~\cite{larisch_topaz_2024}, to \AMLlong{}, which offers high-level abstractions for mathematical optimization~\cite{jusevicius_experimental_2021}. However, even though modern \AML{} provide built-in constructs for recurring structures like network flow~\cite{nethercote_minizinc_2007} and extensions for specialized domains such as stochastic programming~\cite{valente_extending_2009}, to obtain LPs from algorithms, users still must manually specify each set of constraints.
In this setting, LP compilers complement existing modeling languages by automatically generating LPs from algorithms expressed in high-level programming languages.

Although the resulting LPs are polynomial in size, they remain extremely large, creating difficulty for existing solvers because of memory limits, long presolve and solution times, and increased vulnerability to rounding errors and numerical instability. 
Our previous work introduced two LP compiler optimizations, \CSlong{}~\cite{khosravi_smaller_2026} and \HSBlong{}~\cite{khosravi_efficient_2025}, for generating smaller LPs from algorithms. 
At a high level, \CS{} reduces redundancies by sharing constraints across similar statements, while \HSB{} reduces redundancies through hierarchical synchronization based on input size, allowing some program states to become identifiable as unreachable at compile time.

In this paper, we provide a unified cross-benchmark evaluation of the LP compiler optimizations \CS{} and \HSB{} across a broad collection of fundamental combinatorial algorithms, including problems with exponential extension complexity. 
Beyond measuring LP size, we examine how compiler-driven LP size reductions affect solver performance through overall run time, presolve cost, numerical conditioning, and solution accuracy.
This evaluation also reveals a key limitation of \HSB{}, which can become ineffective when user insight yields substantially smaller execution-time bounds for certain code regions than those implied by its default worst-case structural decomposition.
To address this limitation, we propose the \HSBCSlong{} framework, which preserves the hierarchical synchronization of \HSB{} for unaffected code while providing affected regions with a time-abstraction layer through a common timing and variable transition interface. This enables reductions based on smaller local bounds, support localized optimizations that remain agnostic to inter-block time-step transitions, and broadens the applicability of structural LP-size reduction beyond \HSB{}'s worst-case decomposition.

The remainder of this paper is structured as follows. Section~\ref{sec:background} reviews the LP compiler \Sparktope{}~\cite{avis_sparktope_2022} and the optimization methods most relevant to this work. Section~\ref{sec:relatedWork} surveys other LP size reduction approaches and discusses their main limitations. Section~\ref{sec:methodology} presents the proposed \HSBCS{} framework and explains its relationship to \HSB{}. Section~\ref{sec:results} reports experimental results across a broad suite of benchmark problems with a particular focus on maximum matching. Finally, Section~\ref{sec:conclusion} summarizes our findings and outlines directions for future work. 

\section{Background}
\label{sec:background}
The LP compiler~\cite{avis_polynomial_2019, avis_sparktope_2022} translates deterministic algorithms into LP models. The resulting LP is guaranteed to have polynomial size, provided that the algorithm runs in polynomial time. 
The compiler may be regarded as a modeling language that represents subsets of the Boolean hypercube by generating linear constraints for selected 0/1 feasible points expressed in a high-level programming language.
In the terminology of~\cite{khosravi_efficient_2025}, for each polynomial-time computable \( f : \{0,1\}^p \rightarrow \{0,1\}^q \), there is a polynomial-time computable \emph{imperative model} \( P = \{ z \mid Az \leq b \} \subseteq [0,1]^{p+q+r} \) such that, for each input \( x \in \{0,1\}^p \), the polytope contains a \emph{unique} vertex corresponding to the triple \( (x, f(x), s) \in V(P) \cap \{0,1\}^{p+q+r}\), where \( s \) denotes auxiliary variables and \( V(P) \) denotes the set of vertices of \( P \), which may also include non-integral vectors. This is equivalent to the \emph{$x$-0/1 property} described in~\cite{avis_polynomial_2019}. 

\begin{figure}[!b] 
\begin{center}
\includegraphics [width=1\columnwidth]{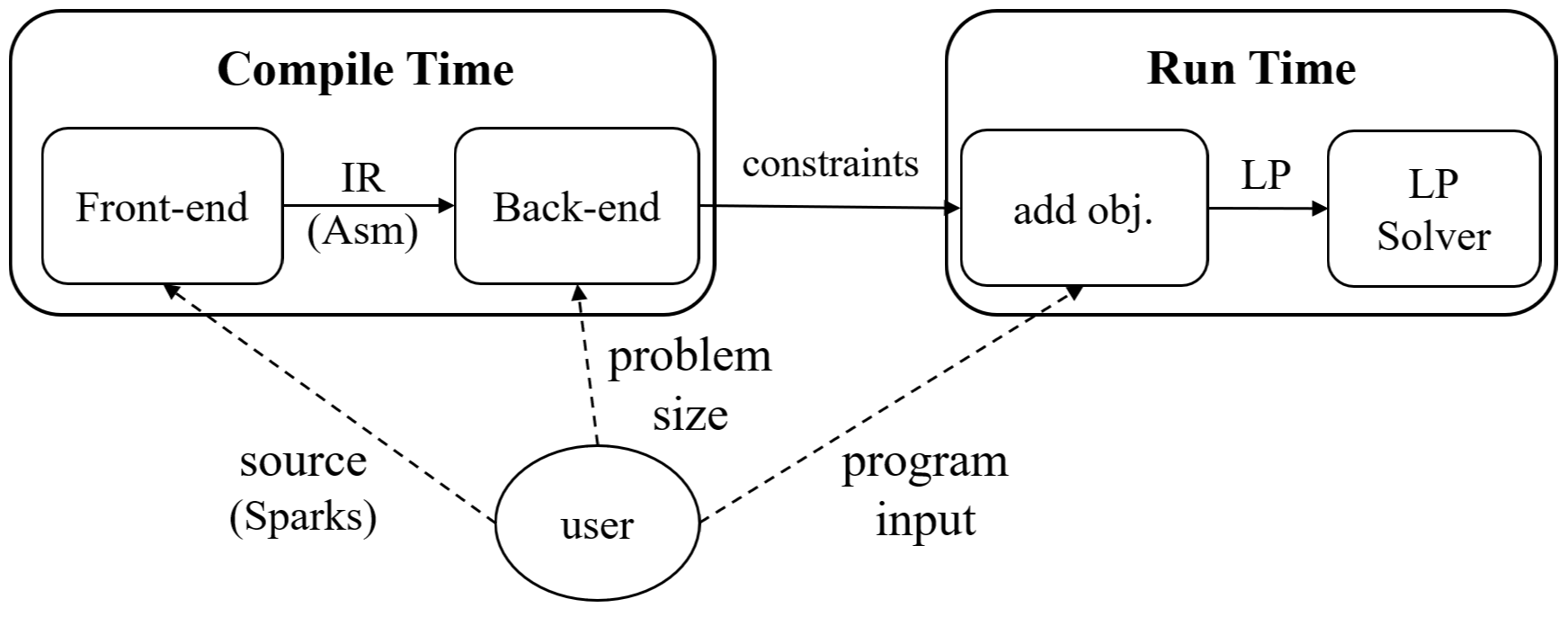}
\caption{Workflow of the LP compiler proposed in~\cite{avis_sparktope_2022}. Figure adapted from~\cite{khosravi_efficient_2025}.}
\label{fig:compiler-workflow}
\end{center}
\end{figure}

Figure~\ref{fig:compiler-workflow} shows the workflow of the LP compiler. 
During the model generation phase, or \emph{compile time}, the user specifies the problem size~$n$ together with an upper bound, denoted by \TBn{}, on the number of execution steps as a function of the input size. 
During the model solution phase, or \emph{run time}, the user provided concrete input values are encoded into the LP through the objective function. As a result, the compiled LP can be solved for any input of size $n$ without requiring recompilation. 
In the intermediate language, \Asm{}, the integer-valued variables are represented in binary form according to a word size provided by the user, where each binary digit is modeled by an LP variable constrained within [0,1] range.
Because the compiler does not know in advance when a statement will execute, it assumes that every statement may run at any execution time step $t$ within \TBn{} duration.
Accordingly, for each \Asm{} statement, it creates copies of the corresponding constraints across \TBn{} duration and attaches time indices to all its mutable \Asm{} variables.
Boolean controller variables $S(l,t)$ encode the program counter (PC) state, where $S(l,t)$ indicates that the implicit PC is at line $l$ at time step $t$, for $t \in \{1,...,\TB(n)\}$. 
The initial PC state is enforced by the equality constraint $S(1,1)=1$, which starts execution on line $1$ at time step $1$.
In addition, to ensure that only one \Asm{} line executes at each time step $t \in 1,\dots,\TB(n)$, the compiler constrains the implicit PC to be on exactly one line at time step $t$. This is forced through the equality constraints $\sum_{(l=1)}^{L} S(l,t)=1$, where $L$ is the total number of \Asm{} lines. 
The compiler generates two main types of constraints for every \Asm{} statement: control flow and data flow constraints. The \emph{control flow} constraints specify the next PC state based on the current one. The data flow constraints fall into two groups: \emph{memory update} constraints, which assign values to variables that are directly changed by the statement, and \emph{carry-forward} constraints, which retain the values of all other variables by propagating their values from the previous time step to the current one.
Integer values are introduced into the formulation through initialization equalities and the objective function, after which they are carried through the model by the data flow constraints. 

Even though imperative LP models guarantee integral solutions whenever the input variables have 0/1 values, floating-point solvers may still return fractional solutions because of numerical rounding error, an issue well known in both theoretical research and applied \IP{}~\cite{jarck_exact_2020}. 
In~\cite{avis_sparktope_2022}, the model correctness was further validated in two ways: by solving the formulation using exact arithmetic and by fixing the inputs to their 0/1 values. 
Under the latter setting, two of the three solvers examined recovered optimal solutions, whereas \gurobi{} did not. 
A more detailed discussion of these solver-precision challenges is given in Section~\ref{sec:results}.

\subsection{LP Compiler Optimizations}
\label{sec:backgroundLpCompilerOptimizations}
Three LP compiler optimizations are relevant to this paper. 
\POlong~\cite{avis_sparktope_2022}, which we use for comparison in Section~\ref{sec:results}, reduces LP size by eliminating redundant constraints based on user-provided execution-time ranges.

\CSlong~\cite{khosravi_smaller_2026}, which we use as a concrete localized optimization within \HSBCS{}, reduces LP size by replacing individual constraint sets generated for \emph{identical} statements with a shared constraint set, activated whenever the implicit Program Counter (PC) is at any of the corresponding lines via aggregation of their $S$ controllers. This is particularly effective for costly array operations. 
\CS{} further increases sharing opportunities by introducing placeholder substitutions, which temporarily turn Operator-Equivalent statements---statements with the same operation but different variables and constant values---into identical ones, enabling their constraints to be shared through controller aggregation. Separate placeholders are introduced for array indices. 

\HSBlong~\cite{khosravi_efficient_2025}, which serves as the synchronization layer within \HSBCS{}, introduces a hierarchical synchronization structure over the code, with transition times between regions statically known as a function of input size, enabling LP size reduction by localizing constraint and variable generation to those durations.
To do this, \HSB{} automatically decomposes the code into a hierarchy of blocks, and semi-automatically computes time related annotations and local \TBn{} values for them, requiring from the user only the maximum number of loop iterations, denoted as $\vMi$. 
During constraint generation, \HSB{} uses these annotations to calculate \ETIlongpl{} of blocks and to generate control flow constraints that simulate synchronization barriers at the end of certain blocks. It then restricts constraint copies to their corresponding \ETI{}s and carry-forward constraints to the union of the \ETI{}s of blocks that access each variable. As a result, redundant constraint copies and variable versions outside the relevant \ETI{}s are eliminated, significantly reducing LP size.

\section{Related Work}
\label{sec:relatedWork}
Methods for reducing LP size can be generally classified into two main groups: modification and reformulation methods.
\emph{Modification} techniques do not rely on problem knowledge and are similar to the methods used at the presolve stage of LP solvers. They simplify and reduce the size of models by a series of operations such as variable substitution and aggregation, fixing variable values, and elimination of singleton rows, redundant constraints, parallel or dominated rows and columns. 
While smaller models often lead to better solver performance, solution time can also improve due to other factors such as increased sparsity and better numerical conditioning, for example, through operations such as scaling~\cite{swietanowski_modular_1995}.
For LP models, the modification procedures are restricted to techniques that allow postsolve recovery of both primal and dual solutions~\cite{achterberg_presolve_2020}.
In the context of simplifying imperative LP models, modification techniques differ from the optimization techniques applied within the LP compiler (Section~\ref{sec:background}) in that they are applied at run time, that is, during model solution. This gives them the advantage of access to the input values encoded in the objective function. 
However, unlike compiler-based optimizations, they have the disadvantage of requiring the model to be loaded into memory before it can be simplified or reduced in size. Moreover, these techniques do not exploit the structure and semantics of the source code.
Nonetheless, LP compiler optimizations and modification techniques are similar in that their reductions in LP size are typically limited to a constant factor and rarely change the asymptotic growth of the formulation. 

\emph{Reformulation} techniques generally use problem-specific knowledge to derive compact models that remain equivalent to the original formulation. They often achieve a reduction in the number of constraints by increasing the problem dimension through the addition of a polynomial number of auxiliary variables. 
Extended Formulations (EFs) are compact formulations in higher dimensions for exponential-size polytopes, in such a way that their images under linear projection recover the original exponential polytope~\cite{conforti_extended_2010,fiorini_smaller_2017,edmonds_paths_1965,fiorini_exponential_2015}. 
Some polynomial-time problems, however, have natural LP formulations with exponential extension complexity. 
Exponential-size \IP{}s can be solved by representing all but a polynomial number of constraints implicitly using one or more cutting plane oracles.
Alternatively, Compact Integer Programs (CIPs) avoid reliance on external oracles, which can be advantageous for problems, such as the contact map overlap problem~\cite{carr_compact_2002}, where the oracle fails to produce quality cuts, degrading solver performance. 
Recent work~\cite{schade_lower_2025} identifies a trade-off in CIPs between the number of constraints and the number of integer variables, and establishes nearly tight lower bounds on how many integer variables are necessary. This is particularly relevant because the performance of \IP{} solvers depends on the number of integer variables.
Although CIPs, like EFs, reduce model size by introducing auxiliary variables, they do not need to project back to the original polytope~\cite{lancia_deriving_2016}. 
In either case, constructing such reformulations depends heavily on human ingenuity. 
WEF~\cite{avis_polynomial_2019}, which forms the foundation of the LP compiler in~\cite{avis_sparktope_2022}, is also a reformulation approach. 
The compiler systematically produces polynomial-size LPs for polynomial-time problems, even including problems with exponential extension complexity. 
However, the resulting LPs are often extremely large and difficult for current solvers to handle.

Some standard compiler optimizations~\cite{muchnick_advanced_1997,schneck_survey_1973} that reduce code size, such as \emph{constant propagation} and \emph{dead code elimination}, may also yield smaller LP formulations when applied within the LP compiler of~\cite{avis_sparktope_2022}. 
In contrast, optimizations which rely on statement execution frequency, such as code motion, may have little direct effect on LP size because the LP model contains the same number of copies of the constraint for every statement, although they may still help indirectly by reducing \TBn{}. 
Techniques that expand code size, including \emph{loop unrolling} and data bandwidth-increasing transformations~\cite{cardoso_compiling_2010}, may even enlarge the LP. This is because each additional statement or data structure contributes to \TBn{} extra sets of constraints or LP variables.  

A notable result on conversions between
problem representations is Dobkin's log-space reduction from Horn-SAT to LP, which established the $\PP$-completeness of LP~\cite{dobkin_linear_1979}. 
In~\cite{valiant_reducibility_1982}, Valiant shows that every polynomial-time decision problem has a polynomial-sized LP formulation. This is done by transforming families of circuits for polynomial-time problems into families of polynomial-size LPs. 
Even though circuit constructions that are highly optimized can lead to smaller LP formulations than those produced by the LP compiler, deriving such circuits (or Horn clauses) from high-level algorithmic descriptions typically demands either significant manual design or sophisticated optimization tools. 

\section{Methodology}
\label{sec:methodology}
In this section, we introduce the \HSBCS{} framework. 
\HSBCS{} preserves the structural synchronization layer (Section~\ref{sec:synchronizationHsb}) of \HSB{} (Section~\ref{sec:backgroundLpCompilerOptimizations}) as the default mechanism for code decomposition and synchronization, while allowing users to select blocks to be treated as \emph{flat blocks}---that is, undecomposed blocks containing one or more complete statements with a single entry point and a single exit point---when problem-specific insight indicates smaller local \TBn{} values than the worst-case structural bounds (Section~\ref{sec:tbnReductionIssue}).
The inherited \HSB{} layer hierarchically decomposes the code, annotates each block with time-distance metadata and local \TBn{} values (Section~\ref{sec:codeDecompositionTbnCalculations}), and uses these annotations during constraint generation to simulate synchronization barriers and generate \ETI{}s  (Section~\ref{sec:synchronizationHsb}). This eliminates constraint copies and variable versions outside the corresponding block \ETI{}s.
\HSBCS{} then uses this synchronization structure as an abstraction layer for localized optimization, shown here with localized \CS{} (Section~\ref{sec:backgroundLpCompilerOptimizations}). Thus, \HSBCS{} preserves \HSB{}'s structural gains while incorporating custom local \TBn{} reductions.

\subsection{Problem-specific \TBn{} Reductions}
\label{sec:tbnReductionIssue}
For some algorithms, problem-specific insight yields substantially smaller \TBn{} values than the worst-case bounds inferred by \HSB{} from code structure and user-provided parameters. 
This occurs when worst-case execution is rare relative to the dominant execution pattern.
Such reductions arise in the implementations of \MSlong{} and \MMlong{} in~\cite{avis_sparktope_2022}, as well as in the quadratic implementation of \LISproblem{}~(\LIS{}) (Section~\ref{sec:results}).
For example, in \MS{}, the two unnested loops each iterate $m$ times, but their inner conditional bodies execute only $m$ times in total. Thus, if one executes $u$ times, the other executes $m-u$ times for some $0 \leq u \leq m$, yielding a smaller overall execution bound and therefore a smaller LP in the base compiler, which generates \TBn{} copies of each statement's constraints. 
By contrast, \HSB{} synchronizes regions according to their local \TBn{} values, so every execution of a region must take the same number of time steps. 
Although this enables compile-time pruning of unreachable PC states, it ties regions to worst-case bounds and prevents \HSB{} from exploiting such execution overlaps.

\subsection{Running Example}
\label{sec:exampleCustomTbnReductions}
\begin{figure}[t]
\centering
\begin{tikzpicture}[
  scale=1,
  block/.style={
    draw,
    rounded corners=2pt,
    line width=0.9pt,
    minimum width=8mm,
    minimum height=6mm,
    inner sep=1.8pt,
    font=\small\bfseries,
    text=black
  },
  flatblock/.style={
    block,
    fill=green!15
  },
  boldblock/.style={
    block,
    line width=1.8pt
  },
  boldflatblock/.style={
    flatblock,
    line width=1.8pt
  },
  codebox/.style={
    draw,
    dashed,
    rounded corners=3pt,
    line width=0.9pt,
    align=center,
    inner sep=3pt,
    text width=25mm,
    font=\small\bfseries,
    text=black
  },
  edge/.style={
    line width=0.9pt,
    draw=black
  },
  dashededge/.style={
    dashed,
    line width=0.9pt,
    draw=black
  }
]

\node[block]         (root) at (0,-0.25) {root};

\node[block]         (A1)   at (2.2,1.25) {\blLone};
\node[flatblock]     (D1)   at (2.2,-0.95) {\blLfour};
\node[block]         (F1)   at (2.2,-2.05) {\blCsix};

\node[flatblock]     (B1)   at (4.45,2.00) {\blLtwo};
\node[block]         (B2)   at (4.45,1.25) {\blCthree};
\node[block]         (B3)   at (4.45,0.40) {\blItwo};

\node[block]         (C1a)  at (6.70,2.65) {\blIone};
\node[block]     (C1)   at (6.70,1.55) {\blCtwo};
\node[boldblock]         (C3)   at (9.00,2.65) {\blLthree};
\node[block]         (C4a)  at (11.10,2.65) {\blCone};

\node[block]         (E1)   at (6.70,0.40) {\blCfour};
\node[boldblock]     (E2)   at (4.45,-0.95) {\blCfive};

\node[codebox]       (bodyC1) at (9.55,1.55) {$\lineOne$: $X[i]\leftarrow z$\\$\lineTwo$: $X[q]\leftarrow y$};
\node[codebox]       (bodyC4) at (7.55,-1.40) {$\lineThree$: $z\leftarrow X[j]$\\$\lineFour$: $y\leftarrow X[q]$\\$\lineFive$: $i\leftarrow 2$};

\draw[edge] (root) -- (A1);
\draw[edge] (root) -- (D1);
\draw[edge] (root) -- (F1);

\draw[edge] (A1) -- (B1);
\draw[edge] (A1) -- (B2);
\draw[edge] (A1) -- (B3);

\draw[edge] (B1) -- (C1a);
\draw[edge] (B1) -- (C1);
\draw[edge] (C1a) -- (C3);
\draw[edge] (C3) -- (C4a);

\draw[edge] (B3) -- (E1);
\draw[edge] (D1) -- (E2);

\draw[dashededge] (C1.east) -- (bodyC1.west);
\draw[dashededge] (E2.east) -- (bodyC4.west);

\end{tikzpicture}
\caption{A simplified code block hierarchy. 
Solid boxes denote blocks, dashed boxes denote the code inside leaf blocks, bold outlines indicate blocks with custom \TBn{} reductions based on user insight, and shaded boxes denote affected blocks identified by the user to be flattened manually.
The main block types are denoted by labels $L_r$, $I_r$, and $C_r$, which correspond to nested loop blocks, nested conditional blocks, and code blocks, respectively. Labels $\lineOne,\dots,\lineFive$ denote the code line numbers inside leaf blocks.}
\label{fig:exampletbnreductionblocks}
\end{figure}
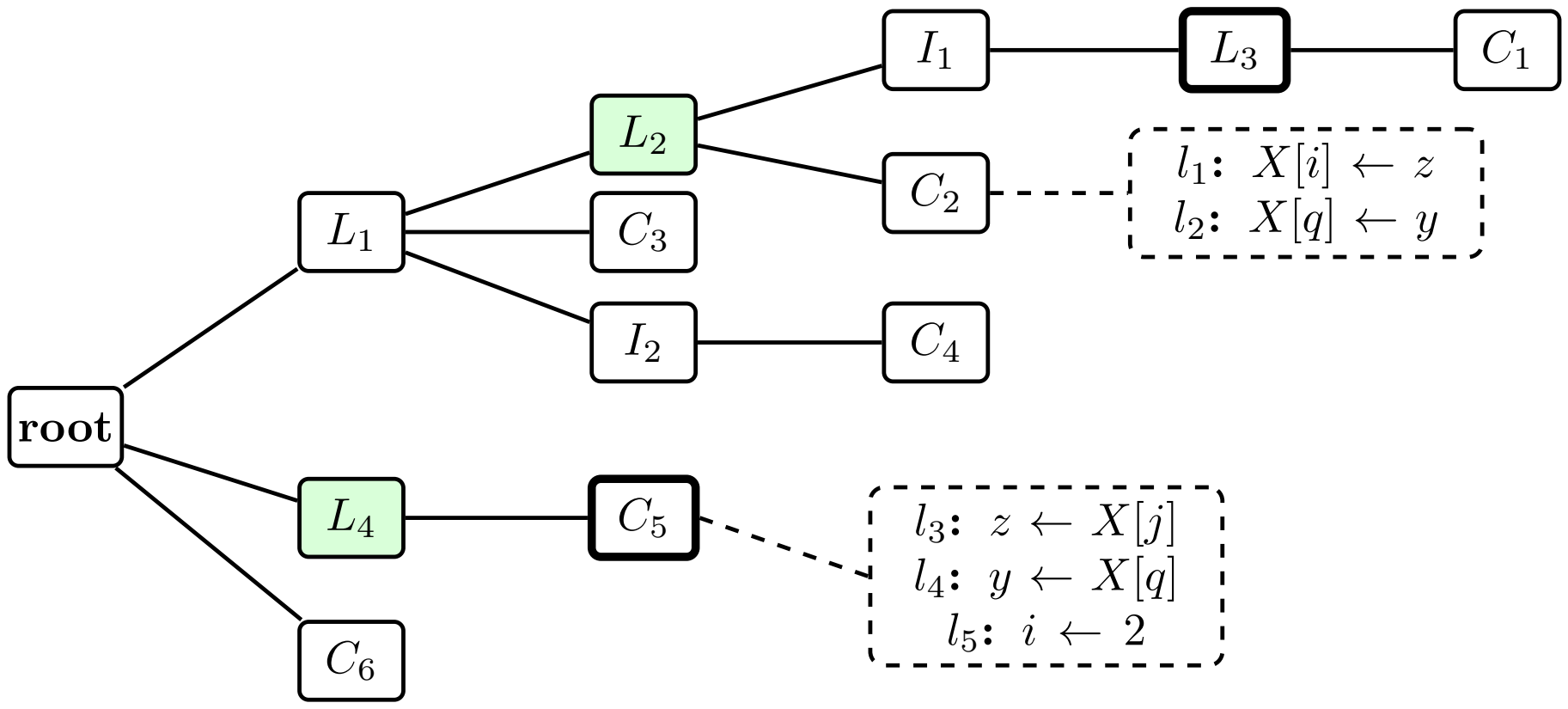

We use the simplified block structure in Figure~\ref{fig:exampletbnreductionblocks} as a running example in the rest of this section. 
In this example, user insight provides two substantial custom \TBn{} reductions. 
In particular, \blLthree{} %\blCtwo{} 
executes only once per iteration of the outermost loop \blLone{}, while \blCfive{} executes only $k$ times within \blLfour{}, where $k \ll \vMi_{\blLfour}$ and $\vMi_{\blLfour}$ is the maximum number of iterations of loop \blLfour{}.

\subsection{Hierarchical Code Decomposition and \TBn{} Calculations}
\label{sec:codeDecompositionTbnCalculations}
\HSBCS{} inherits the default hierarchical code decomposition of \HSB{} on regions that do not have problem-specific \TBn{} reductions.
It recursively decomposes the code until each leaf is either a predefined block or a \emph{sequential block}, that is, a flat block whose control flow advances only to the next line. 
Unlike \HSB{}, however, \HSBCS{} always introduces at least one block, so that other optimizations can be applied over a uniform time-abstraction layer. 
It also allows the user to specify selected regions as flat blocks with custom local \TBn{} values when problem insight yields tighter bounds than the default worst-case analysis.

In the running example of Figure~\ref{fig:exampletbnreductionblocks}, the custom \TBn{} reductions for \blLthree{} and \blCfive{} make \blLtwo{} and \blLfour{} affected blocks that may be flattened by the user. 
This allows the local \TBn{} values of \blLtwo{} and \blLfour{} to capture these reductions directly, namely one execution of \blLthree{} and $k$ executions of \blCfive{}, rather than the structural worst-case bounds $\vMi_{\blLtwo}$ and $\vMi_{\blLfour}$.

Outside such affected regions, \HSBCS{} follows the standard bottom-up \HSB{} recursion for local \TBn{} calculations, proceeding from the leaves to the root, where the \TBn{} of the root becomes the \TBn{} of the entire code.
For sequential blocks, such as \blCfour{} in Figure~\ref{fig:exampletbnreductionblocks}, the local \TBn{} is obtained directly from the number of code lines, since the control flow advances on these lines sequentially. For predefined blocks, which represent the conditional checks and compiler-generated control flow structure of compound statements such as loops and conditionals, the local \TBn{} is determined from the number of simulated time steps required by the corresponding \Asm{} statements as defined in~\cite{avis_sparktope_2022}, together with any relevant code line counts.
At a high level, the local \TBn{} of a loop block \blLone{} is therefore obtained by summing the local \TBn{} values of its children and scaling the resulting per-iteration cost by the iteration bound $\vMi_{\blLone}$. 

At the same time, the compiler annotates blocks with time-related metadata in a single bottom-up pass of the tree for later use during synchronization-aware constraint generation.
These annotations include exact time-offset constants for all blocks and additional symbolic formulas for loop blocks.
At a high level, each block receives an offset within the hierarchy based on its position relative to its preceding siblings and ancestor blocks. Loop blocks are additionally annotated, following the formulations in~\cite{avis_sparktope_2022}, with formulas that, given an iteration index, calculate the time offset at which that iteration begins.

Consequently, in \HSBCS{}, unaffected regions benefit from the LP-size reductions of hierarchical decomposition, while affected regions avoid unnecessary inflation of the overall \TBn{}.
In Figure~\ref{fig:exampletbnreductionblocks}, although \blLtwo{} and \blLfour{} are flattened in \HSBCS{}, unaffected blocks such as \blItwo{} and \blLone{} remain hierarchical. This is particularly beneficial because blocks at the same hierarchical level do not contribute to one another's local \TBn{} values. 
For example, the local \TBn{} values of the root's children---namely \blLone{}, \blLfour{}, and \blCsix{}---are calculated independently rather than cumulatively. The same holds deeper in the hierarchy, for example, among the children of \blLone{}, namely \blLtwo{}, \blCthree{}, and \blItwo{}.

\subsection{Synchronization Across Blocks}
\label{sec:synchronizationHsb}
\HSBCS{} keeps the synchronization mechanism of \HSB{}. 
Using the block annotations, it simulates synchronization barriers by adding idle lines at the end of certain blocks so that, once execution enters a block, the PC remains there for the duration of its local \TBn{}; these intervals are denoted by \ETI{}s.
This makes inter-block transition times statically known as functions of the input size.
For a block nested in $h$ loops, each \ETI{} is determined by a loop index tuple $\Gamma = (\gamma_1, \dots, \gamma_h)$, which is substituted into the annotated formulas for the iteration-start offsets of the enclosing loops, together with the block's annotated constant offset to the root. For example, \blCthree{} in the running example of Figure~\ref{fig:exampletbnreductionblocks} has $\vMi_{\blLone}$ \ETI{}s.

Constraint copies are therefore generated only over the \ETI{}s of enclosing block, while carry-forward constraints for a variable are generated only over the union of the \ETI{}s of the blocks that access it. 
Thus, in the running example, the carry-forward constraints for $i$ are generated over the \ETI{}s of \blLtwo{} and \blLfour{}, which are flattened by the user in \HSBCS{}.

\begin{figure}[!t] 
\centering
\includegraphics [width=1\linewidth]{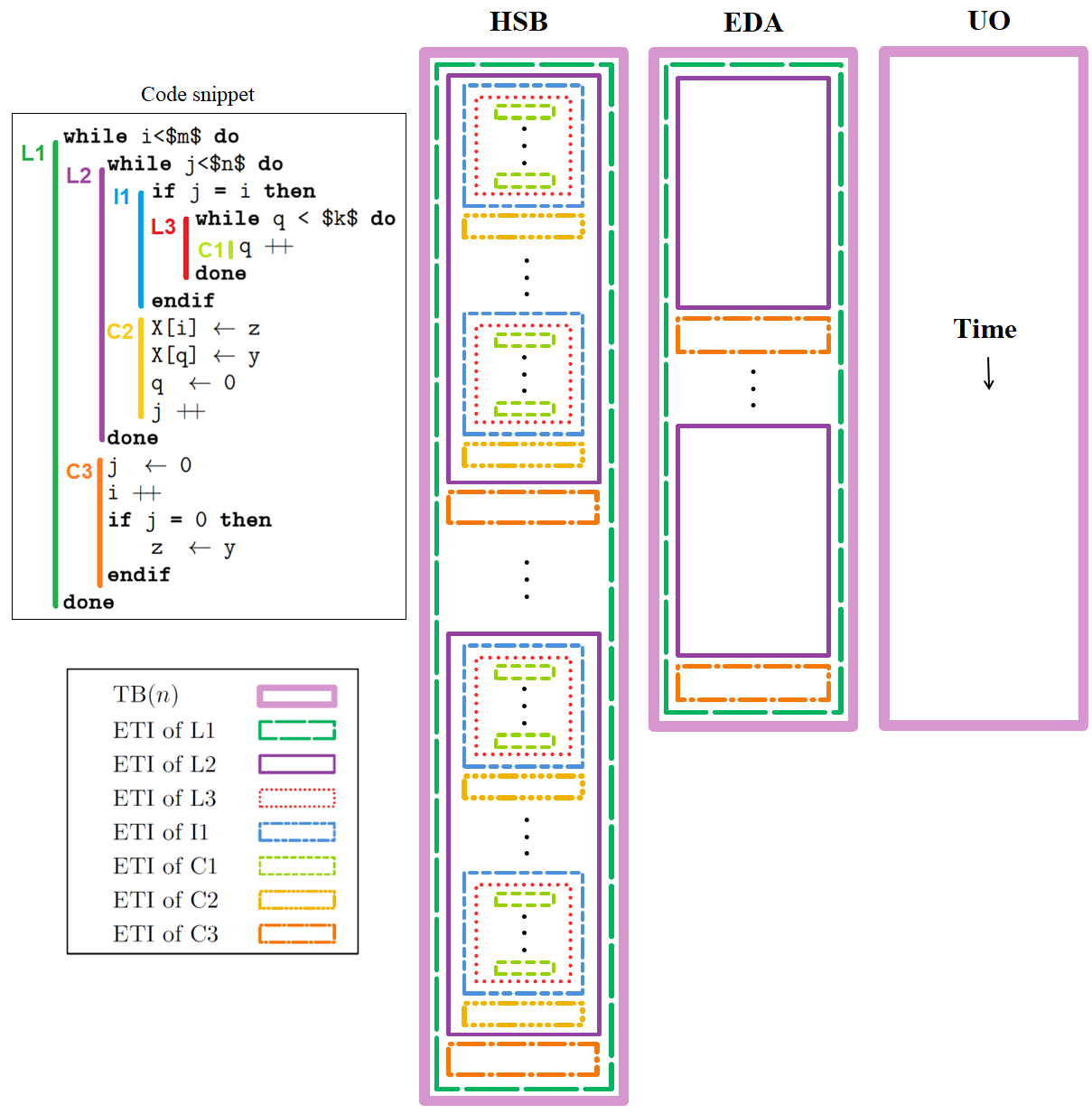} 
\caption{Comparison of the compiler-adjusted control flow under \UO{}, \HSB{} and \HSBCS{} for the simplified \blLone{} block in Figure~\ref{fig:exampletbnreductionblocks}, where \blItwo{} is merged into \blCthree{} and execution returns immediately after the snippet.}
\label{fig:uoHsbEdaComparison}
\end{figure}

Figure~\ref{fig:uoHsbEdaComparison} compares the compiler-adjusted control flow of the \UOlong{}, \HSB{}, and \HSBCS{} for a simplified \blLone{} block, where \blItwo{} is merged into \blCthree{} and execution returns immediately after the snippet. 
\UO{} modifies control flow only to keep the PC in a valid state at each time step up to the global \TBn{}, represented by the single rectangle in the \UO{} panel of Figure~\ref{fig:uoHsbEdaComparison}. Thus, if \blLtwo{} does not execute, the PC jumps directly to \blCthree{}, and once execution reaches the return line, \UO{} keeps the PC there until simulated time reaches \TBn{}.
\HSB{} further modifies the control flow by introducing inter-block synchronization through \ETI{}s, shown as rectangles in the \HSB{} panel, which keep the PC within each entered block for the duration of the corresponding interval. Thus, if \blLtwo{} does not execute, the PC still jumps to the idle line of \blLtwo{} and remains there for the duration of the active \ETI{}, before moving to \blCthree{}. 
\HSBCS{} preserves this synchronization except within the internal structure of affected regions that are flattened by the user, such as \blLtwo{} in Figure~\ref{fig:uoHsbEdaComparison}. Although flattening such a region into a non-sequential block increases the number of constraint copies for statements inside it---from one copy per block \ETI{} in sequential blocks under \HSB{} to a number proportional to the flat block's local \TBn{} times its number of \ETI{}s---unaffected blocks still benefit substantially from synchronization. Moreover, this quantity is typically much smaller than the global \TBn{} in \UO{}.

The current carry-forward implementation uses a moderately optimized strategy, propagating values across each \ETI{} and adding an extra variable version immediately before each interval, to avoid restrictions on other optimizations that require access to variable versions across multiple time steps.

\subsection{Localizing Other Optimizations}
Beyond supporting custom local \TBn{} values, \HSBCS{} provides a synchronization-based framework for localized reductions within eligible blocks. 
In this paper, \CS{} (Section~\ref{sec:backgroundLpCompilerOptimizations}) serves as a concrete example of such localized optimizations.

Within \HSBCS{}, \CS{} is applied independently to each eligible block, namely each flat block with at least two code lines. 
Each such block is treated as an independent program for identifying identical and operator-equivalent statements, array operations, relevant array indices, and placeholder allocations without explicitly managing inter-block timing or synchronization.

In Figure~\ref{fig:exampletbnreductionblocks}, \CS{} is applied independently to \blLtwo{} and \blLfour{}. 
Placeholder substitutions for array $X$ are therefore generated only over the \ETI{}s of \blLtwo{} for assignment statements and over the \ETI{}s of \blLfour{} for reference statements. 
In \blLfour{}, \CS{} restricts $X$ to the LP constraints for a single array-reference assignment to the boolean placeholder $\vPhOne{}$, using row-index placeholder substitutions $\vRow{}=j$ and $\vRow{}=q$, so that $X$ appears only in the constraints for $\vPhOne{} = X[\vRow{}]$. 
The constraints use aggregated $S$ controllers for lines $\lineThree$ and $\lineFour$, activated whenever the implicit PC is on either line. 
All placeholder substitutions occur within one simulated time step by exploiting the simultaneous satisfaction of LP constraints. 
Likewise, placeholder substitutions for the integer variable $i$ are generated only over the \ETI{}s of \blLtwo{}, since $i$ appears as an array index only there.

The current framework does not consider cross-block constraint sharing. Although sub-blocks in different conditional branches may have limited \ETI{} overlap, exploiting it would require constraint generating over \ETI{} unions for only modest benefit.

\section{Results}
\label{sec:results}
This section evaluates the effect of the LP compiler optimization techniques \HSBlong{}~\cite{khosravi_efficient_2025} and \CSlong{}~\cite{khosravi_smaller_2026} on LP size, solver performance, solution accuracy, and numerical conditioning across a range of fundamental combinatorial algorithm benchmarks (Table~\ref{tab:benchmark-config}), including benchmarks with exponential extension complexity. 
We compare these methods against the \UOlong{}, namely the \Sparktope{} compiler of~\cite{david_avis_sparktope_2020} on which our implementation is based.

We then examine the effectiveness of \HSB{} on benchmarks with custom \TBn{} reductions derived from user insight, for which the structural worst-case estimates underlying \HSB{}'s synchronization strategy may no longer be accurate. 
We focus in particular on \MMlong{}, which has exponential extension complexity and sufficiently large \TBn{} reductions to make \HSB{} ineffective. 
On this benchmark, we evaluate the proposed \HSBCSlong{} framework and compare its results against \POlong{}, introduced in~\cite{avis_sparktope_2022}, which eliminates redundant constraints using user-provided execution-time ranges.

Throughout this section, our discussion focuses primarily on the main cross-benchmark trends and percentage changes, while detailed tables for each benchmark are provided in the appendix. Consulting these tables is not necessary for following the results. Nevertheless, we reference them where relevant to help readers who wish to examine the benchmark results in more detail.

\subsection{Experimental Resources}
All experiments were performed on a system running Debian Linux 12.13 and equipped with Intel(R) Xeon(R) Gold 6248 CPUs running at 2.50 GHz. 
Unless specified otherwise, all experiments used the default mode of \scip{} 9.2.1, \gurobi{} 12.0.1, and \cplex{} 22.1.1.0.
Benchmark-specific memory and CPU allocations, as well as the number of runs, are summarized in Table~\ref{tab:benchmark-config}. Each solver run was subject to a 23-hour time limit.

\begin{table}[!b]
\centering
\caption{Benchmark configurations, specifying the amount of memory, the number of runs, and the number of \CPU{} allocations. The time limit is 23 hours for all cases.}
\label{tab:benchmark-config}
\begin{tabular}{c c c c}
\toprule
\makecell{benchmark} &
\makecell{runs} &
\makecell{memory\\(\GB)} &
\makecell{\CPU{}} \\
\midrule
\MSfull{} (\MS{})   & 10 & 32  & 7 \\
\MSTfull{} (\MST{})  & 10 & 32  & 7 \\
\BSfull{} (\BS{})   & 5  & 32  & 7 \\
\KSfull{} (\KS{})   & 5  & 64  & 8 \\
\SPDJfull{} (\SPDJ{}) & 5  & 64  & 8 \\
\SAMfull{} (\SAM{})  & 5  & 64  & 8 \\
\TSfull{} (\TS{})   & 5  & 64  & 8 \\
\LISfull{} (\LIS{})  & 5  & 64  & 8 \\
\LISLfull{} (\LISL{}) & 5  & 64  & 8 \\
\HMMfull{} (\HMM{})  & 3  & 128 & 8 \\
\MMfull{} (\MM{})  & 3  & 372 & 8 \\
\bottomrule
\end{tabular}
\end{table}

\subsection{Linear Program Size Reduction}
\label{sec:resultLpSize}
In this section, we examine the LP sizes generated by \UO{}, \CS{}, and \HSB{} across all benchmarks. 
We discuss overall changes in LP size; detailed tables for each benchmark are given in Tables~\ref{tab:lp-sizes-spdj} to~\ref{tab:lp-sizes-lis} in Appendix~\ref{sec:appendix-lp-size}.
We then focus on \MM{} and its variant, \HMMlong{}, comparing the LPs generated by \PO{}, \CS{}, \HSB{}, and \HSBCS{}.

\begin{figure}[!t]
\centering
\includegraphics[width=1\columnwidth]
{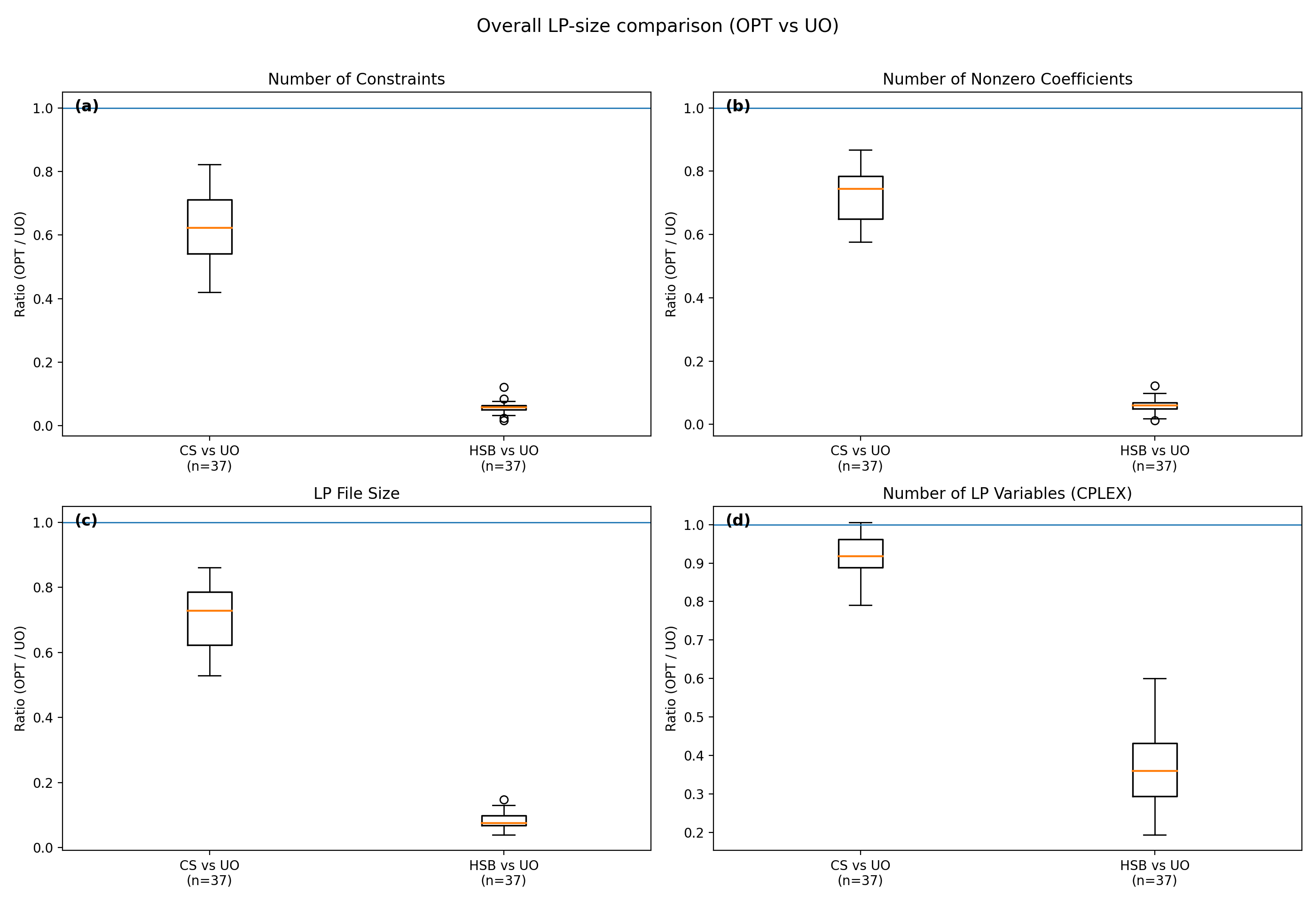}
\caption{
Overall LP-size comparison across all benchmarks (excluding \MM{} variants) and input sizes.
Each plotted value is $\OPT/\UO$ for the corresponding metric, where \OPT{} denotes the optimization method being compared against \UO{}. 
Panels show
(a) number of constraints, 
(b) number of nonzero coefficients, 
(c) LP file size, 
and (d) number of LP variables reported by \cplex{}, which excludes variables appearing only in bound constraints.
Values below 1 indicate improvement relative to \UO{}.
}
\label{fig:lpsize_ratio_overall}
\end{figure}

Figure~\ref{fig:lpsize_ratio_overall} summarizes the overall relative changes in LP size, measured by the number of rows, columns, nonzeros, and LP file size. 
For most benchmark instances, both \CS{} and \HSB{} consistently reduce LP size relative to \UO{}, with the largest reductions in the number of constraints (Figure~\ref{fig:lpsize_ratio_overall}(a)).

\CS{} shows stronger reductions on benchmarks with more code duplication, such as \TSlong{} (Table~\ref{tab:lp-sizes-ts}), particularly those having array operations, such as \SAMlong{} (Table~\ref{tab:lp-sizes-sam}), since these provide more constraint sharing opportunities. 
This dependence on code structure and variation in the amount of array operations across benchmarks produces a wider spread for \CS{} than for \HSB{} in (Figure~\ref{fig:lpsize_ratio_overall}(a)), where \HSB{} is more tightly concentrated around its median except in the number of columns (Figure~\ref{fig:lpsize_ratio_overall}(d)).
Although \CS{} reduces the number of columns for some benchmarks, such as \TS{} with a $21\%$ reduction (Table~\ref{tab:lp-sizes-ts}),
it achieves most of its gains from row reductions, ranging from $17.7\%$ for \BSlong{} (Table~\ref{tab:lp-sizes-bs}) to $58\%$ for \SAM{} (Table~\ref{tab:lp-sizes-sam}); for \SAM{}, this corresponds to a $47.1\%$ reduction in file size relative to \UO{}, from $47$~\GB{} to $25$~\GB{}.
These larger column reductions occur mainly on benchmarks with many front-end generated temporary variables. \CS{} reduces them by reusing temporaries via a simple linear-scan allocation within basic blocks, increasing statement similarity before placeholder allocation.
Accordingly, small benchmarks, particularly those with few or no array sharing opportunities, benefit least from \CS{}, even in cases like \BS{} (Table~\ref{tab:lp-sizes-bs}), which has the largest array sizes.

Overall, \HSB{} achieves larger LP size reductions than \CS{}, as reflected by its lower medians in Figure~\ref{fig:lpsize_ratio_overall}. 
In general its gains increase with the number of loops in different subtrees of the decomposition, as in \MST{} (Table~\ref{tab:lp-sizes-mst}), since such loops do not contribute to one another's local \TBn{}, and with their proximity to the root, because they split larger \TBn{} values across larger descendant regions and suffer less from ancestor-induced multiplicity. Accordingly, benchmarks such as \KSlong{} (Table~\ref{tab:lp-sizes-kd}), whose code consists of a single shallow nested loop show the smallest LP size reductions.

Although \HSB{} achieves substantial reductions in both rows and columns, row reductions are more consistent across benchmarks---ranging from $87.8\%$ to $98.3\%$---while column reductions are more variable---ranging from $39.9\%$ to $80.6\%$, resulting in a wider range in Figure~\ref{fig:lpsize_ratio_overall}(d).

\begin{figure}[!t]
\centering
\includegraphics[width=1\columnwidth]{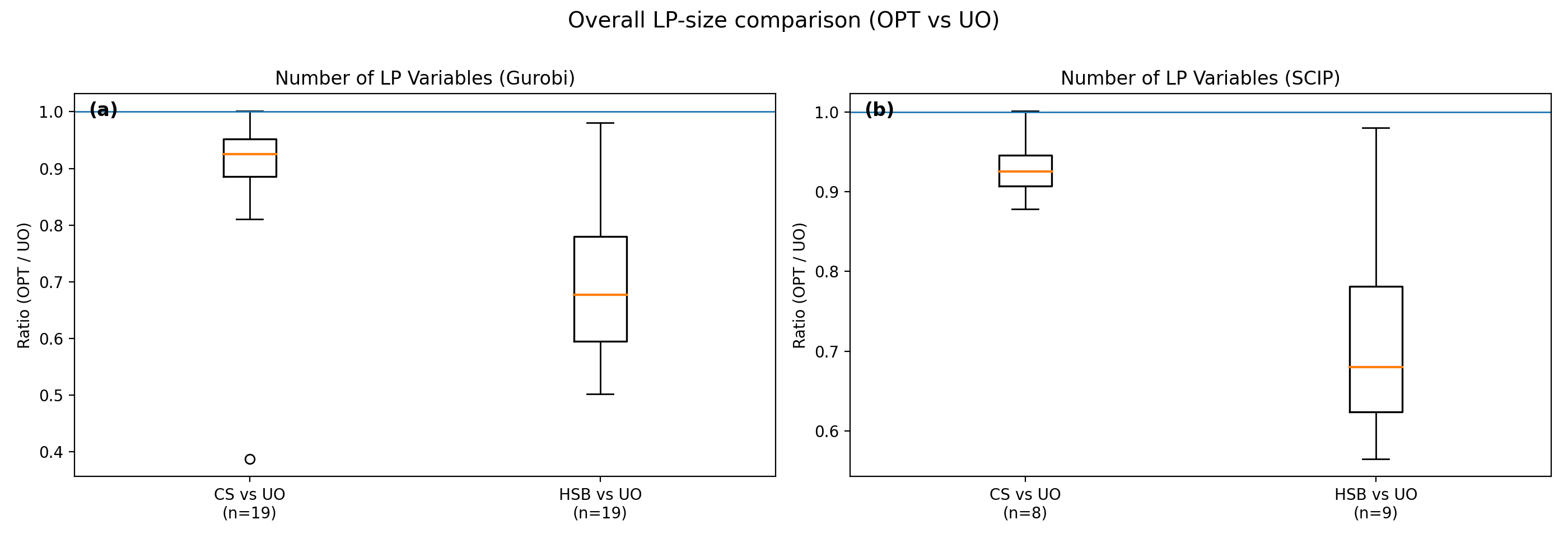}
\caption{Supplementary solver-specific comparison of LP variable counts across benchmarks (excluding \MM{} variants) and input sizes, for the subset of instances with counts available from solver logs. 
Each plotted value is $\OPT/\UO$, where \OPT{} denotes the optimization method being compared against \UO{}.
Panels show the number of LP variables reported by
(a) \gurobi{} and 
(b) \scip{}. 
Unlike \cplex{} in Figure~\ref{fig:lpsize_ratio_overall}(d), both solvers count variables that appear only in bound constraints as columns.
Values below 1 indicate improvement relative to \UO{}.}
\label{fig:lpsize_ratio_overall_gurscip_columns}
\end{figure}

For Figure~\ref{fig:lpsize_ratio_overall}(d), we report column counts from \cplex{}. Unlike \gurobi{} and \scip{}, \cplex{} excludes variables that appear only in bound constraints from its problem-size statistics and solution output.
By contrast, \gurobi{} and \scip{} count these variables as matrix columns, assign them zero coefficients in the objective function~\cite{eli_towle_remove_2019}, and return values for them in the solution.
Figure~\ref{fig:lpsize_ratio_overall_gurscip_columns} highlights this difference for the subset of LP instances with recorded \gurobi{} and \scip{} column counts from solver logs of jointly solved instances, so it covers fewer than \lpnum{} instances. 
This also explains why the \HSB{}/\UO{} column ratio boxplots are higher and wider for \gurobi{} (Figure~\ref{fig:lpsize_ratio_overall_gurscip_columns}(a)) and \scip{} (Figure~\ref{fig:lpsize_ratio_overall_gurscip_columns}(b)) than for \cplex{} (Figure~\ref{fig:lpsize_ratio_overall}(d)), even though they refer to the same \HSB{} models, and \cplex{} results are computed over all \lpnum{} LP instances, whereas the \gurobi{} and \scip{} are based only on jointly solved subset of instances with available recorded column counts. 
We report \cplex{} counts in Figure~\ref{fig:lpsize_ratio_overall}(d) because the discrepancy is due to our implementation, not the optimization methods, and could be removed by a small change to bound-constraint generation or a simple post-processing pass over LPs.

\begin{figure}[!b] 
\centering
\includegraphics [width=0.8\linewidth]{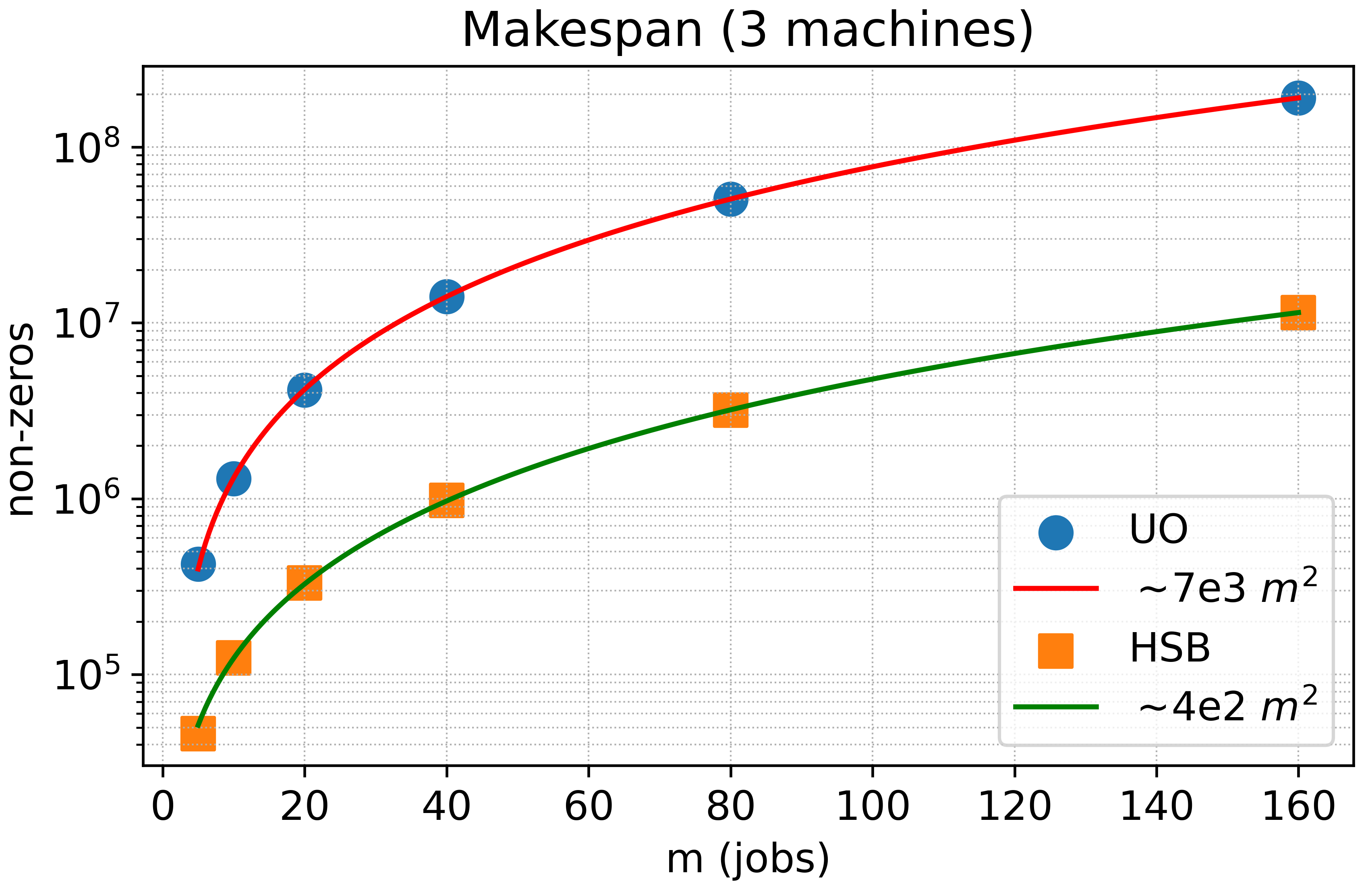} 
\caption{Number of nonzeros in LPs of \UO{} and \HSB{} for the makespan problem with $n=3$ machines.}
\label{fig:funcfit-nonzeros-makespan}
\end{figure}

Figure~\ref{fig:funcfit-nonzeros-makespan}, which fits a quadratic function to the number of nonzero elements for \HSB{} and \UO{} on \MSlong{}, indicates that \HSB{} achieves substantial reductions while preserving the same asymptotic growth rate. This is consistent with compiler optimizations and LP presolve techniques, which typically reduce constant factors rather than asymptotic complexity.

For benchmarks where custom \TBn{} reductions derived from user insight are minor, such as those in \LISlong{} (Table~\ref{tab:lp-sizes-lis}) and \MS{} (Table~\ref{tab:lp-sizes-ms}), \HSB{} still generates substantially smaller LPs than both \UO{} and \CS{}. For example, \HSB{} achieves $94\%$ row and $50\%$ to $78.6\%$ column reductions compared to \CS{} achieving only $22.4\%$ to $34.2\%$ row and $0.4\%$ to $8.6\%$ column reductions.

However, when the custom \TBn{} reductions become significant, \HSB{} fails to improve LP size.
This is the case for \MM{} where in addition to several minor \TBn{} reductions, it has a major one in the blossom algorithm implementation of~\cite{avis_sparktope_2022}, the shrinking operation executes only once per outermost loop iteration. As \HSB{} fails by producing larger LPs than \PO{}, we omit it for benchmark \MM{} in Table~\ref{tab:lp-sizes-mm}.

In contrast, \CS{} benefits from the reduced global \TBn{} and achieves substantial LP size reductions relative to \PO{}, ranging from $55.5\%$ to $62.9\%$ in the number of rows and up to $53.1\%$ in LP file size. To keep solver experiments practical, we use the same input-dependent \TBn{} values as in~\cite{avis_sparktope_2022}. These yield smaller LPs, but they are only valid for inputs of size $n$ whose execution terminates within the chosen \TBn{}. 
Estimated true \TBn{} values for all inputs of size $n$, together with corresponding LP file-size estimates, are shown in parentheses.

To address the shortcomings of \HSB{} on such benchmarks while still exploiting structural gains from the code, we evaluate \HSBCS{} on \MM{}.
In the \HSBCS{} hierarchy, the affected loop structure in \MM{} (Loop B in~\cite{avis_sparktope_2022}) is manually flattened and assigned a user-provided input-dependent local \TBn{} matching the input-dependent reduction of the global \TBn{}. These local \TBn{} values, listed in the caption of Table~\ref{tab:lp-sizes-mm}, were obtained empirically as the maximum number of execution steps of that block over the tested inputs. \HSBCS{} then computes the overall \TBn{} automatically for the entire hierarchy, which is why the final \TBn{} values are not rounded.
As shown in Table~\ref{tab:lp-sizes-mm}, \HSBCS{} generates smaller models than \CS{}, with row reductions of $25.2\%$ to $73.2\%$ relative to \PO{}. 
Applying localized \CS{} within the flat block, denoted \HSBCS{}$^\dagger$, improves this further to $65.8\%$ to $85.7\%$ row reductions and $58.6\%$ to $82.9\%$ file-size reductions relative to \PO{}.

Finally, \MM{} also illustrates that for some benchmarks, the custom \TBn{} reduction issue can be alleviated by a simple source-level modification.  
In the \HMM{} variant (Table~\ref{tab:lp-sizes-mm}), the blossom-shrinking operation in the implementation of~\cite{avis_sparktope_2022} is \emph{hoisted}---that is, moved out of its current loop to the outermost loop---with an extra condition introduced to preserve correctness.
This restores the effectiveness of \HSB{}, which achieves substantial row reductions of $87\%$ to $94.4\%$, nonzero reductions of $82.9\%$ to $89.5\%$, and file-size reductions of $78.2\%$ to $87.3\%$ relative to \PO{}.
Importantly, these gains are obtained using the actual \TBn{}, whereas \PO{}, \CS{}, and \HSBCS{} variants use artificially reduced \TBn{} values. Thus, for \HMM{}, \HSB{} not only generates the smallest model, but also one that can be solved for any input of size $n$ without recompilation. 
In addition, the LP size reduction is not a consequence of the source code modifications, as \HMM{} contains extra lines added to the \MM{} code rather than removing them. In fact, the \PO{} LPs for \HMM{} are slightly larger than those for \MM{}. 
In addition, \HSB{} for \HMM{} remains within a stricter memory limit of 128~\GB{} compared with 372~\GB{} for \PO{}. 
Nevertheless, such algorithmic modifications are not available in every benchmark, which is why \HSBCS{} remains the more generally applicable solution. 
For \HSB{}, file-size reductions are typically smaller than row reductions because some of its rows are denser and many variables appear only in bound constraints. Accordingly, nonzero reductions are smaller than row reductions, indicating that the \HSB{} rows have more coefficients on average. 
In \HMM{}, the much larger true \TBn{} also creates extra variable versions, some appearing only in bound constraints, which increases file size without increasing rows~\cite{ibm_ibm_2025}.
The larger \TBn{} also makes certain base compiler equality constraints denser, particularly those that ensure the PC is on exactly one line at each time step, since each such equality constraint contains \TBn{} variables. By comparison, \PO{} for \MM{} appears less dense only because its \TBn{} is artificially reduced. Although \HSB{} introduces extra summation equalities to assign zero to redundant $S$ controllers in its model compared to \PO{}, this is not necessarily a disadvantage of the method, since all of those variables are fixed to zero in presolve.

\begin{table}[!t]
\centering
\caption{LP sizes for the \MMlong{} problem, using the algorithm from~\cite{avis_sparktope_2022}, and its hoisted variant \HMM{}, with input size $n$.
For \HMM{}, LPs are generated by \HSB{}; 
for \MM{}, LPs are generated by \PO{}, \CS{}, and \HSBCS{}. 
The $^\dagger$ symbol indicates applying the \CS{} optimization within the manually flattened blocks of \HSBCS{}.
The \TBn{} values of the manually flattened block in the \HSBCS{} variants are 700, 1,000, 1,800, and 2,700 for the input sizes, respectively. 
}
\label{tab:lp-sizes-mm}
\resizebox{\columnwidth}{!}{%
\begin{tabular}{ccccccccc}
\toprule
\makecell{$n$} & 
\makecell{benchmark} &
\makecell{\TBn{}} & 
\makecell{\OPT{}} &
\makecell{rows\\$(\times 1000)$} & 
\makecell{columns\\$(\times 1000)$} & 
\makecell{nonzeros\\$(\times 1000)$} & 
\makecell{file\\(\GB{})} \\
\midrule
\multirow{5}{*}{8}
 & \multirow{4}{*}{\MM{}} 
 & \multirow{2}{*}{4,000 (9,747)}   & \PO{}  & 25,334 & 2,856 & 95,125 & 1.81 (3.4) \\ 
 &        &                          & \CS{}   & 11,273 & 2,068 & 54,430 & 0.97 \\
\cmidrule(lr){3-8}
& 
 & \multirow{2}{*}{2,244}   & \HSBCS{}   & 6,800 & 1,112 & 25,592 & 0.50 \\
 &         &                 & \HSBCS{}$^\dagger$  & \textbf{3,632} &  \textbf{915}&  \textbf{16,439} & \textbf{0.31} \\
 \cmidrule(lr){2-8}
  & \HMM{} & 10,748                          & \HSB{} &  1,423 & 3,835 & 10,017 & 0.23 \\ 
\midrule
\multirow{5}{*}{10}
& \multirow{4}{*}{\MM{}} 
 & \multirow{2}{*}{7,000 (19,629)} & \PO{} &  54,809 &  5,355 & 210,573 & 3.98 (11)\\ 
 &      &                             & \CS{} & 22,977 & 3,934 & 115,302 & 2.05 \\
 \cmidrule(lr){3-8}
 & & \multirow{2}{*}{4,396}  & \HSBCS{}   & 17,786 & 2,420 & 67,914 & 1.33 \\
 & &                        & \HSBCS{}$^\dagger$   & \textbf{9,034} & \textbf{1,999} & \textbf{41,749} & \textbf{0.79} \\
 \cmidrule(lr){2-8}
   & \HMM{}  & 23,962 & \HSB{} & 3,756 & 8,943 & 24,395 & 0.58 \\ 
\midrule
\multirow{5}{*}{12}
& \multirow{4}{*}{\MM{}} 
  & \multirow{2}{*}{10,000 (34,771)} & \PO{} & 94,861 & 8,200 & 371,214 & 7.01 (23) \\ 
 &          &                         & \CS{} & 38,026 & 6,110 & 196,911 & 3.48 \\
  \cmidrule(lr){3-8}
  & &  \multirow{2}{*}{9,284} & \HSBCS{}   & 51,011 & 5,833 & 197,132 & 3.85 \\
 & &                     & \HSBCS{}$^\dagger$  & \textbf{24,827} & \textbf{4,824} & \textbf{116,880} & \textbf{2.19} \\
 \cmidrule(lr){2-8}
   & \HMM{}   & 44,596 & \HSB{} & 8,203 & 17,404 & 49,743 & 1.20 \\ 
\midrule
\multirow{5}{*}{16}
& \multirow{4}{*}{\MM{}} 
 & \multirow{2}{*}{16,000 (83,003)} & \PO{} &  238,577 & 15,296 & 955,446 & 18.49 (80)  \\ 
 &  &                               & \CS{}  & 88,634 & \textbf{12,448} & 483,698 & 8.67 \\
   \cmidrule(lr){3-8}
 &  &   \multirow{2}{*}{20,068} & \HSBCS{}   & 178,552 & 17,257 & 703,777 & 14.19 \\
&  &                            & \HSBCS{}$^\dagger$   & \textbf{81,657} & 13,377 & \textbf{398,839} & \textbf{7.67} \\
 \cmidrule(lr){2-8}
  & \HMM{}  & 114,556 & \HSB{} & 31,109 & 51,963 & 163,490 & 4.04 \\ 
\bottomrule
\end{tabular}}
\end{table}

\subsection{Solver Performance}
\label{sec:resultSolverPerformance}
This section evaluates how the LP size reductions achieved by \HSB{} and \CS{} affect solver performance relative to \UO{}.
We report the medians of elapsed time, presolve time, and \RSSlong{} memory usage across jointly solved benchmark instances, computed over the number of runs specified in Table~\ref{tab:benchmark-config}. 
 
\begin{figure}[!t]
\centering
\includegraphics[width=1\columnwidth]{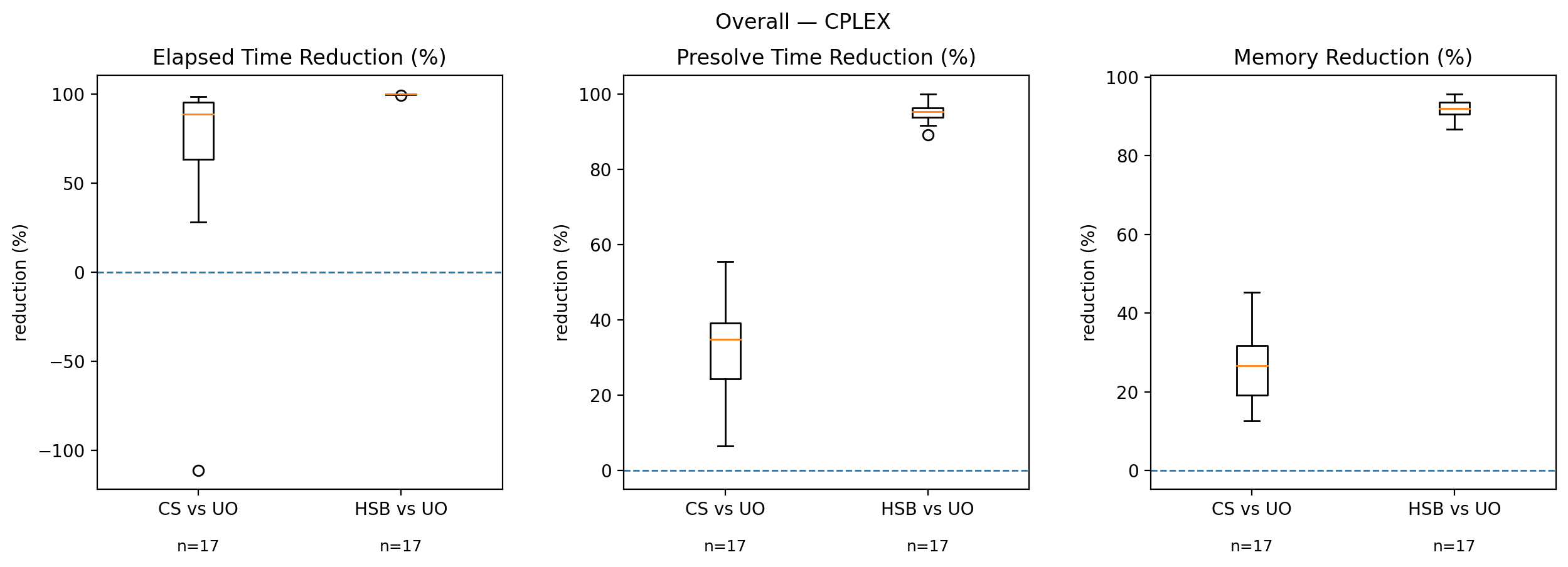}
\caption{Overall percentage reductions in elapsed time, presolve time, and memory usage for \HSB{} and \CS{} relative to \UO{}, as reported by \cplex{}. Positive values indicate improvement relative to \UO{}.}
\label{fig:solver_performance_reduction_overall_cplex}
\end{figure}

\begin{figure}[!b]
\centering
\includegraphics[width=1\columnwidth]{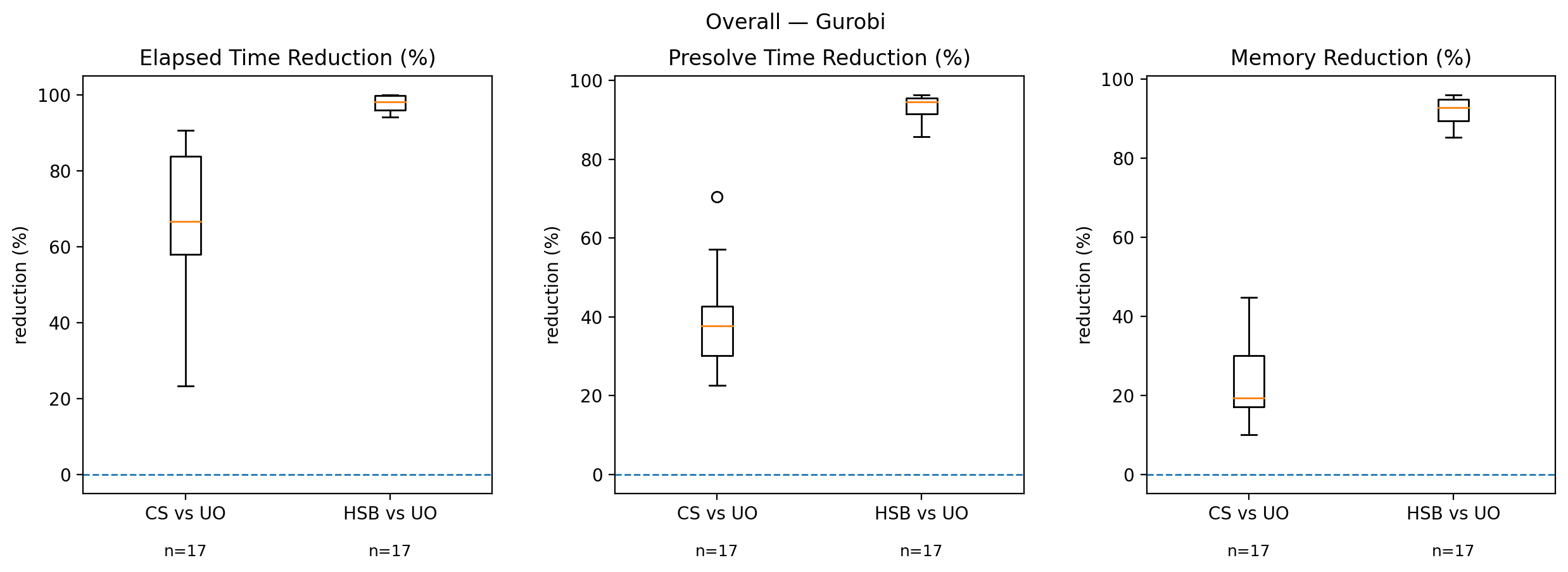}
\caption{Overall percentage reductions in elapsed time, presolve time, and memory usage for \HSB{} and \CS{} relative to \UO{}, as reported by \gurobi{}.
Positive values indicate improvement relative to \UO{}.}
\label{fig:solver_performance_reduction_overall_gurobi}
\end{figure}

\begin{figure}[!t]
\centering
\includegraphics[width=1\columnwidth]{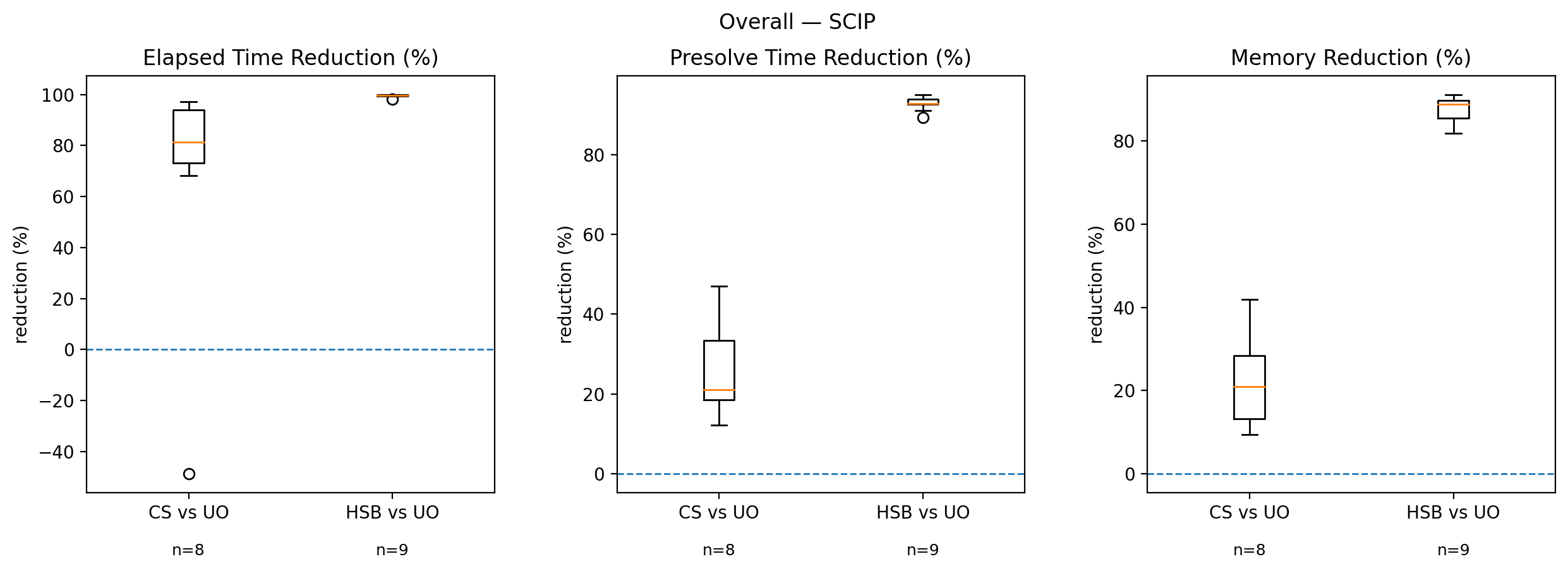}
\caption{Overall percentage reductions in elapsed time, presolve time, and memory usage for \HSB{} and \CS{} relative to \UO{}, as reported by \scip{}.
Positive values indicate improvement relative to \UO{}.}
\label{fig:solver_performance_reduction_overall_scip}
\end{figure}

\begin{figure}[!b]
\centering
\includegraphics[width=1\columnwidth]{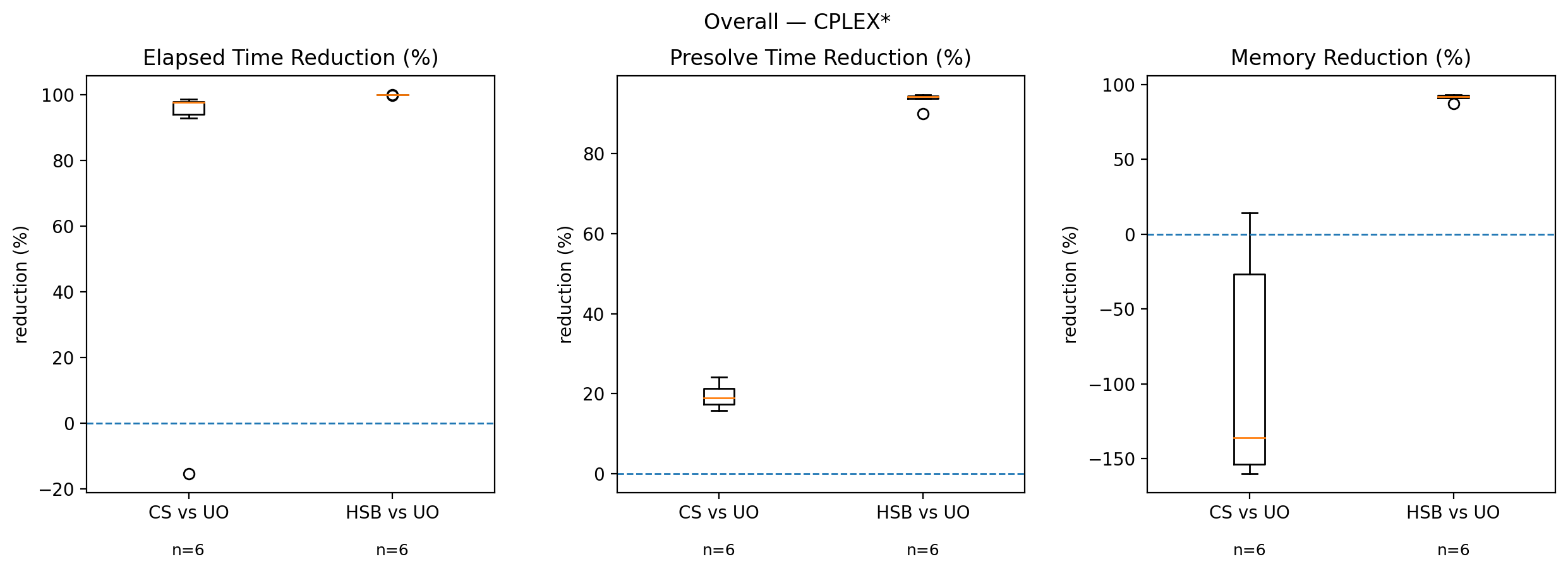}
\caption{Overall percentage reductions in elapsed time, presolve time, and memory usage for \HSB{} and \CS{} relative to \UO{}, as reported by \cplex{}$^*$.
Positive values indicate improvement relative to \UO{}.}
\label{fig:solver_performance_reduction_overall_cplexStar}
\end{figure}

Figures~\ref{fig:solver_performance_reduction_overall_cplex},~\ref{fig:solver_performance_reduction_overall_gurobi},~\ref{fig:solver_performance_reduction_overall_scip}, and~\ref{fig:solver_performance_reduction_overall_cplexStar} summarize the percentage reductions over these jointly solved instances, while Tables~\ref{tab:time-spdj-notforced} to~\ref{tab:time-ms-notforced} in Appendix~\ref{sec:appendix-solver-performance} provide detailed per-benchmark results for all LP instances solved by any optimization method.
For maximum matching, we separately compare \MM{} models generated by \CS{} and the \HSBCS{} variants relative to \PO{}, and we discuss \HSB{} for the hoisted variant \HMM{}.

Both \CS{} and \HSB{} improve solver performance across all tested solvers. However, \HSB{} achieves larger and more consistent gains, as shown by the larger median reduction percentages in the boxplots. The results indicate that both methods generate not only smaller LPs, but also models that are easier to solve.

\begin{figure}[!b]
\centering
\includegraphics[width=0.8\columnwidth]{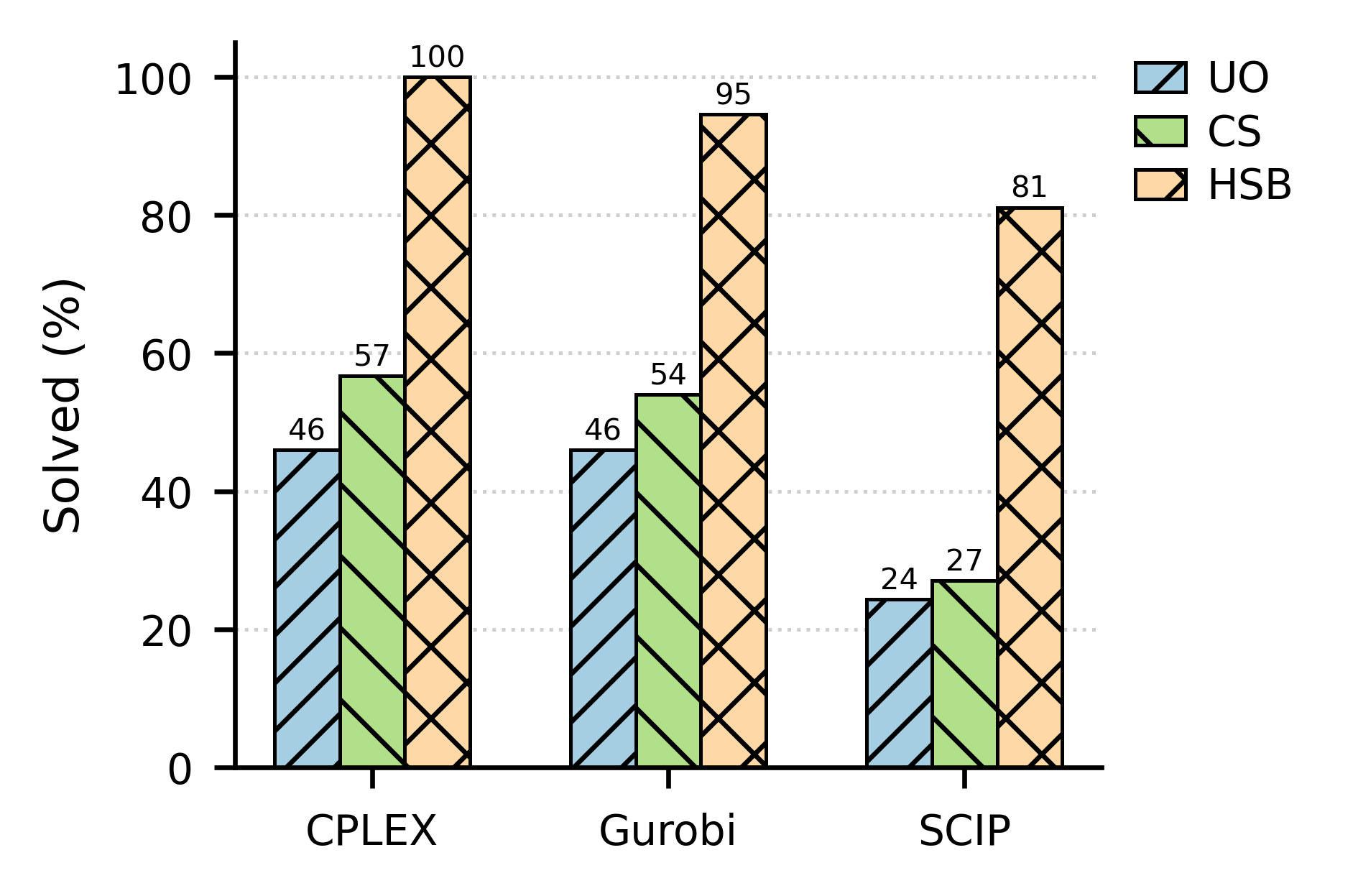}
\caption{Percentage of benchmark LP instances solved by \cplex{}, \gurobi{} and \scip{} for LPs generated by \UO{}, \CS{} and \HSB{}, excluding \MM{} and \HMM{} benchmarks. The total number of benchmark instances is \lpnum{}.}
\label{fig:percentageOfSolvedBenchesByOptAndSolver}
\end{figure}

Furthermore, \HSB{} significantly increases coverage, making more problem instances solvable within the set resource limits compared to \CS{} and \UO{}, as shown in Figure~\ref{fig:percentageOfSolvedBenchesByOptAndSolver}. 
Particularly, \cplex{} slightly outperforms \gurobi{}, solving $100\%$ of \HSB{} instances, compared to $56.8\%$ of \CS{} and $46\%$ of \UO{}. 
\gurobi{} occasionally fails on smaller problem instances compared to \cplex{}, by solving $94.6\%$ of \HSB{}, $54.1\%$ of \CS{}, and $46\%$ of \UO{} instances, mostly due to exceeding the memory limitation specified in Table~\ref{tab:benchmark-config}. 
\scip{}, which is a non-commercial solver, performs worst among solvers, mostly due to exceeding time limits. However, it also manages to solve $81.1\%$ of \HSB{} instances compared to only $27\%$ and $24.3\%$ of \CS{} and \UO{} instances, respectively.

The \IQRlong{} reported in detailed tables in Appendix~\ref{sec:appendix-solver-performance} is reported as a percentage of the median. Most \IQR{} values are below $30\%$, indicating that the solver performance is generally stable across repeated runs. From $681$ reported \IQR{} values, only $13$ exceed this threshold, including $8$ for presolve time, $3$ for elapsed time, and $2$ for memory usage, all arising from \CS{} and \HSB{} runs. 

\CS{} also generally improves solver performance but less consistently than \HSB{}. This is evident from its wider boxplots, which indicate a larger spread in reduction percentages across benchmarks. 
It mainly achieves considerable reductions in elapsed time, with up to $98.8\%$ for \TS{}. 
However, for a few cases the reductions are smaller, or even negative for the case of smallest input size of \KS{} instance, where \cplex{} is $111.1\%$ slower, increasing from $5.4$ minutes for \UO{} to $11.4$ minutes for \CS{} (appearing as a single negative outlier in Figures~\ref{fig:solver_performance_reduction_overall_cplex},~\ref{fig:solver_performance_reduction_overall_scip}, and~\ref{fig:solver_performance_reduction_overall_cplexStar}).

Similarly, \CS{} mainly reduces memory usage relative to \UO{}, with best-case reductions ranging from $10.3\%$ for \KS{} to $45.3\%$ for \SAM{}. 
Solver performance also depends on whether the solver is restricted to a fixed LP algorithm or uses its default mode. For \cplex{} and \gurobi{}, the default mode runs three LP algorithms---primal simplex, dual simplex, and barrier---concurrently, reducing the memory available to each. 
On \MS{}, \cplex{}$^*$, which is restricted to dual simplex, increases the memory usage by $160\%$, contributing to \CS{}'s wider boxplot range in Figure~\ref{fig:solver_performance_reduction_overall_cplexStar}, and also substantially increases the elapsed time for \UO{}, indicating that dual simplex is less favorable on some \MS{} LP instances. However, on the smallest \UO{} instances, the default modes of \cplex{} and \gurobi{} exceed the memory limit because they run three algorithms in parallel, whereas \cplex{}$^*$ solves them by running only one. 
By contrast, \scip{} delegates LP solving to \soplex{}, whose default switches between primal and dual simplex rather than running them in parallel.

Unlike \CS{}, whose efficiency varies more with code duplication and size, \HSB{} remains consistently strong across all solvers and benchmarks relative to \UO{}. 
It reduces elapsed time by $94.2\%$ (Table~\ref{tab:time-kd-notforced}) to $100\%$ on multiple instances, and memory usage by $81.9\%$ (Table~\ref{tab:time-bs-notforced}) to $95.9\%$ (Table~\ref{tab:time-mst-notforced}). 
Moreover, whereas \CS{} shows negligible gains on some benchmarks, \HSB{} almost uniformly dominates \UO{}, aside from a few presolve-time ties, consistent with its stronger LP-size reductions.

Table~\ref{tab:time-mm-notforced} reports solver performance for \MM{} models generated by \PO{}, \CS{}, and \HSBCS{} variants using artificially reduced \TBn{} values (as described in Section~\ref{sec:resultLpSize}). 
Despite these artificial \TBn{} reductions, all solvers fail on \PO{}, whereas \CS{} remains solvable by \cplex{} for all tested input sizes. 
\HSBCS{} further improves on \CS{}, reducing elapsed time by $47.6\%$ to $96.9\%$ and memory usage by $5.6\%$ to $60.4\%$, suggesting that even partial structural synchronization can be competitive with \CS{} and \PO{} on some benchmarks. 
Applying localized \CS{} within the manually flattened blocks of \HSBCS{}, denoted \HSBCS{}$^\dagger$, further improves solver performance, with elapsed-time reductions of $76.3\%$ to $97.7\%$ and memory reductions of $12\%$ to $73.6\%$.

A notable result is that \cplex{} failed to solve the smallest \PO{} instance, even though it is $4.8$ times smaller than the largest solvable \CS{} and \HSBCS{} instances. 
Here, \PO{} exceeds the time limit in Table~\ref{tab:benchmark-config}, and the \cplex{} logs suggest degeneracy, with many iterations but no objective value improvement. 
We also tested \PO{} under a further artificial $2.4$-fold \TBn{} reduction, % to $169,3$
but \cplex{} failed with the same pattern, indicating that LP size reduction alone did not remove the difficulty. 
Although additional experiments on similarly structured benchmarks are needed to confirm this effect, a plausible explanation is that \CS{} and \HSBCS{} mitigate degeneracy by reducing redundant constraints that overdefine primal solutions. 
In \CS{}, this occurs through aggregating constraints for identical statements that differ only in their $S$ controllers, most of which are zero in the solution. 
In \HSBCS{}, redundancy is further reduced outside of the manually flattened block by generating constraints only over the execution intervals of enclosing blocks.

Furthermore, \HSB{} for the manually hoisted variant of \MM{}, denoted by \HMM{} in Table~\ref{tab:time-mm-notforced}, shows substantial improvements in both elapsed time and memory usage relative to \CS{} and \HSBCS{}, despite using the actual \TBn{} values rather than the artificially reduced values used in the other two approaches. 
Solver failures are omitted from Table~\ref{tab:time-mm-notforced} for readability; the full table appears as Table~\ref{tab:time-mm-notforced-complete} in Appendix~\ref{sec:appendix-solver-performance}.

\begin{table}[hbtp] 
\centering
\caption{Run times and memory usage for the \MM{} and \HMM{} benchmarks, with inputs encoded through the objective function. 
For \MM{} we compare \PO{}, \CS{}, \HSBCS{}, and \HSBCS{}$^\dagger$.  
For \HMM{}, we report \HSB{}.
The $^\dagger$ denotes the \HSBCS{} variant in which \CS{} is applied within manually flattened blocks. 
Each result is based on 3~runs with a 23-hour time limit, 372~GB memory, and 8~CPUs, except that \HSBCS{}$^\dagger$ uses a 126~GB memory limit. 
The \IQR{} is reported as a percentage of the median. 
A superscript $^*$ indicates dual simplex algorithm.
\OOM{} indicates out-of-memory, \OOT{} indicates timeout, and \FAIL{} denotes solver failure.} 
\label{tab:time-mm-notforced}
\begin{adjustbox}{max totalsize={\textwidth}{0.76\textheight},center}
\begin{tabular}{cccccccccc}
\toprule
\multicolumn{4}{c}{Info} & \multicolumn{4}{c}{Time (\Hour{})} & \multicolumn{2}{c}{Mem (\GB{})}  \\
\cmidrule(lr){1-4} \cmidrule(lr){5-8} \cmidrule(lr){9-10}
$n$ & solver & problem & \OPT{} &  \multicolumn{2}{c}{Elapsed}  & \multicolumn{2}{c}{Presolve} & \multicolumn{2}{c}{RSS} \\
 &  &  &  &  \Med{}& \IQR{}& \Med{}& \IQR{}& \Med{}& \IQR{}\\
\midrule
\multirow{17}{*}{8}
& \multirow{4}{*}{\cplex$^*$} &  
\multirow{2}{*}{\MM} & 
  \HSBCS{} &  \textbf{0.55} & 0.85\% & 0.00 & 0.57\% & 5.57 & 0.13\%  \\ 
  &   &  & \HSBCS{}$^\dagger$ &  1.57 & 0.95\% & 0.00 & 0.58\% & \textbf{3.14} & 0.37\%  \\ 
\cmidrule(lr){3-10}
  &  &  \multirow{1}{*}{\HMM}   
  & \HSB{}  & 1.81 & 1.74\% & 0.00 & 0.96\% & 2.17 & 0.24\% \\ 
\cmidrule{2-10}
 & \multirow{5}{*}{\cplex} &  
 \multirow{3}{*}{\MM} & 
   \CS{}       & 0.82 & 0.26\% & 0.01 & 0.62\% & 43.86 & 0.22\% \\ 
 \cmidrule(lr){4-10}
  &  &   & \HSBCS{} & 0.43 & 0.47\% & 0.00 & 0.48\% & 17.35 & 0.33\% \\ 
  &  &   & \HSBCS{}$^\dagger$ & \textbf{0.06} & 0.26\% & 0.00 & 0.32\% & \textbf{11.60} & 1.34\%  \\ 
 \cmidrule(lr){3-10}
 &        &  \multirow{1}{*}{\HMM} 
 & \HSB{}  & 0.02 & 0.44\% & 0.00 & 0.64\% & 6.89 & 0.03\% \\ 
\cmidrule{2-10}
  & \multirow{3}{*}{\scip} &  
  \multirow{1}{*}{\MM} & 
      \HSBCS{}$^\dagger$ & \textbf{0.38} & 0.88\% & \textbf{0.02} & 0.06\% & \textbf{16.11} & 0.01\% \\ 
  \cmidrule(lr){3-10}
  &     &  \multirow{1}{*}{\HMM}   
  & \HSB{}  & 0.56 & 1.17\% & 0.01 & 0.93\% & 11.14 & 0.06\% \\ 
\cmidrule{2-10}
  & \multirow{5}{*}{\gurobi} &  
  \multirow{3}{*}{\MM} & 
  \CS{}  & 16.20 & 0.58\% & 0.02 & 0.66\% & 33.31 & 0.01\% \\  
  \cmidrule(lr){4-10}
    &  &   & \HSBCS{} &  0.50 & 0.61\% & 0.01 & 0.42\% & 13.81 & 0.04\% \\ 
  &   &  & \HSBCS{}$^\dagger$ & \textbf{0.38} & 0.43\% & 0.01 & 0.77\% & \textbf{9.27} & 3.06\% \\ 
  \cmidrule(lr){3-10}
    &     &  \multirow{1}{*}{\HMM}   
    & \HSB{}  & 0.05 & 2.43\% & 0.00 & 1.19\% & 8.35 & 22.31\% \\ 
\midrule
\multirow{11}{*}{10}
  & \multirow{1}{*}{\cplex$^*$}  & 
    \multirow{1}{*}{\HMM}   
  & \HSB{}  & 0.92 & 1.36\% & 0.00 & 0.17\% & 5.20 & 0.08\% \\ 
\cmidrule{2-10}
& \multirow{5}{*}{\cplex} & 
\multirow{3}{*}{\MM} & 
\CS{} &  3.69 & 0.29\% & 0.02 & 0.76\% & 88.13 & 0.40\% \\ 
  \cmidrule(lr){4-10}
  &  &   & \HSBCS{} &  0.93 & 0.30\% & 0.01 & 0.19\% & 45.80 & 0.31\%  \\ 
  &  &   & \HSBCS{}$^\dagger$ &  \textbf{0.14} & 0.57\% & 0.01 & 0.08\% & \textbf{29.22} & 0.42\%\\ 
\cmidrule(lr){3-10}
& &  \multirow{1}{*}{\HMM}   
& \HSB{}   & 0.08 & 0.60\% & 0.00 & 0.38\% & 16.53 & 0.42\% \\ 
\cmidrule{2-10}
& \multirow{1}{*}{\scip} &  
   \multirow{1}{*}{\HMM}   
  & \HSB{}   & 2.07 & 1.31\% & 0.03 & 1.04\% & 26.37 & 0.04\% \\
\cmidrule{2-10}
  & \multirow{2}{*}{\gurobi} & 
  \multirow{2}{*}{\MM} & 
    \HSBCS{} & \textbf{1.43} & 0.19\% & 0.03 & 0.86\% & 36.48 & 0.03\% \\ 
   & & & \HSBCS{}$^\dagger$ &  3.34 & 0.27\% & \textbf{0.02} & 0.34\% & \textbf{21.99} & 0.16\% \\ 
  \cmidrule(lr){3-10}
  &       &  \multirow{1}{*}{\HMM}   
  & \HSB{} & 0.21 & 2.43\% & 0.01 & 0.72\% & 31.44 & 0.68\% \\ 
\midrule
\multirow{8}{*}{12}
 & \multirow{1}{*}{\cplex$^*$} &  
   \multirow{1}{*}{\HMM}   
  & \HSB{}  & 3.82 & 6.67\% & 0.01 & 5.80\% & 10.09 & 0.04\% \\ 
\cmidrule{2-10}
& \multirow{4}{*}{\cplex} & 
\multirow{3}{*}{\MM} & 
\CS{} &  3.51 & 0.14\% & 0.03 & 1.14\% & 146.73 & 1.24\% \\ 
    \cmidrule(lr){4-10}
  &  &   & \HSBCS{} & 1.43 & 0.19\% & 0.04 & 0.54\% & 138.58 & 2.22\%  \\ 
  &  &   & \HSBCS{}$^\dagger$ & \textbf{0.45} & 0.85\% & \textbf{0.02} & 0.90\% & \textbf{85.32} & 0.03\%  \\ 
\cmidrule(lr){3-10}
& &  \multirow{1}{*}{\HMM}   
& \HSB{}  & 0.30 & 2.24\% & 0.01 & 0.28\% & 32.31 & 0.02\% \\
\cmidrule{2-10}
& \multirow{1}{*}{\scip} &  
\multirow{1}{*}{\HMM}    
&  \HSB{}  & 4.50 & 3.41\% & 0.08 & 6.23\% & 53.36 & 0.01\% \\ 
\cmidrule{2-10}
  & \multirow{1}{*}{\gurobi} &  
  \multirow{1}{*}{\MM} & 
  \HSBCS{}$^\dagger$ &  \textbf{11.81} & 0.78\% & 0.05 & 0.74\% & \textbf{66.89} & 0.01\% \\ 
\midrule
\multirow{6}{*}{16}
& \multirow{4}{*}{\cplex} & 
\multirow{2}{*}{\MM} & 
\CS{} &  18.68 & 1.06\% & 0.07 & 3.93\% & 320.95 & 0.02\% \\  
    \cmidrule(lr){4-10}
  &  &   & \HSBCS{}$^\dagger$ &  \textbf{4.42} & 0.06\% & 0.07 & 0.62\% & \textbf{282.48} & 0.99\% \\ 
\cmidrule(lr){3-10}
& &  \multirow{1}{*}{\HMM} 
& \HSB{}   & 2.75 & 1.07\% & 0.03 & 2.40\% & 100.33 & 0.05\% \\
\cmidrule{2-10}
  & \multirow{2}{*}{\gurobi} &  
  \multirow{1}{*}{\MM} & 
  \HSBCS{}$^\dagger$ & \textbf{2.96} & 1.22\% & \textbf{0.16} & 1.65\% & \textbf{217.83} & 0.00\% \\ 
  \cmidrule(lr){3-10}
  &    &  \multirow{1}{*}{\HMM}
  & \HSB{}  & 2.18 & 0.17\% & 0.10 & 0.11\% & 64.37 & 0.01\% \\ 
\bottomrule
\end{tabular}
\end{adjustbox}
\end{table}

Algorithmic time complexity also affects the LP size and solver performance. For example, \LIS{}, which is generated from the \LIStimecomp{} algorithm for solving \LISproblem{}, becomes unsolvable at $n=20$ for \UO{} and at $n=40$ for \HSB{}, whereas its \LISLtimecomp{} counterpart, \LISL{}, remains solvable at $n=40$ and $n=80$, respectively.

Note that these reduction comparisons are restricted to the smaller instance sizes for which \UO{} also solves. This limitation is particularly important for \CS{}, since its advantages are more pronounced on larger codes that contain more code duplication and array operations, precisely where \UO{} often becomes too large to solve (e.g., \MM{} variants in Table~\ref{tab:time-mm-notforced}).

\subsection{Solution Accuracy}
\label{sec:resultSolutionAccuracy}
This section examines whether the LP size reductions from \CS{} and \HSB{} also improve solver accuracy relative to \UO{} on jointly solved benchmark instances. 
Because the benchmark inputs are encoded via the objective functions (Section~\ref{sec:background}), most solvers return fractional solutions despite the theoretical guarantee of integer solutions.
We therefore compare each solution by a ground-truth solution obtained by fixing the input variables to their $0/1$ values using equality constraints. 
For jointly solved benchmarks, we report median \MAElong{} and \RMSElong{}, their percentage improvements, and the number and proportion of mismatched variable values. 
The mismatch count is the number of variable values that differ from the ground-truth. It is comparable across methods despite their LP size differences because all methods encode the same underlying problem. Removing redundant execution states and their associated variables also removes variables that may otherwise become fractional due to numerical errors caused by the solver. 
The mismatch percentage normalizes the number of incorrect values by model size.
Together, these metrics indicate whether smaller LPs yield solutions closer to the integer ground truth.
Detailed per-benchmark results are given in Tables~\ref{tab:sol-accuracy-spdj} to~\ref{tab:sol-accuracy-ms} in Appendix~\ref{sec:appendix-solution_accuracy}.

Figure~\ref{fig:mae_rmse_all_boxplots} shows the distributions across LP instances of the \MAE{} differences (left) and \RMSE{} differences (right) for \HSB{} and \CS{} relative to \UO{} across all solvers and benchmarks.
Overall, both \HSB{} and \CS{} improve solution accuracy compared to \UO{}, as their distributions are centered below zero in both metrics.
\HSB{} exhibits larger improvements with substantially lower median than \CS{} for both \MAE{} and \RMSE{}. 
In contrast, \CS{} has a narrower spread, indicating that although its improvements are smaller, they are more consistent across LP instances.
Most \IQR{} values are zero, indicating that the solvers are deterministic and generate the same solutions across repeated runs.
\HSB{} has a few positive outliers in \MAE{}, indicating that for some LP instances its solutions are worse than those of \UO{}. Although the median \RMSE{} difference remains below zero, indicating that improvements persist overall, the upper whisker extends above zero, showing that a small number of extreme errors were large enough to affect the distribution. 
By comparison, \CS{} maintains a narrow distribution for both metrics, indicating it is less affected by extreme errors.

\begin{figure}[!t]
\centering
\includegraphics[width=1\columnwidth]{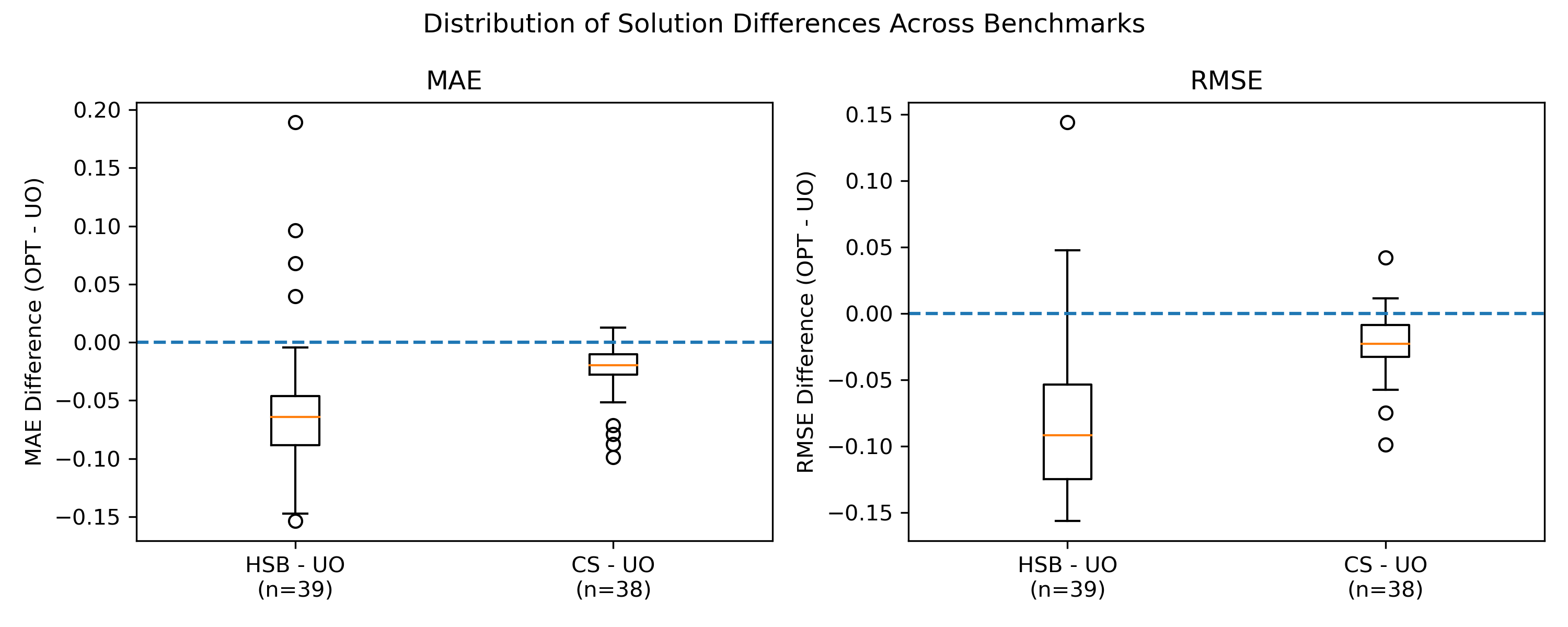}
\caption{Overall solution accuracy differences (\OPT{}$-$\UO{}) across all benchmarks and solvers, where \OPT{} denotes the optimization method compared with \UO{}.
Left: \MAE{} differences. Right: \RMSE{} differences. 
Negative values indicate improvement relative to \UO{}.}
\label{fig:mae_rmse_all_boxplots}
\end{figure}

\begin{figure}[hbtp]
\centering
\includegraphics[width=1\columnwidth]{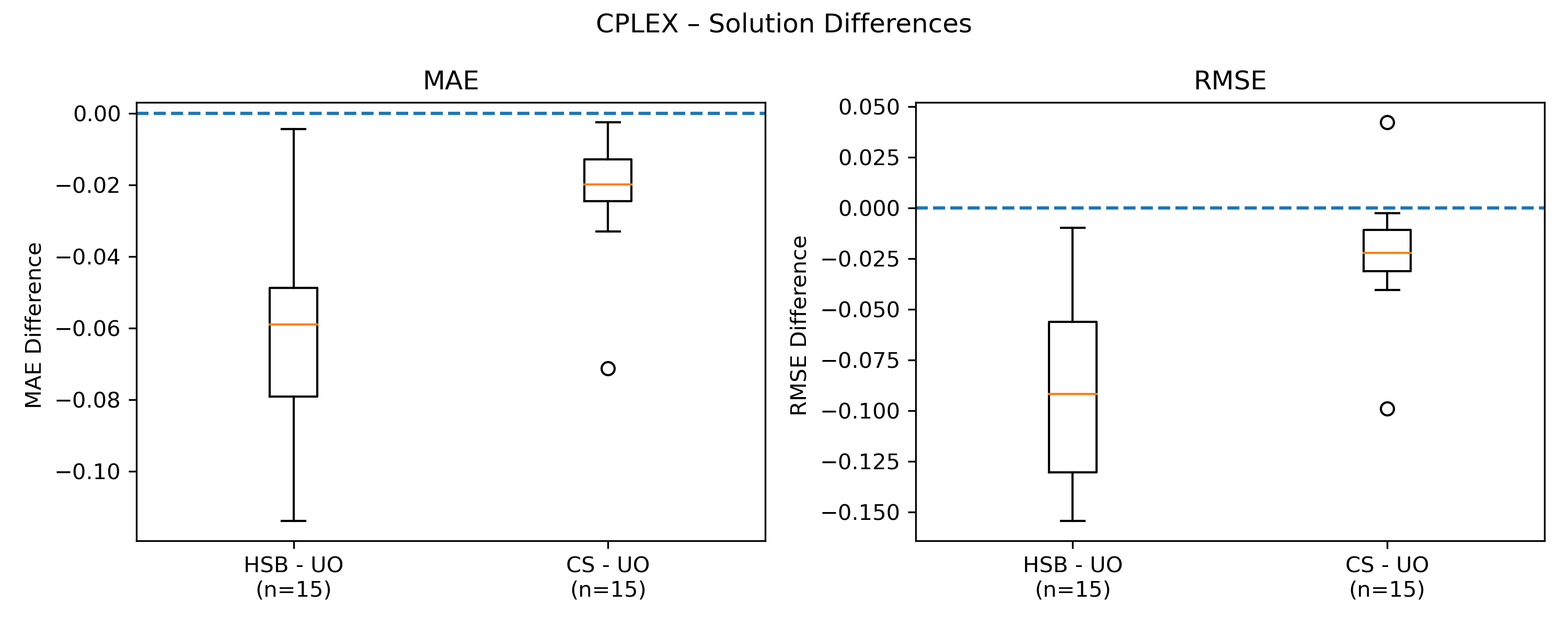}
\caption{\cplex{} solution accuracy differences (\OPT{}$-$\UO{}) across all benchmarks, where \OPT{} denotes the optimization method compared with \UO{}.
Left: \MAE{} differences. Right: \RMSE{} differences. 
Negative values indicate improvement relative to \UO{}.
}
\label{fig:mae_rmse_cplex_boxplots}
\end{figure}

\begin{figure}[!b]
\centering
\includegraphics[width=1\columnwidth]{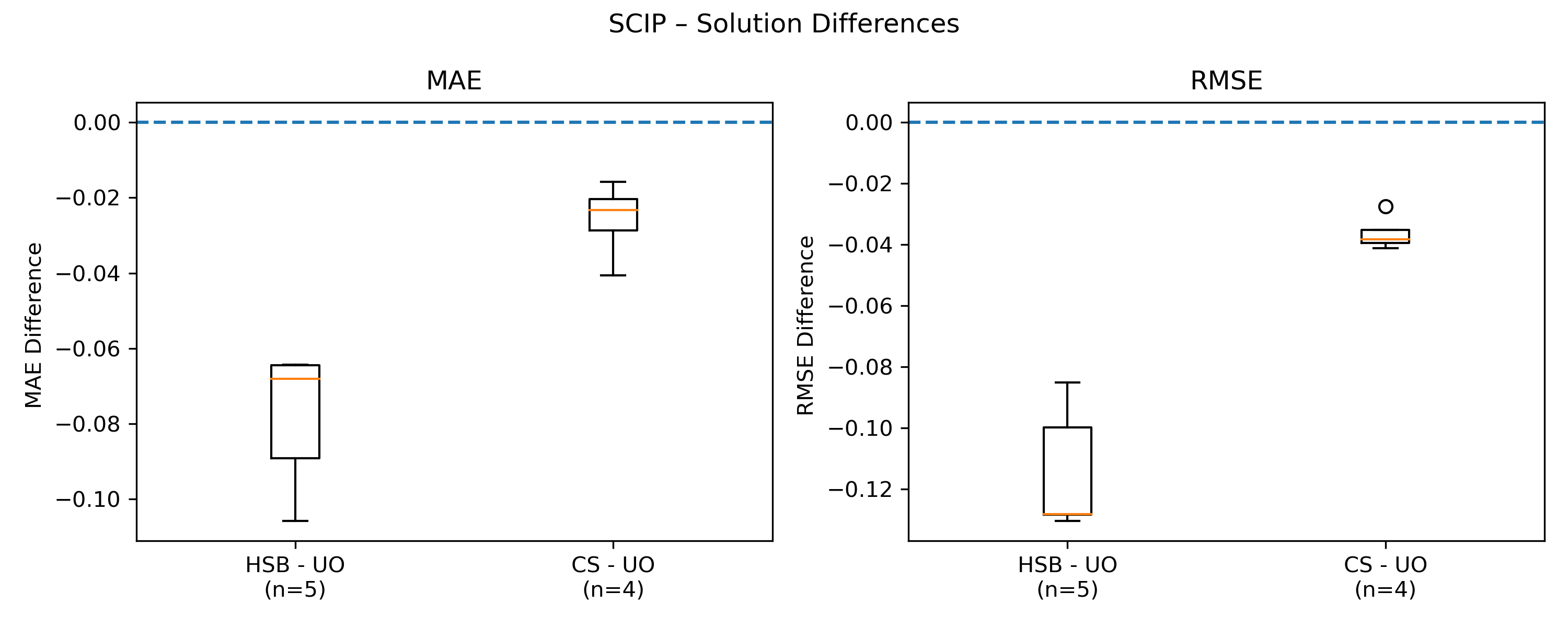}
\caption{\scip{} solution accuracy differences (\OPT{}$-$\UO{}) across all benchmarks, where \OPT{} denotes the optimization method compared with \UO{}.
Left: \MAE{} differences. Right: \RMSE{} differences. 
Negative values indicate improvement relative to \UO{}.
}
\label{fig:mae_rmse_scip_boxplots}
\end{figure}

\begin{figure}[hbtp]
\centering
\includegraphics[width=1\columnwidth]{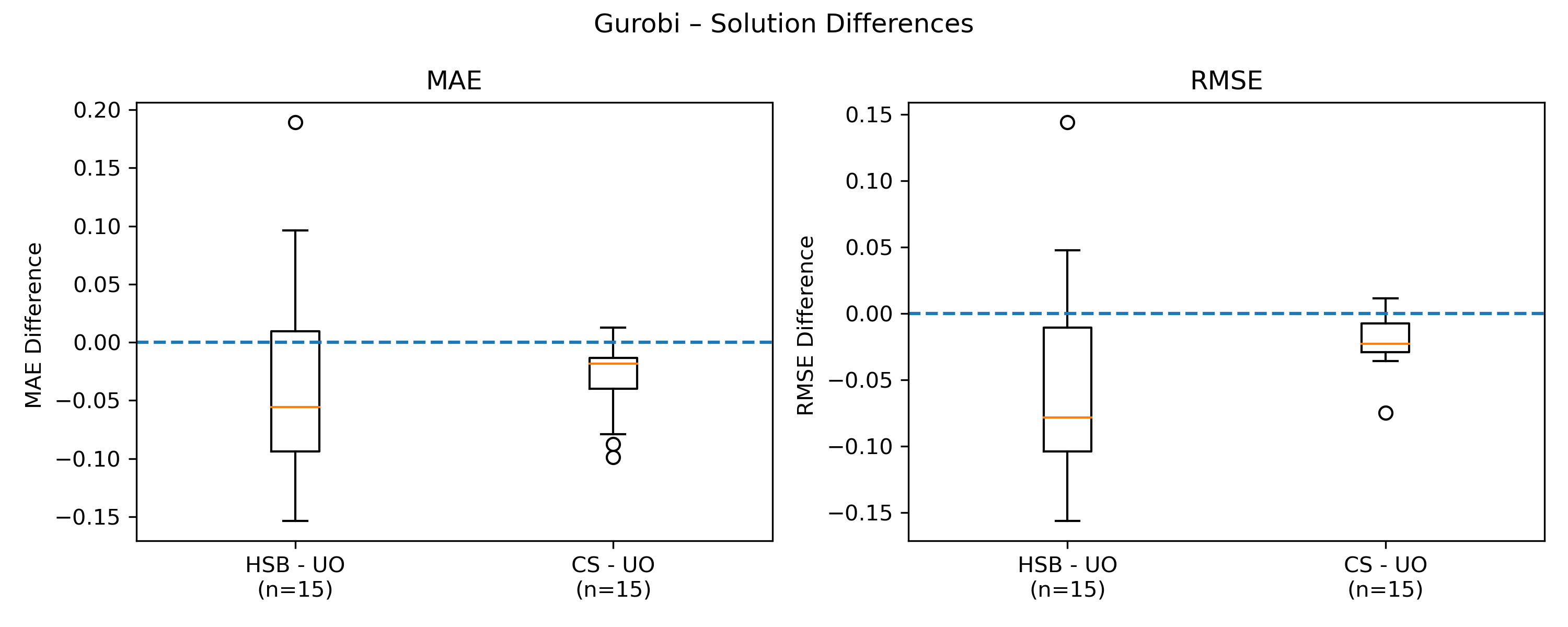}
\caption{\gurobi{} solution accuracy differences (\OPT{}$-$\UO{}) across all benchmarks, where \OPT{} denotes the optimization method compared with \UO{}.
Left: \MAE{} differences. Right: \RMSE{} differences. 
Negative values indicate improvement relative to \UO{}.
}
\label{fig:mae_rmse_gurobi_boxplots}
\end{figure}

To better understand the source of this variability, Figures~\ref{fig:mae_rmse_cplex_boxplots},~\ref{fig:mae_rmse_scip_boxplots}, and~\ref{fig:mae_rmse_gurobi_boxplots} show the distributions broken down by solver. These figures suggest that most extreme errors come from \gurobi{} (Figure~\ref{fig:mae_rmse_gurobi_boxplots}).

\begin{figure}[!b]
\centering
\includegraphics[width=1\columnwidth]{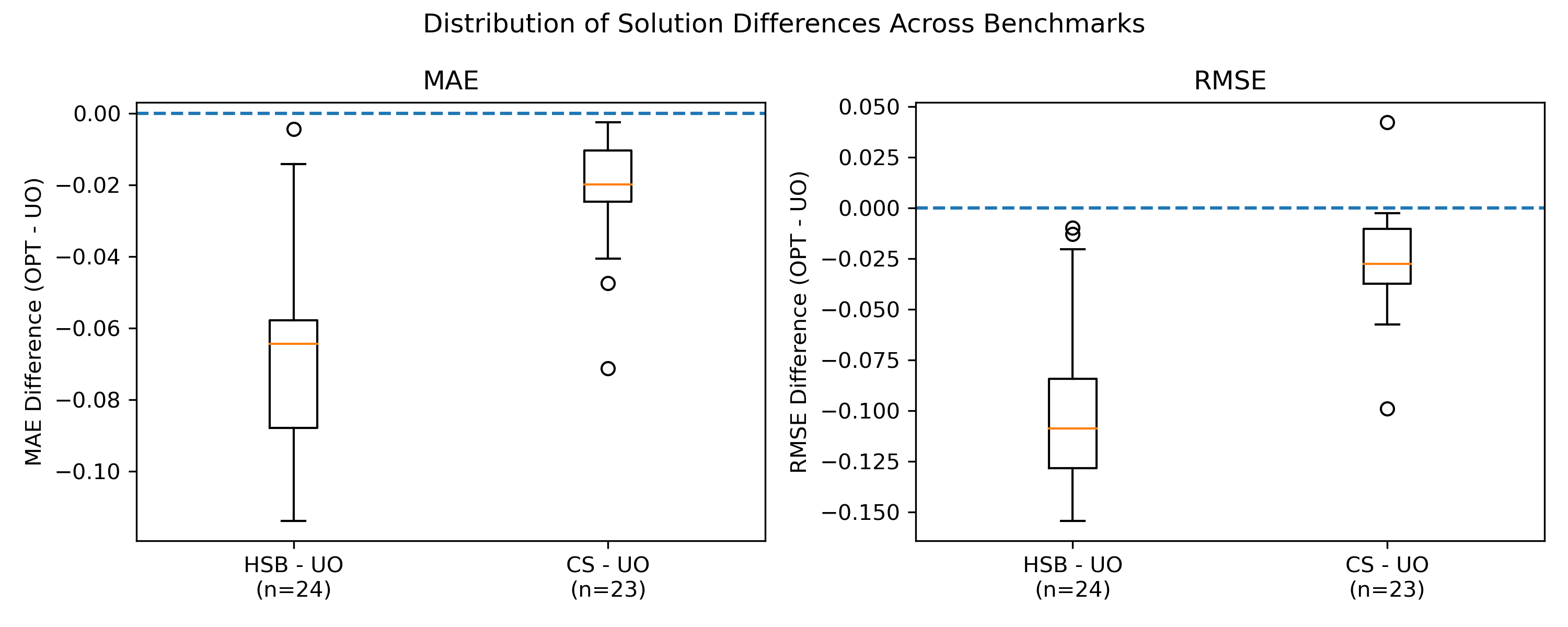}
\caption{Overall solution accuracy differences (\OPT{}$-$\UO{}) across all benchmarks for all solvers except \gurobi{}, where \OPT{} denotes the optimization method compared with \UO{}.
Left: \MAE{} differences. Right: \RMSE{} differences. 
Negative values indicate improvement relative to \UO{}.
}
\label{fig:mae_rmse_all_boxplots_without_gurobi}
\end{figure}

This observation is confirmed by Figure~\ref{fig:mae_rmse_all_boxplots_without_gurobi}, which shows the overall distribution after excluding the \gurobi{}, where all positive \MAE{} outliers for \HSB{} disappear.

It is worth noting that LP variables corresponding to two compiler-introduced classes of variables are excluded from the \MAE{} and \RMSE{} calculations. They are variables representing carries in integer addition operations and array controllers specifying array index values.
Some versions of these variables may exist only in bound constraints, which are treated inconsistently across solvers as mentioned in Section~\ref{sec:resultLpSize}.
Since \gurobi{} and \scip{} include such variables in their returned solutions, their values may vary between the ground truth solution obtained in presolve, therefore artificially distorting the \MAE{} and \RMSE{} distances.

Detailed results for mismatch count and mismatch percentage are given in Tables~\ref{tab:sol-accuracy-spdj} to~\ref{tab:sol-accuracy-ms} in Appendix~\ref{sec:appendix-solution_accuracy}.
For brevity, we do not discuss them extensively here, but they show that both \HSB{} and \CS{} generally reduce mismatch count relative to \UO{}. 
The only exceptions are the \CS{} solutions on \BS{} with $n=160$ under \cplex{} and \cplex{}$^*$. 
Accordingly, Table~\ref{tab:winloss-overall-maermsemismatch} shows that \CS{} yields a larger median mismatch count than \UO{} on 2 of 38 jointly solved LP instances, while improving on the remaining $94.7\%$.

\begin{table}[b]
\centering
\caption{
Percentage of jointly solved benchmark instances across all solvers for which \HSB{} and \CS{} achieved smaller than \UO{} for medians of \MAE{}, \RMSE{}, mismatch count, and mismatch percentage, with medians computed over repeated runs for each LP instance.
}
\label{tab:winloss-overall-maermsemismatch}
\scalebox{0.93}{
\begin{tabular}{c c c c c}
\toprule
\makecell{Comparison} & \makecell{MAE \\ Wins} & \makecell{RMSE \\ Wins} & \makecell{Mismatch Count \\ Wins} & \makecell{Mismatch Percentage \\ Wins} \\
\midrule
\CS{} vs \UO{} & 35/38 (92.1\%) & 34/38 (89.5\%) & 36/38 (94.7\%) & 33/38 (86.8\%)\\
\HSB{} vs \UO{} & 35/39 (89.7\%) & 36/39 (92.3\%) & 39/39 (100\%) & 32/39 (82.1\%)\\
\bottomrule
\end{tabular}
}
\end{table}

Moreover, Table~\ref{tab:winloss-overall-maermsemismatch} shows that \CS{} and \HSB{} achieve smaller mismatch percentages than \UO{} on $86.8\%$ and $82.1\%$ of jointly solved LP instances, respectively. This suggests that the reductions in mismatch count are not due solely to the smaller number of columns in their models.

For \LISL{}, \HSB{} has higher mismatch percentage than \UO{}, worsening by $4.3$ to $14.2$ percentage points under \cplex{} and by $11.4$ to $21.5$ percentage points under \gurobi{}. 
Moreover, under \gurobi{}, \HSB{} also performs worse on \TS{}, consistent with the positive outliers in Figure~\ref{fig:mae_rmse_gurobi_boxplots}. 
In turn, \CS{} has a higher mismatch percentage than \UO{} for \BS{} and \KS{}.
Surprisingly, although \scip{} performs worst in elapsed time and memory usage and solved the fewest benchmark configurations, it often achieves competitive \MAE{} and \RMSE{} values, and in several cases the best overall solution accuracy. 
This is particularly notable for \LISL{}, where \HSB{} worsens solution accuracy under other solvers, but under \scip{} improves \MAE{}, \RMSE{}, mismatch count, and mismatch percentage relative to \UO{}.

To examine whether the obtained reductions in \MAE{} and \RMSE{} are statistically significant, we apply the Wilcoxon signed-rank test, a nonparametric test suitable for paired comparisons when the differences cannot be assumed to be normally distributed.

Tables~\ref{tab:wilcoxon-mae} and~\ref{tab:wilcoxon-rmse} show that the \MAE{} and \RMSE{} reductions are generally statistically significant across solvers for both \HSB{} and \CS{}. This is indicated by the negative median differences and $p$-values smaller than $0.05$, for solvers that have enough (larger than 5) non-tied jointly solved LP instances. 
The main exception is \MAE{} for \HSB{} under \gurobi{}, for which $p$-value, $7.57 \times 10^{-2}$, exceeds the $0.05$ threshold despite a negative median difference. 
The \MAE{} reduction for \HSB{} under \scip{} is statistically significant, although it is based on only five non-tied pairs. Even though the \scip{} results for \CS{} seem promising, the test could not be applied as fewer than 5 non-tied instance pairs were available, due to time limit failures when obtaining the ground-truth solutions for \KS{} with size 10 and \SAM{} with size 20.  

\begin{table}[!b]
\centering
\caption{Wilcoxon signed-rank test results for \MAE{}, where \OPT{} denotes the optimization method compared with \UO{}. 
$N$ denotes the number of non-tied LP instance pairs. 
A negative median \MAE{} difference indicates improvement relative to \UO{}. 
Value \FAIL{} indicates test failure due to insufficient sample size ($N < 5$).
}
\label{tab:wilcoxon-mae}
\resizebox{\textwidth}{!}{
\begin{tabular}{c c c c c c}
\toprule
\makecell{\OPT{}} & \makecell{solver} & \makecell{$N$} & \makecell{median diff\\(\OPT{} - \UO{})} & \makecell{win-rate\\(\OPT{} $<$ \UO{})} & \makecell{$p$-value} \\
\midrule
\multirow{4}{*}{HSB} & \cplex{} & 15 & -0.0590 & 100.00\% & $3.05 \times 10^{-5}$ \\
 & \cplex{}$^*$ & 4 & -0.0767 & 100.00\% & \textbf{---} \\
 & \gurobi{} & 15 & -0.0558 & 73.33\% & $7.57 \times 10^{-2}$ \\
 & \scip{} & 5 & -0.0681 & 100.00\% & $3.12 \times 10^{-2}$ \\
\midrule
\multirow{4}{*}{CS} & \cplex{} & 15 & -0.0199 & 100.00\% & $3.05 \times 10^{-5}$ \\
 & \cplex{}$^*$ & 4 & -0.0092 & 100.00\% & \textbf{---} \\
 & \gurobi{} & 15 & -0.0186 & 80.00\% & $5.80 \times 10^{-4}$ \\
 & \scip{} & 4 & -0.0233 & 100.00\% & \textbf{---} \\
\bottomrule
\end{tabular}
}
\end{table}

\begin{table}[hbtp]
\centering
\caption{
Wilcoxon signed-rank test results for \RMSE{}, where \OPT{} denotes the optimization method compared with \UO{}. 
$N$ denotes the number of non-tied LP instance pairs. 
A negative median \RMSE{} difference indicates improvement relative to \UO{}. 
Value \FAIL{} indicates test failure due to insufficient sample size ($N < 5$).
}
\label{tab:wilcoxon-rmse}
\resizebox{\textwidth}{!}{
\begin{tabular}{c c c c c c}
\toprule
\makecell{\OPT{}} & \makecell{solver} & \makecell{$N$} & \makecell{median diff\\(\OPT{} - \UO{})} & \makecell{win-rate\\(\OPT{} $<$ \UO{})} & \makecell{$p$-value} \\
\midrule
\multirow{4}{*}{HSB} & \cplex{} & 15 & -0.0916 & 100.00\% & $3.05 \times 10^{-5}$ \\
 & \cplex{}$^*$ & 4 & -0.1150 & 100.00\% & \textbf{---} \\
 & \gurobi{} & 15 & -0.0782 & 80.00\% & $1.77 \times 10^{-2}$ \\
 & \scip{} & 5 & -0.1281 & 100.00\% & $3.12 \times 10^{-2}$ \\
\midrule
\multirow{4}{*}{CS} & \cplex{} & 15 & -0.0222 & 93.33\% & $3.36 \times 10^{-3}$ \\
 & \cplex{}$^*$ & 4 & -0.0089 & 100.00\% & \textbf{---} \\
 & \gurobi{} & 15 & -0.0227 & 80.00\% & $1.01 \times 10^{-3}$ \\
 & \scip{} & 4 & -0.0383 & 100.00\% & \textbf{---} \\
\bottomrule
\end{tabular}
}
\end{table}

Beyond these trends in the selected metrics, the results also show some notable solver-specific behavior. 
The non-commercial solver \scip{} is now able to find an integer solution for the \HSB{} LP of \MS{} at input size $5$.
When the input variables are fixed to their integer values, \cplex{} and \gurobi{} return optimal integral solutions whenever they do not exceed the imposed resource limits. This is encouraging, as it suggests improvements in \gurobi{}'s presolving capabilities, whereas earlier reports on older versions reported non-optimal fractional solutions~\cite{avis_sparktope_2022}.
However, \scip{} still returns fractional solutions on a few large \HSB{} LP instances. This is unlikely to indicate a weakness of \HSB{}, since \scip{} exceeds memory on the corresponding \UO{} models at much smaller instance sizes while solving smaller \HSB{} instances successfully. The more plausible explanation is that, at larger scales, \scip{} may struggle with either the size or specific structure of \HSB{} models. 

As part of our long-term goal of embedding these LPs within \IP{} formulations, a possible approach to mitigating these solver limitations is to declare the input variables as integers when using MIP solvers.

\subsection{Condition Number (Kappa) of \textsc{CPLEX}}
\label{sec:resultConditionNumberKappa}
In this section, we examine the basis condition number, referred to by \cplex{} as kappa, as a solver-specific measure of how the LP size reductions achieved by \HSB{} and \CS{} affect the numerical conditioning of the resulting models relative to \UO{}.
Larger kappa values indicate poorer numerical conditioning. Accordingly, we compare the optimization methods using both the relative change in kappa with respect to \UO{} across jointly solved LP instances and the conditioning classes (Table~\ref{tab:mip-kappa-classes}) used by \cplex{}~\cite{ibm_mip_2022} to interpret kappa values for all solved instances.
Detailed per-benchmark report of \cplex{} kappa values for the unscaled basis are given in Tables~\ref{tab:kappaRes_spdj} to~\ref{tab:kappaRes_ms} in Appendix~\ref{sec:appendix-solution_kappa}. 
\begin{table}[!b]
\centering
\caption{\cplex{} MIP kappa basis classes and their condition-number ranges~\cite{ibm_mip_2022}.}
\label{tab:mip-kappa-classes}
\begin{tabular}{l l}
\toprule
\makecell{Basis class} & \makecell{Condition number range ($\kappa$)} \\
\midrule
Stable      & $\kappa < 10^{7}$ \\
Suspicious  & $10^{7} \le \kappa < 10^{10}$ \\
Unstable    & $10^{10} \le \kappa < 10^{14}$ \\
Ill-posed   & $\kappa \ge 10^{14}$ \\
\bottomrule
\end{tabular}
\end{table}

\begin{figure}[!t]
\centering
\includegraphics[width=0.7\columnwidth]{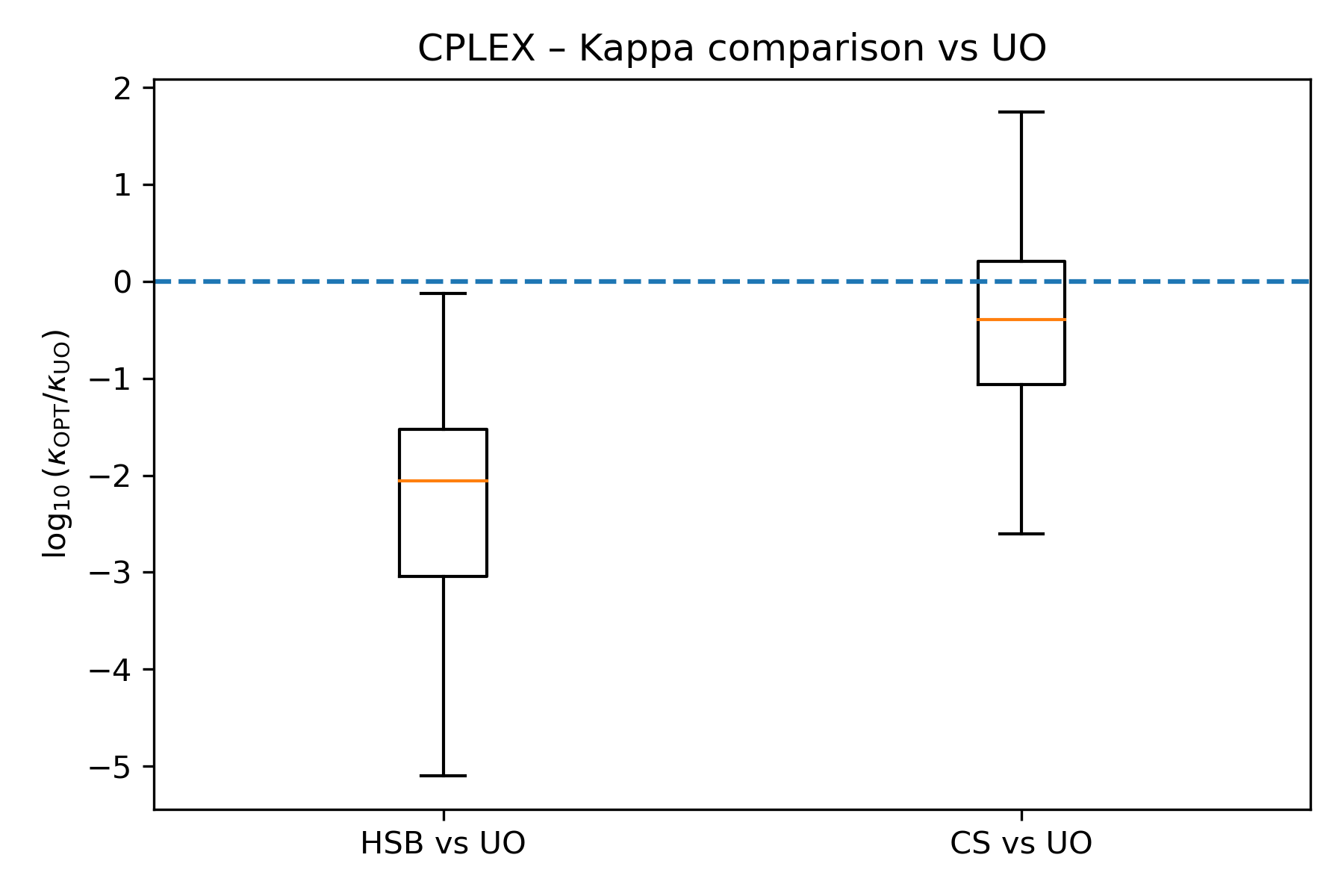}
\caption{Boxplots of $\log_{10}(\kappa_{\mathrm{\OPT}}/\kappa_{\mathrm{\UO}})$, where \OPT{} denotes the optimization method compared with \UO{}, and $\kappa$ denotes the basis condition number reported by \cplex{}.
Negative values indicate improvement relative to \UO{}. }
\label{fig:cplex_kappa_boxplots}
\end{figure}

Figure~\ref{fig:cplex_kappa_boxplots} shows the distributions of $\log_{10}(\kappa_{\mathrm{\OPT{}}}/\kappa_{\mathrm{UO}})$ for \HSB{} and \CS{} relative to \UO{}, where $\kappa$ denotes the kappa value reported by \cplex{}. Thus, negative values indicate improved conditioning relative to \UO{}. 
Overall, both \HSB{} and \CS{} improve conditioning compared to \UO{}, as indicated by the boxplots, which are centered below zero. 
However, \HSB{} yields larger and more consistent reductions in relative kappa across instances, as shown by its lower median and by its entire interquartile range below zero.
In contrast, \CS{} has smaller and less uniform improvements. Although its median also lies below zero, suggesting better conditioning for most instances, some instances have worse conditioning than \UO{}, as indicated by the upper whisker extending above zero.

\begin{figure}[!t]
\centering
\includegraphics[width=1\columnwidth]{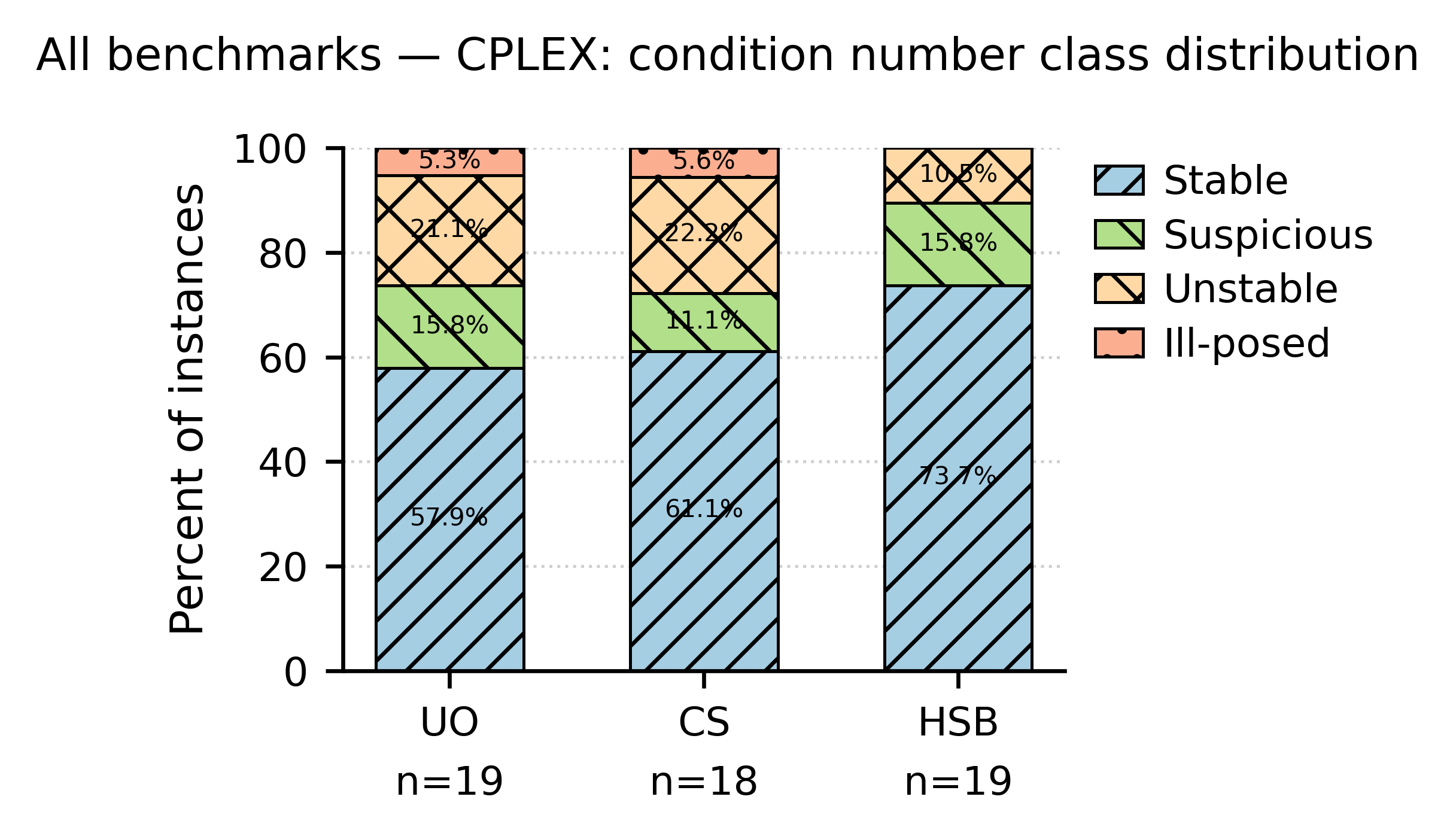}
\caption{Distribution of LP instances solved by \cplex{} across its kappa classes for \UO{}, \CS{}, and \HSB{}, shown as percentages of all benchmark instances. }
\label{fig:kappa_barchart_classes_overall_cplex}
\end{figure}

Figure~\ref{fig:kappa_barchart_classes_overall_cplex} shows the share of solved LP instances in each \cplex{} kappa classes by optimization method.
\HSB{} shows the largest improvement relative to \UO{}, increasing \emph{stable} instances from $57.9\%$ to $73.7\%$, reducing \emph{unstable} ones from $21.1\%$ to $10.5\%$, and entirely eliminating the \emph{ill-posed} class. 
In contrast, \CS{} changes the class distributions only slightly relative to \UO{}. It slightly increases the share of \emph{stable} instances from $57.9\%$ to $61.1\%$, and reduces the share of \emph{suspicious} ones from $15.8\%$ to $11.1\%$, but the share of \emph{unstable} instances also increases slightly from $21.1\% $to $22.2\%$, while the \emph{ill-posed} class remains nearly unchanged.
These results suggest that \HSB{} achieves substantial enough LP size reductions to improve overall numerical conditioning.

\section{Conclusion}
\label{sec:conclusion}
An LP compiler translates high-level polynomial-time algorithmic descriptions into polynomial-size Linear Programs (LPs), even for problems in \PP{} that have exponential extension complexity, making it a good alternative for \AMLlong{}. 
Through extensive evaluation on a broad benchmark suite, we showed that both \CSlong{} and \HSBlong{} consistently generate smaller LPs, with file-size reductions of up to 53\% and 96\%, respectively. These reductions also improve numerical conditioning and solver performance, yielding elapsed-time reductions of up to 98.8\% and 99.9\%, and increasing the range of instances that can be solved under fixed resource limits. 
Among the two, \HSB{} is the more effective and more consistent optimization across benchmarks. 
At the same time, our experiments reveal an important limitation of \HSB{}, since its hierarchical synchronization is based on worst-case execution-time assumptions overall, even though certain code regions can benefit from smaller execution-time estimates through user insight. 
To address this limitation, we introduced the \HSBCSlong{} framework, which preserves the structural advantages of \HSB{} while providing an abstraction layer on which localized reductions can be applied to the affected regions. This broadens the class of algorithms that can benefit from \HSB{}'s hierarchical reductions.
More broadly, our results show that the resulting LPs benefit from both exploited code structure and design of compiler generated LPs in ways that conventional compiler optimizations and solver presolve techniques cannot achieve in isolation. 
Given the observed variability in solver behavior even across models of similar size, a particularly promising direction for future work is to study these techniques in systematically integrated Compact Integer Programs (CIPs) of exponential Integer Programs, where their polynomial-time separation oracles would no longer be required.

% \bibliographystyle{plain}   
% \bibliography{hsbcs}

\begin{thebibliography}{99}

\bibitem{achterberg_presolve_2020}
Achterberg,~T., Bixby,~R.~E., Gu,~Z., Rothberg,~E. and Weninger,~D.,
Presolve reductions in mixed integer programming,
\textit{INFORMS Journal on Computing},
\textbf{32} (2020), no.~2, 473--506.


\bibitem{aprile_smaller_2021}
Aprile,~M., Fiorini,~S., Huynh,~T., Joret,~G. and Wood,~D.~R.,
Smaller extended formulations for spanning tree polytopes in minor-closed classes and beyond,
\textit{The Electronic Journal of Combinatorics},
\textbf{28} (2021), P4.47.


\bibitem{david_avis_sparktope_2020}
Avis,~D. and Bremner,~D.,
\textit{Sparktope},
2020,
\url{https://gitlab.com/sparktope/sparktope},
Additional source: \url{https://zenodo.org/records/3818420}.


\bibitem{avis_sparktope_2022}
Avis,~D. and Bremner,~D.,
Sparktope: linear programs from algorithms,
\textit{Optimization Methods and Software},
\textbf{37} (2022), no.~3, 954--981.


\bibitem{avis_polynomial_2019}
Avis,~D., Bremner,~D., Tiwary,~H.~R. and Watanabe,~O.,
Polynomial size linear programs for problems in P,
\textit{Discrete Applied Mathematics},
\textbf{265} (2019), 22--39.


\bibitem{berrett_optimal_2023}
Berrett,~T.~B. and Samworth,~R.~J.,
Optimal nonparametric testing of missing completely at random and its connections to compatibility,
\textit{The Annals of Statistics},
\textbf{51} (2023), no.~5, 2170--2193.


\bibitem{cardoso_compiling_2010}
Cardoso,~J.~M.~P., Diniz,~P.~C. and Weinhardt,~M.,
Compiling for reconfigurable computing: A survey,
\textit{ACM Computing Surveys},
\textbf{42} (2010), no.~4, 13:1--13:65.


\bibitem{carr_compact_2002}
Carr,~R.~D. and Lancia,~G.,
Compact vs. exponential-size LP relaxations,
\textit{Operations Research Letters},
\textbf{30} (2002), no.~1, 57--65.


\bibitem{comaneci_tropical_2024}
Com\u{a}neci,~A. and Joswig,~M.,
Tropical medians by transportation,
\textit{Mathematical Programming},
\textbf{205} (2024), no.~1, 813--839.


\bibitem{conforti_extended_2010}
Conforti,~M., Cornu\'ejols,~G. and Zambelli,~G.,
Extended formulations in combinatorial optimization,
\textit{4OR},
\textbf{8} (2010), no.~1, 1--48.

\bibitem{dobkin_linear_1979}
Dobkin,~D., Lipton,~R.~J. and Reiss,~S.,
Linear programming is log-space hard for P,
\textit{Information Processing Letters},
\textbf{8} (1979), no.~2, 96--97.


\bibitem{edmonds_maximum_1965}
Edmonds,~J.,
Maximum matching and a polyhedron with 0,1-vertices,
\textit{Journal of Research of the National Bureau of Standards Section B Mathematics and Mathematical Physics},
\textbf{69B} (1965), no.~1--2, 125--130.


\bibitem{edmonds_paths_1965}
Edmonds,~J.,
Paths, trees, and flowers,
\textit{Canadian Journal of Mathematics},
\textbf{17} (1965), 449--467.


\bibitem{eli_towle_remove_2019}
Towle,~E.,
Remove variables with 0 coefficient in the LP file,
\textit{Gurobi Help Center},
2019,
\url{http://support.gurobi.com/hc/en-us/community/posts/360048150752-Remove-variables-with-0-coefficient-in-the-LP-file}.


\bibitem{fiorini_smaller_2017}
Fiorini,~S., Huynh,~T., Joret,~G. and Pashkovich,~K.,
Smaller extended formulations for the spanning tree polytope of bounded-genus graphs,
\textit{Discrete \& Computational Geometry},
\textbf{57} (2017), no.~3, 757--761.


\bibitem{fiorini_exponential_2015}
Fiorini,~S., Massar,~S., Pokutta,~S., Tiwary,~H.~R. and de Wolf,~R.,
Exponential lower bounds for polytopes in combinatorial optimization,
\textit{Journal of the ACM},
\textbf{62} (2015), no.~2, 17:1--17:23.

\bibitem{grotschel_ellipsoid_1981}
Gr\"otschel,~M., Lov\'asz,~L. and Schrijver,~A.,
The ellipsoid method and its consequences in combinatorial optimization,
\textit{Combinatorica},
\textbf{1} (1981), no.~2, 169--197.


\bibitem{ibm_ibm_2025}
IBM,
\textit{CPXXgetnumrows and CPXgetnumrows},
2025,
IBM Documentation for IBM CPLEX Optimizer for z/OS,
\url{https://www.ibm.com/docs/en/cofz/22.1.2?topic=SS9UKU_22.1.2/com.ibm.cplex.zos.help/refcallablelibrary/cpxapi/getnumrows.htm}.


\bibitem{ibm_mip_2022}
IBM,
\textit{MIP kappa: detecting and coping with ill-conditioned MIP models},
2022,
\url{https://www.ibm.com/docs/en/icos/22.1.1?topic=tmpp-mip-kappa-detecting-coping-ill-conditioned-mip-models}.


\bibitem{jarck_exact_2020}
Jarck,~K.,
\textit{Exact Mixed-Integer Programming},
Ph.D. thesis,
Technische Universit\"at Berlin, 2020,
\url{https://depositonce.tu-berlin.de/handle/11303/11067}.


\bibitem{jusevicius_experimental_2021}
Jusevi\v{c}ius,~V., Oberdieck,~R. and Paulavi\v{c}ius,~R.,
Experimental analysis of algebraic modelling languages for mathematical optimization,
\textit{Informatica},
\textbf{32} (2021), no.~2, 283--304.


\bibitem{khosla_soft_2024}
Khosla,~M. and Williams,~A.~H.,
Soft matching distance: A metric on neural representations that captures single-neuron tuning,
\textit{Proceedings of UniReps: the First Workshop on Unifying Representations in Neural Models},
PMLR, 2024, pp.~326--341.


\bibitem{khosravi_efficient_2025}
Khosravi,~S. and Bremner,~D.,
Efficient compilation of algorithms into compact linear programs,
\textit{2025 IEEE International Conference on Collaborative Advances in Software and COmputiNg (CASCON)},
2025, pp.~361--366.


\bibitem{khosravi_smaller_2026}
Khosravi,~S. and Bremner,~D.,
Smaller linear programs from algorithms via constraint sharing,
\textit{Proceedings of the 15th International Conference on Operations Research and Enterprise Systems -- ICORES},
SciTePress, 2026, pp.~278--285.


\bibitem{lancia_deriving_2016}
Lancia,~G. and Serafini,~P.,
Deriving compact extended formulations via LP-based separation techniques,
\textit{Annals of Operations Research},
\textbf{240} (2016), no.~1, 321--350.


\bibitem{larisch_topaz_2024}
Larisch,~J., Thijm,~T.~A., Ahmad,~S., Wu,~P., Arnfeld,~T. and Fayed,~M.,
Topaz: Declarative and verifiable authoritative DNS at CDN-scale,
\textit{Proceedings of the ACM SIGCOMM 2024 Conference},
Association for Computing Machinery, 2024, pp.~891--903.


\bibitem{loera_combinatorics_2013}
De Loera,~J.~A. and Kim,~E.~D.,
Combinatorics and geometry of transportation polytopes: An update,
\textit{Discrete Geometry and Algebraic Combinatorics},
\textit{Contemporary Mathematics},
\textbf{625} (2013), American Mathematical Society, 37--76.


\bibitem{martin_using_1991}
Martin,~R.~K.,
Using separation algorithms to generate mixed integer model reformulations,
\textit{Operations Research Letters},
\textbf{10} (1991), no.~3, 119--128.


\bibitem{muchnick_advanced_1997}
Muchnick,~S.,
\textit{Advanced Compiler Design and Implementation},
1st edition, Morgan Kaufmann, 1997.


\bibitem{nethercote_minizinc_2007}
Nethercote,~N., Stuckey,~P.~J., Becket,~R., Brand,~S., Duck,~G.~J. and Tack,~G.,
MiniZinc: Towards a standard CP modelling language,
\textit{Principles and Practice of Constraint Programming -- CP 2007},
Springer, 2007, pp.~529--543.


\bibitem{ted_integer_2025}
Ralphs,~T.~K.,
\textit{Integer Programming (Lecture 3). ISE 418},
\textit{Scribd},
\url{https://coral.ise.lehigh.edu/~ted/files/ie418/lectures/Lecture3.pdf}.


\bibitem{rothvoss_01_2013}
Rothvoss,~T.,
Some 0/1 polytopes need exponential size extended formulations,
\textit{Mathematical Programming},
\textbf{142} (2013), no.~1, 255--268.


\bibitem{rothvoss_matching_2017}
Rothvoss,~T.,
The matching polytope has exponential extension complexity,
\textit{Journal of the ACM},
\textbf{64} (2017), no.~6, 41:1--41:19.


\bibitem{schade_lower_2025}
Schade,~J., Sinha,~M. and Weltge,~S.,
Lower bounds on the complexity of mixed-integer programs for stable set and knapsack,
\textit{Mathematical Programming},
2025.


\bibitem{schneck_survey_1973}
Schneck,~P.~B.,
A survey of compiler optimization techniques,
\textit{Proceedings of the ACM Annual Conference},
Association for Computing Machinery, 1973, pp.~106--113.


\bibitem{stanley_two_1986}
Stanley,~R.~P.,
Two poset polytopes,
\textit{Discrete \& Computational Geometry},
\textbf{1} (1986), no.~1, 9--23.


\bibitem{swietanowski_modular_1995}
Swietanowski,~A.,
A modular presolve procedure for large scale linear programming,
\textit{International Institute for Applied Systems Analysis},
(1995), no.~wp95113, 1--29.


\bibitem{tiwary_extension_2020}
Tiwary,~H.~R., Verdugo,~V. and Wiese,~A.,
On the extension complexity of scheduling polytopes,
\textit{Operations Research Letters},
\textbf{48} (2020), no.~4, 472--479.


\bibitem{valente_extending_2009}
Valente,~C., Mitra,~G., Sadki,~M. and Fourer,~R.,
Extending algebraic modelling languages for stochastic programming,
\textit{INFORMS Journal on Computing},
\textbf{21} (2009), no.~1, 107--122.

\bibitem{valiant_reducibility_1982}
Valiant,~L.~G.,
Reducibility by algebraic projections,
\textit{Enseignement Math\'ematique},
\textbf{28} (1982), no.~3-4, 253--268.

\end{thebibliography}
%-----------------------------------------------
\appendix
\section{Additional LP Size Results}
\label{sec:appendix-lp-size}
This appendix reports detailed LP size results for benchmarks in Table~\ref{tab:benchmark-config}, complementing the overall LP size reductions discussed in Section~\ref{sec:resultLpSize}.
Tables~\ref{tab:lp-sizes-spdj} to~\ref{tab:lp-sizes-lis} report the LP sizes, including the number of rows, columns and nonzeros, together with their LP file sizes. In these tables, rows denote constraints, columns denote variables, and nonzeros denote the number of nonzero elements in the coefficient matrix. The tables compare \UOlong{}, \CSlong{}, and \HSBlong{}.

The baseline LP compiler of~\cite{david_avis_sparktope_2020} lacks support for certain basic data structures and operators, including two-dimensional integer arrays and the subtraction, division and multiplication operators. In some benchmarks, we therefore implemented these operations using input lookup tables, rather than loop-based implementations, to avoid increasing algorithm running time and code size.

\begin{table}[hbtp] 
\centering
\caption{LP sizes for the \SPDJlong{} problem, using Dijkstra’s algorithm with adjacency lists and input size $n$.}
\label{tab:lp-sizes-spdj}
\scalebox{1}{
\begin{tabular}{c c c c c c c}
\toprule
\makecell{$n$} & 
\makecell{\TBn{}} &
\makecell{\OPT{}} & 
\makecell{rows\\$(\times 1000)$} & 
\makecell{columns\\$(\times 1000)$} & 
\makecell{nonzeros\\$(\times 1000)$} & 
\makecell{file\\(\GB{})} \\
\midrule
\multirow{3}{*}{10}
& \multirow{3}{*}{1,960}
     & \UO{}  & 6,684 & 617 & 24,398 & 0.48 \\
&    & \CS{}  & 4,447 & 533 &18,348 & 0.36 \\
&    & \HSB{} & \textbf{385} & \textbf{250} & \textbf{1,445} & \textbf{0.04} \\  
\midrule
\multirow{3}{*}{20}
& \multirow{3}{*}{6,510 }
     & \UO{}  & 42,986 & 2,851 & 160,878 & 3.25 \\
&    & \CS{}  & 26,772 & 2,539 & 114,963 & 2.29 \\
&    & \HSB{} & \textbf{2,369} & \textbf{1,232} & \textbf{8,345} & \textbf{0.23} \\  
\midrule
\multirow{3}{*}{30}
& \multirow{3}{*}{13,660 }
     & \UO{}  & 147,682 & 8,019 & 559,612 & 11.52 \\
&    & \CS{}  & 90,314 & 7,295 & 394,338 & 7.98 \\
&    & \HSB{} & \textbf{7,853} & \textbf{3,644} & \textbf{26,910} & \textbf{0.80} \\  
\midrule
\multirow{3}{*}{40}
& \multirow{3}{*}{23,410 }
     & \UO{}  & 324,261 & 16,084 & 1,234,591 &  25.90 \\
&    & \CS{}  & 195,513 & 14,843 & 860,518 & 17.70 \\
&    & \HSB{} & \textbf{17,155} & \textbf{7,654} & \textbf{58,082} & \textbf{1.76} \\  
\bottomrule
\end{tabular}
}
\end{table}

\begin{table}[hbtp]
\centering
\caption{LP sizes for the \KSlong{} problem, using a dynamic programming algorithm with input size $n$.}
\label{tab:lp-sizes-kd}
\scalebox{1}{
\begin{tabular}{c c c c c c c}
\toprule
\makecell{$n$} & 
\makecell{\TBn{}} &
\makecell{\OPT{}} & 
\makecell{rows\\$(\times 1000)$} & 
\makecell{columns\\$(\times 1000)$} & 
\makecell{nonzeros\\$(\times 1000)$} & 
\makecell{file\\(\GB{})} \\
\midrule
\multirow{3}{*}{10}
& \multirow{3}{*}{760}
     & \UO{}  & 1,198 & 133 & 4,316 & 0.08 \\
&    & \CS{}  & 873 & 113 & 3,484 & 0.06 \\
&    & \HSB{} & \textbf{145} & \textbf{80} & \textbf{524} & \textbf{0.01} \\  
\midrule
\multirow{3}{*}{40}
& \multirow{3}{*}{3,650}
     & \UO{}  & 14,483 & 909 & 54,699 & 1.05 \\
&    & \CS{}  & 9,488 & 788 & 40,724 & 0.76 \\
&    & \HSB{} & \textbf{1,218} & \textbf{407} & \textbf{4,464} & \textbf{0.11} \\  
\midrule
\multirow{3}{*}{80}
& \multirow{3}{*}{7,290}
     & \UO{}  & 52,256 & 2,282 & 201,266 & 3.86 \\
&    & \CS{}  & 32,437 & 2,020& 144,142 & 2.68 \\
&    & \HSB{} & \textbf{4,061} & \textbf{939} & \textbf{15,180} & \textbf{0.39} \\  
\midrule
\multirow{3}{*}{160}
& \multirow{3}{*}{14,570}
     & \UO{}  & 204,099 & 6,078 & 797,133 & 15.73 \\
&    & \CS{}  & 121,508 & 5,509 & 554,647 & 10.50 \\
&    & \HSB{} & \textbf{14,945} & \textbf{2,243} & \textbf{56,924} & \textbf{1.51} \\  
\bottomrule
\end{tabular}
}
\end{table}

\begin{table}[hbtp] 
\centering
\caption{LP sizes for the \MSTlong{} problem, using Prim’s algorithm with input size $n$. } 
\label{tab:lp-sizes-mst}
\scalebox{1}{
\begin{tabular}{c c c c c c c}
\toprule
\makecell{$n$} & 
\makecell{\TBn{}} &
\makecell{\OPT{}} & 
\makecell{rows\\$(\times 1000)$} & 
\makecell{columns\\$(\times 1000)$} & 
\makecell{nonzeros\\$(\times 1000)$} & 
\makecell{file\\(\MB{})} \\
\specialrule{.1em}{.05em}{.05em}
\multirow{3}{*}{3}
 & \multirow{3}{*}{573} &\UO{} & 1,277 & 182 & 4,342 &  84 \\
 & & \CS{}     & 782    & 145   & 3,229   & 61 \\
 &  &\HSB{} & \textbf{55}    & \textbf{73}    & \textbf{279}   & \textbf{6}  \\
\midrule
\multirow{3}{*}{5}
  & \multirow{3}{*}{1,753} & \UO{} & 8,059 & 911 & 28,726 &  564 \\
  & & \CS{}  & 4,359   & 624   & 19,468  &  374 \\
  &  & \HSB{} & \textbf{263}   & \textbf{249}   & \textbf{1,185} & \textbf{32}  \\
\midrule
\multirow{3}{*}{8}
  & \multirow{3}{*}{5,323} & \UO{} & 39,569 & 4,221 & 146,047 &  2,926 \\
  & & \CS{}  & 19,292  & 4,013   & 91,390  & 1,769 \\
  &  & \HSB{} & \textbf{978}   & \textbf{780}   & \textbf{4,266} & \textbf{139}  \\
\midrule
\multirow{3}{*}{12}
  & \multirow{3}{*}{14,703} & \UO{} & 247,862 & 10,337 & 941,313 & 19,395  \\
  & & \CS{}  &  108,722  & 9,734  & 549,475 &  10,815 \\
 &  & \HSB{} & \textbf{4,307} & \textbf{2,418} & \textbf{17,685} & \textbf{771}  \\
\bottomrule
\end{tabular}
}
\end{table}

\begin{table}[hbtp] 
\centering
\caption{LP sizes for the integer array search problem, using \BSlong{} algorithm with input size $n$. }
\label{tab:lp-sizes-bs}
\scalebox{0.9}{
\begin{tabular}{c c c c c c c}
\toprule
\makecell{$n$} & 
\makecell{\TBn{}} &
\makecell{\OPT{}} & 
\makecell{rows\\$(\times 1000)$} & 
\makecell{columns\\$(\times 1000)$} & 
\makecell{nonzeros\\$(\times 1000)$} & 
\makecell{file\\(\GB{})} \\
\midrule
\multirow{3}{*}{160}
& \multirow{3}{*}{144}
     & \UO{}  & 2,483 & 73 & 9,688 & 0.18 \\
&    & \CS{}  & 2,042 & 73 & 8,395 & 0.16 \\
&    & \HSB{} & \textbf{164} & \textbf{20} & \textbf{632} & \textbf{0.02} \\ 
\midrule
\multirow{3}{*}{640}
& \multirow{3}{*}{178}
     & \UO{}  & 14,286 & 280 & 56,277 & 1.08 \\
&    & \CS{}  & 11,749 & 281 & 48,711 & 0.93 \\
&    & \HSB{} & \textbf{916} & \textbf{62} & \textbf{3,561} & \textbf{0.12} \\ 
\midrule
\multirow{3}{*}{1,280}
& \multirow{3}{*}{195}
     & \UO{}  & 33,830 & 581 & 133,566 & 2.63 \\
&    & \CS{}  & 27,817 & 582 & 115,578 & 2.25 \\
&    & \HSB{} & \textbf{2,159} & \textbf{120} & \textbf{8,419} & \textbf{0.29} \\ 
\midrule
\multirow{3}{*}{2,560}
& \multirow{3}{*}{212}
     & \UO{}  & 79,270 & 1,225 & 313,438 & 6.27 \\
&    & \CS{}  & 65,162 & 1,226 & 271,176 & 5.35 \\
&    & \HSB{} & \textbf{5,050} & \textbf{243} & \textbf{19,736} & \textbf{0.68} \\ 
\midrule
\multirow{3}{*}{5,120}
& \multirow{3}{*}{229}
     & \UO{}  & 183,847 & 2,601 & 727,722 & 14.70 \\
&    & \CS{}  & 151,086 & 2,603 & 629,511 & 12.52 \\
&    & \HSB{} & \textbf{11,703} & \textbf{504} & \textbf{45,830} & \textbf{1.59} \\ 
\bottomrule
\end{tabular}
}
\end{table}

\begin{table}[hbtp]
\centering
\caption{LP sizes for the \MakeLowercase{\LISproblem{}} problem, using a patience-sorting algorithm with time complexity \LISLtimecomp{} (\LISL{}), with input size $n$.}
\label{tab:lp-sizes-lisl}
\scalebox{1}{
\begin{tabular}{c c c c c c c}
\toprule
\makecell{$n$} & 
\makecell{\TBn{}} &
\makecell{\OPT{}} & 
\makecell{rows\\$(\times 1000)$} & 
\makecell{columns\\$(\times 1000)$} & 
\makecell{nonzeros\\$(\times 1000)$} & 
\makecell{file\\(\GB{})} \\
\midrule
\multirow{3}{*}{10}
& \multirow{3}{*}{726}
     & \UO{} & 1,693 & 173 & 6,138 & 0.11  \\
&    & \CS{} &  1,205 & 156 & 4,810 & 0.09 \\
&    & \HSB{} & \textbf{114} & \textbf{67} & \textbf{428} & \textbf{0.01} \\
\midrule
\multirow{3}{*}{20}
& \multirow{3}{*}{1,726}
     & \UO{} &  7,795 & 602 & 29,053 & 0.55  \\
&    & \CS{} & 5,447 & 559 & 22,306 & 0.42  \\
&    & \HSB{} & \textbf{ 476} & \textbf{209} & \textbf{1,737} & \textbf{0.04}  \\ 
\midrule
\multirow{3}{*}{40}
& \multirow{3}{*}{4,006}
     & \UO{} &  37,179 & 2,280 & 141,130 & 2.71  \\
&    & \CS{} &  25,688 & 2,171 & 106,963 & 2.04 \\
&    & \HSB{} & \textbf{ 2,119} & \textbf{719} & \textbf{7,615} & \textbf{0.21}  \\ 
\midrule
\multirow{3}{*}{80}
& \multirow{3}{*}{9,126}
     & \UO{} & 180,375 & 9,392 & 692,354 & 13.45  \\
&    & \CS{} &  123,828 & 9,127 & 520,887 & 10.02  \\
&    & \HSB{} & \textbf{ 9,845} & \textbf{2,758} & \textbf{35,173} & \textbf{0.99} \\ 
\bottomrule
\end{tabular}
}
\end{table}

\begin{table}[hbtp]
\centering
\caption{LP sizes for the \SAMlong{} problem, using the standard merge algorithm, with input sizes $n$ and $m$. }
\label{tab:lp-sizes-sam}
\scalebox{0.85}{
\begin{tabular}{c c c c c c c c}
\toprule
\makecell{$n$} & 
\makecell{$m$} & 
\makecell{\TBn{}} &
\makecell{\OPT{}} & 
\makecell{rows\\$(\times 1000)$} & 
\makecell{columns\\$(\times 1000)$} & 
\makecell{nonzeros\\$(\times 1000)$} & 
\makecell{file\\(\GB{})} \\
\midrule
\multirow{3}{*}{10}
&  \multirow{3}{*}{7}
&  \multirow{3}{*}{403} 
     & \UO{} & 654 & 58 & 2,325 & 0.04 \\
& &    & \CS{} & 277 & 50 & 1,468 & 0.02 \\
& &    & \HSB{} & \textbf{38} & \textbf{29} & \textbf{162} & \textbf{0.00}  \\ 
\midrule
\multirow{3}{*}{20}
&  \multirow{3}{*}{15}
& \multirow{3}{*}{817}
     & \UO{} & 2,227 & 138 & 8,192 & 0.14 \\
& &    & \CS{} & 936 & 121 & 4,991 & 0.08  \\
& &    & \HSB{} & \textbf{121} & \textbf{64} & \textbf{493} & \textbf{0.01} \\ 
\midrule
\multirow{3}{*}{40}
&  \multirow{3}{*}{32}
& \multirow{3}{*}{1,668}
     & \UO{} & 8,392 & 337 & 31,779 & 0.57 \\
& &    & \CS{} & 3,552 & 302 & 18,874 & 0.31 \\
& &    & \HSB{} & \textbf{432} & \textbf{143} & \textbf{1,717} & \textbf{0.04} \\ 
\midrule
\multirow{3}{*}{80}
&  \multirow{3}{*}{75}
& \multirow{3}{*}{3,577}
     & \UO{} & 37,470 & 913 & 145,027 & 2.63 \\
& &    & \CS{} &  16,046 & 838 & 84,725 & 1.42  \\
& &    & \HSB{} & \textbf{1,853} & \textbf{346} & \textbf{7,296} & \textbf{0.18}  \\ 
\midrule
\multirow{3}{*}{160}
&  \multirow{3}{*}{143}
& \multirow{3}{*}{6,981}
     & \UO{} & 148,285 & 2,431 & 580,776 & 10.75 \\
& &    & \CS{} &  63,946 & 2,285 & 336,497 & 5.72  \\
& &    & \HSB{} & \textbf{ 7,189} & \textbf{792} & \textbf{28,250} & \textbf{0.72} \\ 
\midrule
\multirow{3}{*}{320}
&  \multirow{3}{*}{285}
& \multirow{3}{*}{13,927}
     & \UO{} & 626,898 & 7,261 & 2,472,866 & 47.18  \\
& &    & \CS{} &  271,521 & 6,969 & 1,426,146 & 24.98 \\
& &    & \HSB{} & \textbf{ 30,046} & \textbf{2,002} & \textbf{118,147} & \textbf{3.13}  \\ 
\bottomrule
\end{tabular}
}
\end{table}

\begin{table}[hbtp] 
\centering
\caption{LP sizes for the \TSlong{} problem using Kahn's algorithm, with input size $n$.}
\label{tab:lp-sizes-ts}
\scalebox{0.85}{
\begin{tabular}{c c c c c c c}
\toprule
\makecell{$n$} & 
\makecell{\TBn{}} &
\makecell{\OPT{}} & 
\makecell{rows\\$(\times 1000)$} & 
\makecell{columns\\$(\times 1000)$} & 
\makecell{nonzeros\\$(\times 1000)$} & 
\makecell{file\\(\GB{})} \\
\midrule
\multirow{3}{*}{10}
& \multirow{3}{*}{928}
     & \UO{} & 2,980 & 318 & 10,783 & 0.21\\
&    & \CS{} &  1,733 & 251 & 7,455 & 0.14 \\
&    & \HSB{} & \textbf{161} & \textbf{120} & \textbf{640} & \textbf{0.02} \\
\midrule
\multirow{3}{*}{20}
& \multirow{3}{*}{1,848}
     & \UO{} &  12,144 & 946 & 45,307 & 0.89 \\
&    & \CS{} & 6,780 & 792 & 30,155 & 0.57 \\
&    & \HSB{} & \textbf{ 614} & \textbf{325} & \textbf{576} & \textbf{0.06} \\ 
\midrule
\multirow{3}{*}{40}
& \multirow{3}{*}{3,688}
     & \UO{} & 50,623 & 3,120 & 192,382 & 3.88 \\
&    & \CS{} &  27,494 & 2,773 & 124,843 & 2.42 \\
&    & \HSB{} & \textbf{2,485} & \textbf{1,016} & \textbf{8,833} & \textbf{0.24} \\ 
\midrule
\multirow{3}{*}{80}
& \multirow{3}{*}{7,368}
     & \UO{} & 219,722 & 11,531 & 844,380 & 17.23  \\
&    & \CS{} & 117,527 & 10,758 & 540,255 & 10.57 \\
&    & \HSB{} & \textbf{10,601} & \textbf{3,645} & \textbf{36,961} & \textbf{1.03} \\ 
\bottomrule
\end{tabular}
}
\end{table}

\begin{table}[hbtp] 
\centering
\caption{LP sizes for the \MSlong{} problem, using the algorithm from~\cite{avis_sparktope_2022}, with $m$ jobs on $n=3$ machines. }
\label{tab:lp-sizes-ms}
\scalebox{1}{
\begin{tabular}{c c c c c c c}
\toprule
\makecell{$m$} & 
\makecell{\TBn{}} &
\makecell{\OPT{}} & 
\makecell{rows\\$(\times 1000)$} & 
\makecell{columns\\$(\times 1000)$} & 
\makecell{nonzeros\\$(\times 1000)$} & 
\makecell{file\\(MB)} \\
\midrule
\multirow{3}{*}{5}
& \multirow{2}{*}{201} & \UO{}  &  157&	30	&521   & 10\\ 
&  & \CS{} &  122&	27	&443  & 8\\ 
 & 194 & \HSB{} & \textbf{9}&	\textbf{14}&	\textbf{51}  & \textbf{1} \\ 
\midrule
\multirow{3}{*}{10}
  & \multirow{2}{*}{321}  &\UO{} & 380 & 58 & 1,304  & 23  \\ 
  &  & \CS{} & 281 & 55 & 1,074  & 19  \\
  & 374 & \HSB{} & \textbf{25} & \textbf{39} & \textbf{124}  &\textbf{3}  \\
\midrule
\multirow{3}{*}{20}
  & \multirow{2}{*}{631}  &\UO{} & 1,170 & 148 & 4,160 &  75  \\ 
  &  & \CS{} & 831 & 143 & 3,330 &  59 \\ 
 & 734 & \HSB{} & \textbf{74} & \textbf{92} & \textbf{331}   &\textbf{7}  \\
\midrule
\multirow{3}{*}{40}
  & \multirow{2}{*}{1,251} &\UO{} & 3,845 & 411 & 14,099  & 255  \\ 
  &  & \CS{} & 2,636 & 402 & 11,030  & 196  \\ 
 &1,454 &  \HSB{} & \textbf{236} & \textbf{232} & \textbf{976}  & \textbf{21} \\
\midrule
\multirow{3}{*}{80}
 & \multirow{2}{*}{2,491} &\UO{} & 13,483 & 1,250 & 50,634  & 936  \\ 
 &  & \CS{} & 9,008 & 1,237 & 38,998  & 701  \\
 & 2,894 & \HSB{} & \textbf{813} & \textbf{636} & \textbf{3,196}  &\textbf{69}  \\
\midrule
\multirow{3}{*}{160}
  &\multirow{2}{*}{4,971}  &\UO{} & 49,869 & 4,150 & 190,407 &  3,595 \\ 
  &  & \CS{} & 32,836 & 4,135 & 145,422  & 2,648 \\
 & 5,774 & \HSB{} & \textbf{2,999} & \textbf{1,923} & \textbf{11,445}  & \textbf{249} \\ 
\bottomrule
\end{tabular}
}
\end{table}

\begin{table}[hbtp] 
\centering
\caption{LP sizes for the \LISproblem{} (\LIS{}) problem, using a quadratic-time dynamic programming algorithm, with input size $n$.}
\label{tab:lp-sizes-lis}
\scalebox{1}{
\begin{tabular}{c c c c c c c}
\toprule
\makecell{$n$} & 
\makecell{\TBn{}} &
\makecell{\OPT{}} & 
\makecell{rows\\$(\times 1000)$} & 
\makecell{columns\\$(\times 1000)$} & 
\makecell{nonzeros\\$(\times 1000)$} & 
\makecell{file\\(\GB{})} \\
\midrule
\multirow{3}{*}{5}
& \multirow{2}{*}{251}
     & \UO{} &  212 & 36 & 714 & 0.01 \\
&    & \CS{} &  166 & 34 & 615 & 0.01 \\
\cmidrule(lr){2-7}
&  366 & \HSB{} & \textbf{29} & \textbf{26} & \textbf{121} & \textbf{0.00} \\
\midrule
\multirow{3}{*}{10}
& \multirow{2}{*}{976}
     & \UO{} & 1,480 & 189 & 1,378 & 0.10 \\
&    & \CS{} & 1,077 & 180 & 4,207 & 0.08 \\
\cmidrule(lr){2-7}
& 1,661   & \HSB{} & \textbf{234} & \textbf{151} & \textbf{890} & \textbf{0.02} \\
\midrule
\multirow{3}{*}{20}
& \multirow{2}{*}{3,851}
     & \UO{} &  11,157 & 1,086 & 10,518 & 0.78 \\
&    & \CS{} &  7,600 & 1,055 & 7,041 & 0.60 \\
\cmidrule(lr){2-7}
&  7,101  & \HSB{} & \textbf{ 1,892} & \textbf{899} & \textbf{6,855} & \textbf{0.16} \\
\midrule
\multirow{3}{*}{30}
& \multirow{2}{*}{8,626}
     & \UO{} &  33,792 & 2,950 & 123,303 & 2.40 \\
&    & \CS{} & 22,373 & 2,881 & 93,323 & 1.81 \\
\cmidrule(lr){2-7}
&  16,341  & \HSB{} & \textbf{5,906} & \textbf{2,502} & \textbf{21,117} & \textbf{0.52} \\
\midrule
\multirow{3}{*}{40}
& \multirow{2}{*}{15,301}
     & \UO{} & 90,239 & 7,038 & 332,255 & 6.61  \\
&    & \CS{} & 58,492 & 6,931 & 247,937 & 4.90 \\
\cmidrule(lr){2-7}
&  29,381  & \HSB{} & \textbf{15,912} & \textbf{5,958} & \textbf{56,213} & \textbf{1.42} \\
\bottomrule
\end{tabular}
}
\end{table}

\clearpage
\section{Additional Solver Performance Results}
\label{sec:appendix-solver-performance}
This appendix provides detailed solver run time and memory usage results for all benchmarks in Table~\ref{tab:benchmark-config}, to complement the overall comparison presented in Section~\ref{sec:resultSolverPerformance}.
Tables~\ref{tab:time-spdj-notforced} to~\ref{tab:time-ms-notforced} report the median elapsed time, presolve time, and RSS memory usage across runs specified in Table~\ref{tab:benchmark-config}. 
\IQR{} is reported as a percentage of the median. 
\OOM{} indicates an out-of-memory error, \OOT{} a timeout, and \FAIL{} solver failure.

\begin{table}[hbtp]
\centering
\caption{
Solver run times and memory usage for the \SPDJlong{} problem, using Dijkstra’s algorithm with adjacency lists and input size $n$, where problem inputs are encoded through the objective function.
Each result is based on 5 runs, with a 23-hour time limit, 64~GB memory, and 8~CPUs.
}
\label{tab:time-spdj-notforced}
\scalebox{0.85}{
\begin{tabular}{ccccccccc}
\toprule
\multicolumn{3}{c}{Info} & \multicolumn{4}{c}{Time (\Hour{})} & \multicolumn{2}{c}{Memory (\GB{})} \\
\cmidrule(lr){1-3} \cmidrule(lr){4-7} \cmidrule(lr){8-9}
$n$ & solver & \OPT{} & \multicolumn{2}{c}{Elapsed}  & \multicolumn{2}{c}{Presolve} & \multicolumn{2}{c}{RSS} \\
 &  &   &  \Med{}& \IQR{}& \Med{}& \IQR{}& \Med{}& \IQR{}\\
\midrule
\multirow{9}{*}{10}
 & \multirow{3}{*}{\cplex} & \UO{}  & 6.09 & 0.38\% & 0.00 & 22.15\% & 16.27 & 0.12\% \\
 &  & \CS{}  & 2.23 & 1.64\% & 0.00 & 16.34\% & 11.55 & 0.25\% \\
 &  & \HSB{} & \textbf{0.00} & 1.16\% & 0.00 & 0.00\% & \textbf{0.93} & 0.55\% \\
\cmidrule{2-9}
 & \multirow{3}{*}{\scip} & \UO{}  & \OOT{} & \FAIL{} & \FAIL{} & \FAIL{} & \FAIL{} & \FAIL{}   \\
 &  & \CS{}  & \OOT{} & \FAIL{} & \FAIL{} & \FAIL{} & \FAIL{} & \FAIL{}  \\
 &  & \HSB{}  &  0.05 & 0.11\% & 0.00 & 0.57\% & 1.92 & 0.13\%  \\
\cmidrule{2-9}
 & \multirow{3}{*}{\gurobi} & \UO{}  & 0.53 & 7.20\% & 0.01 & 8.88\% & 13.14 & 0.06\%  \\
 &  & \CS{}  &   0.24 & 17.67\% & 0.01 & 15.39\% & 10.00 & 0.08\%   \\
 &  & \HSB{}  &  \textbf{0.01} & 2.39\% & \textbf{0.00} & 3.80\% & \textbf{0.63} & 0.51\% \\
\midrule
\multirow{7}{*}{20}
 &  \multirow{2}{*}{\cplex} & \UO{}  &  \OOM{} & \FAIL{} & \FAIL{} & \FAIL{} & \FAIL{} & \FAIL{}  \\
 &  & \CS{}  &  \OOM{} & \FAIL{} & \FAIL{} & \FAIL{} & \FAIL{} & \FAIL{}   \\
 &  & \HSB{}  & 0.02 & 3.30\% & 0.00 & 1.43\% & 5.12 & 0.06\%  \\ 
\cmidrule{2-9}
 & \multirow{1}{*}{\scip} &  \HSB{}  &  0.68 & 0.45\% & 0.01 & 0.90\% & 10.49 & 0.03\% \\
\cmidrule{2-9}
 & \multirow{3}{*}{\gurobi} & \UO{}  & \OOM{} & \FAIL{} & \FAIL{} & \FAIL{} & \FAIL{} & \FAIL{}  \\
 &  & \CS{}  &  \OOM{} & \FAIL{} & \FAIL{} & \FAIL{} & \FAIL{} & \FAIL{}  \\
 &  & \HSB{}  &  0.11 & 0.79\% & 0.00 & 5.36\% & 3.58 & 0.28\%   \\
 \midrule
\multirow{3}{*}{30}
 &  \multirow{1}{*}{\cplex} & \HSB{} & 0.07 & 0.97\% & 0.00 & 0.69\% & 16.04 & 0.04\%  \\
\cmidrule{2-9}
 & \multirow{1}{*}{\scip} & \HSB{}  & 7.52 & 1.40\% & 0.05 & 0.71\% & 35.92 & 0.08\% \\
\cmidrule{2-9}
 & \multirow{1}{*}{\gurobi} &  \HSB{}  &  0.41 & 0.71\% & 0.02 & 9.07\% & 11.22 & 0.12\%  \\
 \midrule
\multirow{3}{*}{40}
 &  \multirow{1}{*}{\cplex} &  \HSB{}  & 0.25 & 0.80\% & 0.01 & 3.34\% & 32.90 & 0.02\%   \\
\cmidrule{2-9}
 & \multirow{1}{*}{\scip} &  \HSB{}  & \OOM{} & \FAIL{} & \FAIL{} & \FAIL{} & \FAIL{} & \FAIL{}  \\ 
\cmidrule{2-9}
 & \multirow{1}{*}{\gurobi} & \HSB{}  &   0.85 & 0.43\% & 0.04 & 4.92\% & 24.14 & 0.14\%  \\
 \bottomrule
\end{tabular}
}
\end{table}

\begin{table}[hbtp]
\centering
\caption{Solver run times and memory usage for the \KSlong{} problem, using a dynamic programming algorithm with input size $n$, where problem inputs are encoded through the objective function.
Each result is based on 5 runs, with a 23-hour time limit, 64~GB memory, and 8~CPUs.
}
\label{tab:time-kd-notforced}
\scalebox{1}{
\begin{tabular}{ccccccccc}
\toprule
\multicolumn{3}{c}{Info} & \multicolumn{4}{c}{Time (\Hour{})} & \multicolumn{2}{c}{Memory (\GB{})} \\
\cmidrule(lr){1-3} \cmidrule(lr){4-7} \cmidrule(lr){8-9}
$n$ & solver & \OPT{} & \multicolumn{2}{c}{Elapsed}  & \multicolumn{2}{c}{Presolve} & \multicolumn{2}{c}{RSS} \\
 &  &   &  \Med{}& \IQR{}& \Med{}& \IQR{}& \Med{}& \IQR{}\\
\midrule
\multirow{9}{*}{10}
 & \multirow{3}{*}{\cplex} & \UO{}  & 0.09 & 2.20\% & 0.00 & 15.94\% & 3.18 & 3.11\%  \\
 &  & \CS{} & 0.19 & 3.10\% & 0.00 & 24.88\% & 2.78 & 0.67\%  \\
 &  & \HSB{} & \textbf{0.00} & 4.06\% & 0.00 & 5.00\% & \textbf{0.42} & 0.63\%  \\
\cmidrule{2-9}
 & \multirow{3}{*}{\scip} & \UO{} & 1.54 & 0.99\% & 0.01 & 2.52\% & 4.89 & 0.12\%  \\
& & \CS{} & 2.29 & 2.32\% & \textbf{0.00} & 3.55\% & 3.85 & 0.08\%  \\
 & & \HSB{} & \textbf{0.01} & 4.48\% & \textbf{0.00} & 4.42\% & \textbf{0.71} & 0.22\% \\
\cmidrule{2-9}
 & \multirow{3}{*}{\gurobi} & \UO{}  & 0.13 & 0.97\% & 0.00 & 12.32\% & 2.24 & 0.34\%  \\
 &  & \CS{}  & 0.02 & 4.42\% & 0.00 & 7.52\% & 2.01 & 0.37\%  \\
 &  & \HSB{} & \textbf{0.00} & 5.30\% & 0.00 & 1.11\% & \textbf{0.29} & 0.54\%  \\
\midrule
\multirow{9}{*}{40}
 &  \multirow{2}{*}{\cplex} & \UO{} & 9.28 & 0.17\% & 0.01 & 3.02\% & 31.34 & 0.49\%   \\
 &  & \CS{} & 6.65 & 0.65\% & 0.01 & 12.97\% & 23.11 & 0.34\%  \\
 &  & \HSB{} & \textbf{0.01} & 1.01\% & \textbf{0.00} & 2.02\% & \textbf{2.97} & 0.60\%  \\ 
\cmidrule{2-9}
 & \multirow{3}{*}{\scip} & \UO{} & \OOM{}   & \FAIL{} & \FAIL{} & \FAIL{} & \FAIL{} & \FAIL{}  \\
 & & \CS{} & \OOT{} & \FAIL{} & \FAIL{} & \FAIL{} & \FAIL{} & \FAIL{}  \\
 & & \HSB{} & \textbf{0.11} & 1.85\% & \textbf{0.01} & 4.69\% & \textbf{5.63} & 0.06\%  \\ 
\cmidrule{2-9}
 & \multirow{3}{*}{\gurobi} & \UO{} & 1.03 & 0.22\% & 0.03 & 2.21\% & 22.48 & 0.06\%  \\
 &  & \CS{}  & 0.26 & 2.58\% & 0.02 & 17.91\% & 18.24 & 0.04\%  \\
 &  & \HSB{} & \textbf{0.06} & 0.37\% & \textbf{0.00} & 9.24\% & \textbf{3.31} & 0.22\%  \\
 \midrule
\multirow{7}{*}{80}
 &  \multirow{2}{*}{\cplex} & \UO{} & \OOM{} & \FAIL{} & \FAIL{} & \FAIL{} & \FAIL{} & \FAIL{}   \\
 & & \CS{} & \OOM{} & \FAIL{} & \FAIL{} & \FAIL{} & \FAIL{} & \FAIL{} \\
 &  & \HSB{} & \textbf{0.03} & 1.26\% & \textbf{0.00} & 1.05\% & \textbf{7.76} & 0.12\%   \\
\cmidrule{2-9}
 & \multirow{1}{*}{\scip} & \HSB{}  & \textbf{0.48} & 0.41\% & \textbf{0.02} & 1.09\% & \textbf{21.58} & 0.02\%   \\ 
\cmidrule{2-9}
 & \multirow{3}{*}{\gurobi} & \UO{} & \OOM{} & \FAIL{} & \FAIL{} & \FAIL{} & \FAIL{} & \FAIL{}   \\
 &  & \CS{} & \OOM{} & \FAIL{} & \FAIL{} & \FAIL{} & \FAIL{} & \FAIL{}  \\
 &  & \HSB{} & \textbf{0.31} & 0.40\% & \textbf{0.01} & 2.68\% & \textbf{9.92} & 0.05\%  \\
 \midrule
\multirow{3}{*}{160}
 &  \multirow{1}{*}{\cplex} & \HSB{} & \textbf{0.60} & 0.31\% & \textbf{0.02} & 0.85\% & \textbf{29.08} & 0.02\% \\
\cmidrule{2-9}
 & \multirow{1}{*}{\scip} &  \HSB{}  & \OOT{} & \FAIL{} & \FAIL{} & \FAIL{} & \FAIL{} & \FAIL{}  \\ 
\cmidrule{2-9}
 & \multirow{1}{*}{\gurobi} & \HSB{} & \textbf{1.38} & 0.36\% & \textbf{0.03} & 0.86\% & \textbf{31.57} & 0.24\%  \\
 \bottomrule
\end{tabular}}
\end{table}

\begin{table}[hbtp] 
\centering
\caption{Solver run times and memory usage for the \MSTlong{} problem, using Prim’s algorithm with input size $n$, with inputs encoded through the objective function. 
Each result is based on 10 runs, with a 23-hour time limit, 32~GB memory, and 7~CPUs.
A superscript $^*$ indicates the dual simplex algorithm.  
} 
\label{tab:time-mst-notforced}
\scalebox{1}{
\begin{tabular}{ccccccccc}
\toprule
\multicolumn{3}{c}{Info} & \multicolumn{4}{c}{Time (\Sec{})} & \multicolumn{2}{c}{Memory (\MB{})} \\
\cmidrule(lr){1-3} \cmidrule(lr){4-7} \cmidrule(lr){8-9}
$n$ & solver & \OPT{} & \multicolumn{2}{c}{Elapsed}  & \multicolumn{2}{c}{Presolve} & \multicolumn{2}{c}{RSS} \\
 &  &   &  \Med{}& \IQR{}& \Med{}& \IQR{}& \Med{}& \IQR{}\\
\midrule
\multirow{12}{*}{3}
 & \multirow{3}{*}{\cplex$^*$} & \UO{}  & \OOT{} & \FAIL{}& \FAIL{}&\FAIL{} & \FAIL{}&\FAIL{} \\
 &                             & \CS{} & 5186.13 & 5.15\% & 1.73 & 4.91\% & 835 & 0.69\% \\
 &                             & \HSB{} & \textbf{0.85} & 9.71\% & \textbf{0.10} & 0.00\% & \textbf{74} & 2.03\% \\
 \cmidrule{2-9}
 & \multirow{3}{*}{\cplex} & \UO{}  & 4514.53 & 14.43\% &  3.32 & 13.25\% & 3557 & 1.15\% \\
 &                         & \CS{} & 509.96 & 26.85\% & 2.38 & 47.69\% & 2626 & 4.30\% \\
 &                         & \HSB{} & \textbf{1.04} & 6.04\% & \textbf{0.12} & 0.00\% & \textbf{227} & 2.60\% \\
 \cmidrule{2-9}
  & \multirow{3}{*}{\scip} & \UO{}  & \OOM{} & \FAIL{} & \FAIL{} & \FAIL{} & \FAIL{} & \FAIL{} \\
 &                         & \CS{} & 2869.61 & 1.04\% & 17.95 & 1.67\% & 3750 & 0.09\% \\ 
 &  & \HSB{}  & \textbf{3.39} & 1.40\% &  \textbf{0.87} & 3.45\% & \textbf{376} & 0.39\% \\
 \cmidrule{2-9}
  & \multirow{3}{*}{\gurobi} & \UO{}  & 699.04 & 1.07\% &  8.00 & 0.91\% & 2785 & 0.49\% \\ 
  &  & \CS{}  & 232.46 & 11.85\% & 4.93 & 5.38\% & 2103 & 2.34\% \\
 &  & \HSB{}  & \textbf{0.81} & 39.88\% &  \textbf{0.36} & 2.08\% & \textbf{131} & 32.18\% \\
\midrule
\multirow{10}{*}{5}
& \multirow{1}{*}{\cplex$^*$} & \HSB{} & \textbf{4.62} & 5.47\% & \textbf{0.53} & 7.62\% & \textbf{264} & 0.90\% \\
 \cmidrule{2-9}
 & \multirow{3}{*}{\cplex} & \UO{}  & 8968.60 & 13.40\% & 21.85 & 3.36\% & 18,784 & 0.32\% \\ 
    &  & \CS{}  & 4274.74 & 13.36\% & 14.24 & 96.33\% & 11973 & 1.29\% \\ 
 &  & \HSB{} & \textbf{4.69} & 7.79\% & \textbf{0.62} & 13.31\% & \textbf{811} & 0.45\% \\ 
 \cmidrule{2-9}
  & \multirow{2}{*}{\scip} &  \CS{} & \OOT{} & \FAIL{} & \FAIL{} & \FAIL{} & \FAIL{} & \FAIL{} \\
 &  & \HSB{}  & \textbf{19.52} & 2.52\% & \textbf{5.38} & 1.95\% & \textbf{1883} & 0.08\% \\
 \cmidrule{2-9}
  & \multirow{3}{*}{\gurobi} & \UO{}  & 2977.82 & 12.57\% &  61.95 & 7.91\% & 15,229 & 0.04\% \\
 &                          & \CS{}  & 1694.73 & 23.45\% & 35.45 & 32.93\% & 10,077 & 0.97\% \\
 &                          & \HSB{}  & \textbf{4.61} & 2.33\% & \textbf{2.25} & 2.33\% & \textbf{619} & 0.50\% \\
\midrule
\multirow{7}{*}{8}
& \multirow{1}{*}{\cplex$^*$} & \HSB{} & \textbf{15.36} & 3.73\% & \textbf{2.27} & 4.07\% & \textbf{844} & 0.82\% \\ 
 \cmidrule{2-9}
 & \multirow{2}{*}{\cplex} & \UO{}  & \OOM{} & \FAIL{} & \FAIL{} & \FAIL{} & \FAIL{} & \FAIL{} \\
   &                       &\HSB{} & \textbf{23.60} & 17.40\% & \textbf{2.59} & 22.74\% & \textbf{2606} & 0.80\% \\
 \cmidrule{2-9}
  & \multirow{1}{*}{\scip} & \HSB{}  & \textbf{93.75} & 9.23\% & \textbf{23.77} & 2.70\% & \textbf{8522} & 0.03\% \\
 \cmidrule{2-9}
  & \multirow{2}{*}{\gurobi} & \UO{}  & \OOM{} & \FAIL{} & \FAIL{} & \FAIL{} & \FAIL{} & \FAIL{} \\
    &                         & \HSB{}  & \textbf{22.91} & 2.22\% &  \textbf{9.79} & 1.38\% & \textbf{2858} & 0.21\% \\
\midrule
\multirow{4}{*}{12}
& \multirow{1}{*}{\cplex$^*$} & \HSB{} & \textbf{67.07} & 3.21\% & \textbf{12.21} & 2.25\% & \textbf{3246} & 0.33\% \\ 
 \cmidrule{2-9}
  & \multirow{1}{*}{\cplex} & \HSB{} & \textbf{72.89} & 8.58\% & \textbf{13.70} & 14.00\% & \textbf{10,079} & 0.22\% \\ 
 \cmidrule{2-9}
  & \multirow{1}{*}{\scip} &  \HSB{}  & \OOM{} & \FAIL{} & \FAIL{} & \FAIL{} & \FAIL{} & \FAIL{} \\
 \cmidrule{2-9}
  & \multirow{1}{*}{\gurobi} &  \HSB{}  & \textbf{320.10} & 10.10\% & \textbf{44.72} & 6.74\% & \textbf{15,784} & 0.03\% \\
\bottomrule
\end{tabular}}
\end{table}

\begin{table}[hbtp]
\centering
\caption{Solver run times and memory usage for the integer array search problem, using \BSlong{} algorithm with input size $n$, where problem inputs are encoded through the objective function.
Each result is based on 5 runs, with a 23-hour time limit, 32~GB memory, and 7~CPUs.
A superscript $^*$ indicates the dual simplex algorithm.
}
\label{tab:time-bs-notforced}
\scalebox{0.8}{
\begin{tabular}{ccccccccc}
\toprule
\multicolumn{3}{c}{Info} & \multicolumn{4}{c}{Time (\Hour{})} & \multicolumn{2}{c}{Memory (\GB{})} \\
\cmidrule(lr){1-3} \cmidrule(lr){4-7} \cmidrule(lr){8-9}
$n$ & solver & \OPT{} & \multicolumn{2}{c}{Elapsed}  & \multicolumn{2}{c}{Presolve} & \multicolumn{2}{c}{RSS} \\
 &  &   &  \Med{}& \IQR{}& \Med{}& \IQR{}& \Med{}& \IQR{}\\
\midrule
\multirow{12}{*}{160}
 & \multirow{3}{*}{\cplex$^*$} & \UO{} & 2.78 & 3.09\% & 0.00 & 3.81\% & 2.14 & 0.25\%  \\
 & & \CS{}  & 0.20 & 1.31\% & 0.00 & 1.14\% & 1.84 & 0.47\% \\
 & & \HSB{}  & \textbf{0.00} & 14.58\% & 0.00 & 5.43\% & \textbf{0.19} & 0.33\% \\
\cmidrule{2-9}
 & \multirow{3}{*}{\cplex} & \UO{}  & 1.05 & 2.06\% & 0.00 & 0.20\% & 6.50 & 0.45\% \\
 &  & \CS{}   & 0.17 & 1.02\% & 0.00 & 16.59\% & 4.77 & 0.21\%  \\
 &  & \HSB{}  &\textbf{0.00} & 7.05\% & 0.00 & 1.09\% & \textbf{0.48} & 0.36\% \\
\cmidrule{2-9}
 & \multirow{3}{*}{\scip} & \UO{} & 2.29 & 0.50\% & 0.01 & 0.58\% & 7.50 & 0.06\% \\
& & \CS{} & 0.73 & 0.66\% & 0.01 & 3.32\% & 6.50 & 0.04\%\\
 & & \HSB{}  & \textbf{0.04} & 0.34\% & \textbf{0.00} & 0.74\% & \textbf{1.36} & 0.19\% \\
\cmidrule{2-9}
 & \multirow{3}{*}{\gurobi} & \UO{}  & 0.50 & 7.19\% & 0.00 & 2.48\% & 4.66 & 0.14\% \\ 
 &  & \CS{}  & 0.21 & 3.35\% & 0.00 & 22.81\% & 3.94 & 0.17\% \\
 &  & \HSB{} & \textbf{0.02} & 8.45\% & 0.00 & 0.46\% & \textbf{0.60} & 1.59\% \\
\midrule
\multirow{12}{*}{640}
 &  \multirow{2}{*}{\cplex$^*$} & \UO{}  & \OOT{} & \FAIL{} & \FAIL{} & \FAIL{} & \FAIL{} & \FAIL{} \\
 &  & \CS{}  & 2.24 & 1.61\% & 0.02 & 0.25\% & 9.78 & 0.09\% \\
 &  & \HSB{}  & \textbf{0.08} & 1.43\% & \textbf{0.00} & 1.36\% & \textbf{0.95} & 0.37\% \\ 
\cmidrule{2-9}
 &  \multirow{2}{*}{\cplex} & \UO{} & \OOM{} & \FAIL{} & \FAIL{} & \FAIL{} & \FAIL{} & \FAIL{} \\ 
 &  & \CS{}  & 5.03 & 0.51\% & 0.03 & 8.80\% & 23.69 & 0.06\% \\
 &  & \HSB{} & \textbf{0.03} & 4.13\% & \textbf{0.00} & 0.22\% & \textbf{2.56} & 0.15\% \\ 
\cmidrule{2-9}
 & \multirow{3}{*}{\scip} & \UO{} & \OOM{} & \FAIL{} & \FAIL{} & \FAIL{} & \FAIL{} & \FAIL{} \\
 & & \CS{} & \OOM{} & \FAIL{} & \FAIL{} & \FAIL{} & \FAIL{} & \FAIL{}  \\
 & & \HSB{} & \textbf{0.25} & 0.14\% & \textbf{0.00} & 0.44\% & \textbf{7.60} & 0.05\% \\ 
\cmidrule{2-9}
 & \multirow{3}{*}{\gurobi} & \UO{} & \OOM{} & \FAIL{} & \FAIL{} & \FAIL{} & \FAIL{} & \FAIL{} \\ 
 &  & \CS{}  & 2.07 & 0.91\% & 0.02 & 8.41\% & 18.58 & 0.44\% \\
 &  & \HSB{} & \textbf{0.12} & 7.81\% & \textbf{0.00} & 0.43\% & \textbf{2.65} & 0.08\% \\
 \midrule
\multirow{9}{*}{1280}
 &  \multirow{2}{*}{\cplex$^*$} & \CS{} & 3.56 & 0.95\% & 0.10 & 0.14\% & 22.54 & 0.03\% \\
 &  & \HSB{} & \textbf{0.16} & 1.56\% & \textbf{0.01} & 0.47\% & \textbf{2.17} & 0.64\% \\
\cmidrule{2-9}
 &  \multirow{3}{*}{\cplex} & \UO{} & \OOT{} & \FAIL{} & \FAIL{} & \FAIL{} & \FAIL{} & \FAIL{} \\
 & & \CS{} & \OOM{} & \FAIL{} & \FAIL{} & \FAIL{} & \FAIL{} & \FAIL{} \\
 &  & \HSB{}  & \textbf{0.07} & 4.92\% & \textbf{0.01} & 0.29\% & \textbf{4.82} & 2.72\% \\ 
\cmidrule{2-9}
 & \multirow{1}{*}{\scip} & \HSB{}  & \textbf{0.57} & 3.79\% & \textbf{0.01} & 17.61\% & \textbf{17.94} & 0.03\% \\ 
\cmidrule{2-9}
 & \multirow{2}{*}{\gurobi} &  \CS{} & \OOM{} & \FAIL{} & \FAIL{} & \FAIL{} & \FAIL{} & \FAIL{} \\
 &  & \HSB{} & \textbf{0.30} & 4.00\% & \textbf{0.00} & 0.65\% & \textbf{5.49} & 0.11\% \\
\midrule
 \multirow{5}{*}{2560}
 &  \multirow{2}{*}{\cplex$^*$} &  \CS{} & \OOM{} & \FAIL{} & \FAIL{} & \FAIL{} & \FAIL{} & \FAIL{} \\
 &  & \HSB{} & \textbf{8.80} & 2.80\% & \textbf{0.04} & 0.50\% & \textbf{4.71} & 0.20\% \\ 
\cmidrule{2-9}
 &  \multirow{1}{*}{\cplex} &  \HSB{}  & \textbf{0.48} & 3.99\% & \textbf{0.04} & 0.04\% & \textbf{11.06} & 0.19\% \\
\cmidrule{2-9}
 & \multirow{1}{*}{\scip} &  \HSB{}  & \OOM{} & \FAIL{} & \FAIL{} & \FAIL{} & \FAIL{} & \FAIL{} \\
\cmidrule{2-9}
 & \multirow{1}{*}{\gurobi} &  \HSB{} & \textbf{0.71} & 5.76\% & \textbf{0.01} & 0.68\% & \textbf{12.24} & 0.03\% \\ 
  \midrule
  \multirow{3}{*}{5120}
 &  \multirow{1}{*}{\cplex$^*$} &  \HSB{} & \textbf{2.70} & 1.17\% & \textbf{0.19} & 0.91\% & \textbf{10.97} & 0.06\% \\ 
\cmidrule{2-9}
 &  \multirow{1}{*}{\cplex} & \HSB{}  & \textbf{1.02} & 1.05\% & \textbf{0.19} & 0.30\% & \textbf{23.34} & 0.13\% \\ 
\cmidrule{2-9}
 & \multirow{1}{*}{\gurobi} & \HSB{}  & \OOM{} & \FAIL{} & \FAIL{} & \FAIL{} & \FAIL{} & \FAIL{} \\
\bottomrule
\end{tabular}}
\end{table}

\begin{table}[hbtp]
\centering
\caption{Solver run times and memory usage for the \MakeLowercase{\LISproblem{}} problem, using a patience-sorting algorithm with time complexity \LISLtimecomp{} (\LISL{}), with input size $n$, where problem inputs are encoded through the objective function. 
Each result is based on 5 runs, with a 23-hour time limit, 64~GB memory, and 8~CPUs.
}
\label{tab:time-lisl-notforced}
\scalebox{1}{
\begin{tabular}{ccccccccc}
\toprule
\multicolumn{3}{c}{Info} & \multicolumn{4}{c}{Time (\Hour{})} & \multicolumn{2}{c}{Memory (\GB{})} \\
\cmidrule(lr){1-3} \cmidrule(lr){4-7} \cmidrule(lr){8-9}
$n$ &  solver & \OPT{} & \multicolumn{2}{c}{Elapsed}  & \multicolumn{2}{c}{Presolve} & \multicolumn{2}{c}{RSS} \\
 &  &   &  \Med{}& \IQR{}& \Med{}& \IQR{}& \Med{}& \IQR{}\\
\midrule
\multirow{9}{*}{10} 
 & \multirow{3}{*}{\cplex} & \UO{}  &  0.19 & 0.87\% & 0.00 & 7.28\% & 4.40 & 2.06\%  \\
 &  & \CS{}  &  0.02 & 10.64\% & 0.00 & 8.30\% & 3.30 & 3.53\%   \\
 &  & \HSB{}  & \textbf{0.00} & 3.56\% & 0.00 & 7.50\% & \textbf{0.39} & 1.96\% \\
\cmidrule{2-9}
 & \multirow{3}{*}{\scip} & \UO{}  & 5.33 & 0.88\% & 0.01 & 2.23\% & 6.42 & 0.08\% \\
 &  & \CS{}  & 0.87 & 29.45\% & 0.01 & 24.24\% & 4.63 & 0.06\%\\
 &  & \HSB{}  & \textbf{0.01} & 3.20\% & \textbf{0.00} & 3.66\% & \textbf{0.61} & 0.87\% \\
\cmidrule{2-9}
 & \multirow{3}{*}{\gurobi} & \UO{}   &  0.11 & 0.70\% & 0.00 & 6.78\% & 3.28 & 0.23\% \\
 &  & \CS{}    &  0.04 & 5.00\% & 0.00 & 13.89\% & 2.72 & 0.36\% \\
 &  & \HSB{}   & \textbf{0.00} & 0.38\% & 0.00 & 1.70\% & \textbf{0.28} & 0.87\%  \\
\midrule
\multirow{9}{*}{20} 
 &  \multirow{2}{*}{\cplex} & \UO{}   & 2.92 & 0.48\% & 0.01 & 1.56\% & 18.65 & 0.23\%  \\
 &  & \CS{}    &  0.25 & 2.09\% & \textbf{0.00} & 3.41\% & 13.51 & 0.14\%  \\
 &  & \HSB{}    & \textbf{0.01} & 11.81\% & \textbf{0.00} & 9.69\% & \textbf{1.21} & 2.96\% \\
\cmidrule{2-9}
 & \multirow{3}{*}{\scip} & \UO{}  &  \OOT{} &  \FAIL{} & \FAIL{} & \FAIL{} & \FAIL{} & \FAIL{} \\
 &  & \CS{}    & \OOT{} & \FAIL{} & \FAIL{} & \FAIL{} & \FAIL{} & \FAIL{}  \\
 &  & \HSB{}    & \textbf{0.06} & 1.19\% & \textbf{0.00} & 0.90\% & \textbf{2.38} & 0.10\% \\
\cmidrule{2-9}
 & \multirow{3}{*}{\gurobi}  & \UO{} & 0.57 & 0.79\% & 0.01 & 2.76\% & 13.23 & 0.11\%  \\
 &  & \CS{}    &  0.21 & 0.27\% & 0.01 & 0.73\% & 10.77 & 0.05\% \\
 &  & \HSB{}    & \textbf{0.03} & 0.69\% & \textbf{0.00} & 3.35\% & \textbf{0.89} & 0.58\% \\
 \midrule
 \multirow{7}{*}{40}
 &  \multirow{2}{*}{\cplex} & \UO{}   & \OOM{} & \FAIL{} & \FAIL{} & \FAIL{} & \FAIL{} & \FAIL{}  \\
 &  & \CS{}    &  3.07 & 40.30\% & 0.02 & 42.67\% & 60.79 & 3.45\%  \\
 &  & \HSB{}    & \textbf{0.03} & 1.38\% & \textbf{0.00} & 16.26\% & \textbf{5.09} & 2.95\%  \\
\cmidrule{2-9}
 & \multirow{1}{*}{\scip} & \HSB{}    & 0.32 & 2.93\% & 0.01 & 3.82\% & 10.79 & 0.04\%  \\
\cmidrule{2-9}
 & \multirow{3}{*}{\gurobi}  & \UO{} & \OOM{} &   \FAIL{} & \FAIL{} & \FAIL{} & \FAIL{} & \FAIL{}   \\
 &  & \CS{}    & \OOM{}  &   \FAIL{} & \FAIL{} & \FAIL{} & \FAIL{} & \FAIL{}  \\
 &  & \HSB{}    & 0.15 & 0.93\% & 0.01 & 19.22\% & 3.66 & 0.28\%  \\
 \midrule
 \multirow{5}{*}{80}
 &  \multirow{2}{*}{\cplex} & \CS{}    & \OOM{} & \FAIL{} &   \FAIL{} & \FAIL{} & \FAIL{} & \FAIL{}  \\
 &  & \HSB{}    & 0.25 & 3.00\% & 0.01 & 1.91\% & 17.22 & 0.06\% \\
\cmidrule{2-9}
 & \multirow{1}{*}{\scip} & \HSB{}    &\OOM{}  & \FAIL{} &   \FAIL{} & \FAIL{} & \FAIL{} & \FAIL{}  \\
\cmidrule{2-9}
 & \multirow{1}{*}{\gurobi}  & \HSB{}    &  0.73 & 0.31\% & 0.03 & 2.38\% & 15.24 & 0.02\%  \\
\bottomrule
\end{tabular}}
\end{table}

\begin{table}[hbtp]
\centering
\caption{Solver run times and memory usage for the \SAMlong{} problem, using the standard merge algorithm, with input sizes $n$ and $m$, where problem inputs are encoded through the objective function. 
Each result is based on 5 runs, with a 23-hour time limit, 64~GB memory, and 8~CPUs.
}
\label{tab:time-sam-notforced}
\scalebox{0.75}{
\begin{tabular}{cccccccccc}
\toprule
\multicolumn{4}{c}{Info} & \multicolumn{4}{c}{Time (\Min{})} & \multicolumn{2}{c}{Memory (\GB{})} \\
\cmidrule(lr){1-4} \cmidrule(lr){5-8} \cmidrule(lr){9-10}
$n$ & $m$ &  solver & \OPT{} & \multicolumn{2}{c}{Elapsed}  & \multicolumn{2}{c}{Presolve} & \multicolumn{2}{c}{RSS} \\
 &  &  & &  \Med{}& \IQR{}& \Med{}& \IQR{}& \Med{}& \IQR{}\\
\midrule
\multirow{9}{*}{10} &
\multirow{9}{*}{7} & 
\multirow{3}{*}{\cplex} & \UO{} & 32.31 & 6.62\% & 0.02 & 10.64\% & 2.01 & 1.37\% \\
 &  & & \CS{}  & 0.72 & 1.65\% & 0.01 & 1.43\% & 1.10 & 0.94\%  \\
 &  &  & \HSB{} & \textbf{ 0.02} & 5.56\% & \textbf{0.00} & 21.43\% & \textbf{0.20} & 5.53\% \\
\cmidrule{3-10}
 &  & \multirow{3}{*}{\scip} & \UO{}  & 108.62 & 3.58\% & 0.20 & 3.95\% & 2.31 & 0.13\%   \\
 &  &   & \CS{}  & 3.00 & 3.49\% & 0.11 & 0.60\% & 1.62 & 0.29\%  \\
 &  &  & \HSB{} & \textbf{ 0.25} & 5.81\% & \textbf{0.01} & 7.63\% & \textbf{0.24} & 0.53\% \\
\cmidrule{3-10}
 &  & \multirow{3}{*}{\gurobi} & \UO{}  &  3.29 & 5.14\% & 0.07 & 3.02\% & 1.28 & 0.46\%  \\
 &  & & \CS{}  &  0.53 & 1.91\% & 0.03 & 0.25\% & 0.82 & 0.56\%  \\
 &  &  & \HSB{} & \textbf{ 0.07} & 12.47\% & \textbf{0.01} & 0.00\% & \textbf{0.16} & 4.39\%  \\
\midrule
\multirow{9}{*}{20} &
\multirow{9}{*}{15} & 
\multirow{3}{*}{\cplex}  & \UO{}  &  23.46 & 4.74\% & 0.09 & 21.57\% & 5.08 & 3.03\%  \\
 &  &  & \CS{}  &  1.10 & 1.34\% & 0.04 & 0.57\% & 2.89 & 0.29\%  \\
 &  &  & \HSB{} & \textbf{0.06} & 2.86\% & \textbf{0.00} & 1.72\% & \textbf{0.45} & 2.13\%  \\
\cmidrule{3-10}
 &  & \multirow{3}{*}{\scip}  & \UO{}  &  144.69 & 9.16\% & 0.66 & 7.64\% & 8.08 & 0.06\% \\
 &  &  & \CS{}  & 4.73 & 3.87\% & 0.35 & 4.33\% & 4.70 & 0.09\%  \\
 &  &  & \HSB{} & \textbf{0.73} & 6.26\% & \textbf{0.04} & 1.17\% & \textbf{0.72} & 0.24\%  \\
\cmidrule{3-10}
 &  & \multirow{3}{*}{\gurobi}  & \UO{}  &  7.99 & 11.04\% & 0.25 & 7.42\% & 4.07 & 0.05\%  \\
 &  &  & \CS{}  &  2.05 & 6.62\% & 0.12 & 4.56\% & 2.25 & 0.33\% \\
 &  &  & \HSB{} & \textbf{0.37} & 8.22\% & \textbf{0.01} & 1.87\% & \textbf{0.43} & 0.44\% \\
\midrule

\multirow{9}{*}{40} &
\multirow{9}{*}{32} & 
\multirow{3}{*}{\cplex}  & \UO{}  &  1085.37 & 11.08\% & 0.40 & 20.28\% & 17.08 & 0.21\% \\
 &  &  & \CS{}  &  13.55 & 8.60\% & 0.19 & 22.35\% & 9.81 & 0.68\%  \\
 &  &  & \HSB{} & \textbf{ 0.30} & 3.14\% & \textbf{0.02} & 0.45\% & \textbf{1.33} & 4.10\% \\
\cmidrule{3-10}
 &  & \multirow{3}{*}{\scip} & \UO{}  & \OOT{} & \FAIL{} & \FAIL{} & \FAIL{} & \FAIL{} & \FAIL{}  \\
 &  &  & \CS{}  &  358.03 & 1.45\% & 1.27 & 2.52\% & 16.67 & 0.03\% \\
 &  &  & \HSB{} & \textbf{ 3.13} & 0.81\% & \textbf{0.15} & 2.97\% & \textbf{2.60} & 0.09\% \\
\cmidrule{3-10}
 &  & \multirow{3}{*}{\gurobi}  & \UO{}  &  37.35 & 4.87\% & 1.36 & 4.40\% & 12.54 & 0.58\% \\
 &  & & \CS{}  & 3.49 & 23.67\% & 0.40 & 33.61\% & 7.68 & 0.05\%  \\
 &  &  & \HSB{} & \textbf{ 1.51} & 3.94\% & \textbf{0.06} & 0.91\% & \textbf{1.17} & 0.30\% \\
\midrule

\multirow{8}{*}{80} &
\multirow{8}{*}{75} & 
\multirow{3}{*}{\cplex} & \UO{}  &  \OOM{} & \FAIL{} & \FAIL{} & \FAIL{} & \FAIL{} & \FAIL{}   \\
 &  &   & \CS{}  & 25.08 & 6.03\% & 0.78 & 10.49\% & 37.34 & 0.08\% \\
 &  &  & \HSB{} & \textbf{ 2.08} & 12.44\% & \textbf{0.09} & 9.28\% & \textbf{4.74} & 0.30\% \\
\cmidrule{3-10}
 &  & \multirow{2}{*}{\scip} &\CS{}  & \OOT{} & \FAIL{} & \FAIL{} & \FAIL{} & \FAIL{} & \FAIL{}   \\
 &  &  & \HSB{} & 20.11 & 2.13\% & 0.69 & 3.30\% & 11.65 & 0.03\%  \\
\cmidrule{3-10}
 &  & \multirow{3}{*}{\gurobi} & \UO{}  & \OOM{} & \FAIL{} & \FAIL{} & \FAIL{} & \FAIL{} & \FAIL{}  \\
 &  &  & \CS{}  & 17.56 & 1.61\% & 1.94 & 4.04\% & 30.57 & 0.03\%  \\
 &  &  & \HSB{} & \textbf{8.72} & 6.29\% & \textbf{0.22} & 10.39\% & \textbf{4.34} & 0.33\% \\
\midrule

\multirow{5}{*}{160} &
\multirow{5}{*}{143} & 
\multirow{2}{*}{\cplex}  & \CS{}  &  \OOM{} & \FAIL{} & \FAIL{} & \FAIL{} & \FAIL{} & \FAIL{}  \\
 &  &  & \HSB{} &  11.01 & 17.17\% & 0.45 & 4.73\% & 15.71 & 0.43\%  \\
\cmidrule{3-10}
 &  & \multirow{1}{*}{\scip} &\HSB{} & 54.50 & 2.80\% & 2.39 & 5.81\% & 44.29 & 0.02\% \\
\cmidrule{3-10}
 &  & \multirow{2}{*}{\gurobi} &\CS{}  & \OOM{} & \FAIL{} & \FAIL{} & \FAIL{} & \FAIL{} & \FAIL{}   \\
 &  &  & \HSB{} &  35.51 & 1.28\% & 0.89 & 1.48\% & 15.81 & 0.41\% \\
\midrule

\multirow{3}{*}{320} &
\multirow{3}{*}{285} & 
\multirow{1}{*}{\cplex}  & \HSB{} &  71.27 & 10.17\% & 2.47 & 42.87\% & 59.35 & 0.47\%  \\
\cmidrule{3-10}
 &  & \multirow{1}{*}{\scip} &  \HSB{} &  \OOM{} & \FAIL{} & \FAIL{} & \FAIL{} & \FAIL{} & \FAIL{}   \\
\cmidrule{3-10}
 &  & \multirow{1}{*}{\gurobi} & \HSB{} & \OOM{} & \FAIL{} & \FAIL{} & \FAIL{} & \FAIL{} & \FAIL{}   \\
\bottomrule
\end{tabular}}
\end{table}

\begin{table}[hbtp]
\centering
\caption{Solver run times and memory usage for the \TSlong{} problem using Kahn's algorithm, with input size $n$, where problem inputs are encoded through the objective function. 
Each result is based on 5 runs, with a 23-hour time limit, 64~GB memory, and 8~CPUs.
}
\label{tab:time-ts-notforced}
\scalebox{1}{
\begin{tabular}{ccccccccc}
\toprule
\multicolumn{3}{c}{Info} & \multicolumn{4}{c}{Time (\Hour{})} & \multicolumn{2}{c}{Memory (\GB{})} \\
\cmidrule(lr){1-3} \cmidrule(lr){4-7} \cmidrule(lr){8-9}
$n$ & solver & \OPT{} & \multicolumn{2}{c}{Elapsed}  & \multicolumn{2}{c}{Presolve} & \multicolumn{2}{c}{RSS} \\
 &  &   &  \Med{}& \IQR{}& \Med{}& \IQR{}& \Med{}& \IQR{}\\
\midrule
\multirow{9}{*}{10}
 & \multirow{3}{*}{\cplex} & \UO{}  & 3.99 & 15.14\% & 0.00 & 16.00\% & 7.63 & 1.94\%   \\
 &  & \CS{}  &  0.12 & 20.87\% & 0.00 & 7.94\% & 5.26 & 4.04\%  \\
 &  & \HSB{}  & \textbf{ 0.00} & 3.31\% & \textbf{0.00} & 0.00\% & \textbf{0.45} & 4.90\% \\
\cmidrule{2-9}
 & \multirow{3}{*}{\scip} & \UO{}  & \OOT{} & \FAIL{} & \FAIL{} &  \FAIL{} & \FAIL{} & \FAIL{}  \\
 &  & \CS{}  & \OOT{} &   \FAIL{} & \FAIL{} &  \FAIL{} & \FAIL{} & \FAIL{}   \\
 &  & \HSB{}  &  0.01 & 2.55\% & 0.00 & 3.22\% & 0.79 & 0.31\%   \\
\cmidrule{2-9}
 & \multirow{3}{*}{\gurobi} & \UO{} & 0.31 & 6.26\% & 0.01 & 9.37\%  &  5.29 & 0.07\%   \\
 &  & \CS{}    &  0.03 & 20.49\% & \textbf{0.00} & 6.57\% & 4.27 & 0.79\%  \\
 &  & \HSB{}   & \textbf{ 0.00} & 9.66\% & \textbf{0.00} & 6.07\% & \textbf{0.35} & 0.40\%  \\
\midrule
\multirow{7}{*}{20}
 &  \multirow{3}{*}{\cplex} & \UO{} & 4.82 & 21.62\% & 0.01 & 23.54\% & 26.30 & 0.20\% \\
 &  & \CS{}    &  2.84 & 16.96\% & 0.01 & 11.77\% & 17.95 & 0.24\%  \\
 &  & \HSB{}    & \textbf{ 0.00} & 1.12\% & \textbf{0.00} & 1.26\% & \textbf{1.42} & 1.22\% \\
\cmidrule{2-9}
 & \multirow{1}{*}{\scip} & \HSB{}    & 0.06 & 0.96\% & 0.00 & 3.20\% & 2.57 & 0.11\%  \\
\cmidrule{2-9}
 & \multirow{3}{*}{\gurobi}  & \UO{} &  0.80 & 3.12\% & 0.02 & 6.35\% & 20.77 & 0.04\% \\
 &  & \CS{}    &  0.27 & 4.25\% & 0.01 & 3.38\% & 14.51 & 0.04\%  \\
 &  & \HSB{}    &  \textbf{0.02} & 4.76\% & \textbf{0.00} & 0.72\% & \textbf{0.97} & 0.31\%\\
 \midrule
 \multirow{7}{*}{40}
 &  \multirow{3}{*}{\cplex} & \UO{}  & \OOM{} &   \FAIL{} & \FAIL{} &  \FAIL{} & \FAIL{} & \FAIL{}   \\
 &  & \CS{}    & \OOM{} &  \FAIL{} & \FAIL{} &  \FAIL{} & \FAIL{} & \FAIL{}   \\
 &  & \HSB{}    &  0.02 & 15.70\% & 0.00 & 5.29\% & 5.39 & 5.72\%  \\
\cmidrule{2-9}
 & \multirow{1}{*}{\scip} &  \HSB{}    &  0.20 & 0.19\% & 0.01 & 1.22\% & 9.85 & 0.06\% \\
\cmidrule{2-9}
 & \multirow{3}{*}{\gurobi}  & \UO{} & \OOM{} &  \FAIL{} & \FAIL{} &  \FAIL{} & \FAIL{} & \FAIL{}   \\
 &  & \CS{}    & \OOM{} & \ \FAIL{} & \FAIL{} &  \FAIL{} & \FAIL{} & \FAIL{}  \\
 &  & \HSB{}    &  0.07 & 3.28\% & 0.00 & 6.57\% & 3.76 & 0.26\% \\
 \midrule
 \multirow{3}{*}{80}
 &  \multirow{1}{*}{\cplex} &  \HSB{}    &  0.19 & 0.39\% & 0.01 & 1.55\% & 19.42 & 0.07\%  \\
\cmidrule{2-9}
 & \multirow{1}{*}{\scip} & \HSB{}    &  0.87 & 0.29\% & 0.06 & 1.41\% & 40.24 & 0.01\% \\
\cmidrule{2-9}
 & \multirow{1}{*}{\gurobi}  &  \HSB{}    &  0.38 & 0.33\% & 0.02 & 2.35\% & 13.71 & 0.04\% \\
\bottomrule
\end{tabular}}
\end{table}

\begin{table}[hbtp]
\centering
\caption{Solver run times and memory usage for the \MSlong{} problem, using the algorithm from~\cite{avis_sparktope_2022}, with $m$ jobs on $n=3$ machines, where problem inputs are encoded through the objective function. 
Each result is based on 10 runs, with a 23-hour time limit, 32~GB memory, and 7~CPUs.
A superscript $^*$ indicates the dual simplex algorithm.
}
\label{tab:time-ms-notforced}
\scalebox{0.75}{
\begin{tabular}{ccccccccc}
\toprule
\multicolumn{3}{c}{Info} & \multicolumn{4}{c}{Time (\Sec{})} & \multicolumn{2}{c}{Memory (\MB{})} \\
\cmidrule(lr){1-3} \cmidrule(lr){4-7} \cmidrule(lr){8-9}
$m$ & solver & \OPT{} & \multicolumn{2}{c}{Elapsed}  & \multicolumn{2}{c}{Presolve} & \multicolumn{2}{c}{RSS} \\
 &  &   &  \Med{}& \IQR{}& \Med{}& \IQR{}& \Med{}& \IQR{}\\
\midrule
\multirow{12}{*}{20}
 & \multirow{3}{*}{\cplex$^*$} & \UO{} & 2561.90 & 2.34\% & 2.28 & 1.75\% & 1155 & 0.51\% \\
 &  & \CS{} & 52.80 & 18.47\% & 1.92 & 8.83\% & 2746 & 2.51\% \\
 &  & \HSB{} &  \textbf{1.04} & 2.40\% & \textbf{0.12} & 8.00\% & \textbf{85} & 2.71\% \\
\cmidrule{2-9}
 & \multirow{3}{*}{\cplex} & \UO{} & 1234.65 & 1.12\% & 2.63 & 14.26\% & 3359 & 2.10\% \\
 &  & \CS{} & 57.97 & 4.74\% & 1.92 & 9.09\% & 2738 & 0.77\% \\
 &  & \HSB{} & \textbf{1.18} & 6.57\% &  \textbf{0.15} & 13.33\% & \textbf{291} & 1.19\% \\
\cmidrule{2-9}
 & \multirow{3}{*}{\scip} & \UO{} & 1506.72 & 1.61\% & 18.38 & 1.29\% & 4030 & 0.07\% \\
 &  & \CS{} & 378.46 & 3.22\% & 14.68 & 0.80\% & 3202 & 0.11\% \\
 &  & \HSB{} & \textbf{2.94} & 7.99\% & \textbf{1.34} & 7.46\% & \textbf{448} & 0.49\% \\ 
\cmidrule{2-9}
 & \multirow{3}{*}{\gurobi} & \UO{} & 282.47 & 0.71\% & 6.20 & 0.44\% & 2619 & 0.46\% \\
 &  & \CS{} & 50.64 & 20.61\% & 4.63 & 4.59\% & 2356 & 0.32\% \\
 &  & \HSB{} & \textbf{1.29} & 3.68\% & \textbf{0.54} & 3.70\% & \textbf{191} & 0.28\% \\
\midrule
\multirow{12}{*}{40}
 & \multirow{3}{*}{\cplex$^*$} & \UO{} & 10,231.66 & 1.50\% & 8.18 & 2.08\% & 3330 & 1.43\% \\
 & & \CS{} & 209.28 & 21.46\% & 6.21 & 11.24\% & 7812 & 1.27\% \\
 & & \HSB{} & \textbf{4.22} & 0.95\% & \textbf{0.46} & 2.72\% & \textbf{243} & 0.62\% \\
\cmidrule{2-9}
 & \multirow{3}{*}{\cplex} & \UO{} & 872.27 & 5.20\% & 9.23 & 9.15\% & 9773 & 1.53\% \\
 &  & \CS{}  & 267.12 & 5.89\% & 6.99 & 10.09\% & 7904 & 0.77\% \\
 &  & \HSB{}  & \textbf{5.58} & 11.25\% &  \textbf{0.57} & 6.14\% & \textbf{778} & 16.66\% \\
\cmidrule{2-9}
 & \multirow{3}{*}{\scip} & \UO{} & 8579.44 & 2.65\% & 60.92 & 2.64\% & 12,591 & 0.04\% \\
& & \CS{} & \OOT{} & \FAIL{} & \FAIL{} & \FAIL{} & \FAIL{} & \FAIL{} \\
 & & \HSB{} & \textbf{10.04} & 5.08\% & \textbf{4.48} & 5.86\% & \textbf{1298} & 0.23\% \\
\cmidrule{2-9}
 & \multirow{3}{*}{\gurobi} & \UO{} & 990.38 & 1.55\% & 20.63 & 1.36\% & 7375 & 0.09\% \\
 &  & \CS{} & 113.36 & 20.95\% & 14.41 & 2.10\% & 6543 & 3.38\% \\
 &  & \HSB{} & \textbf{5.37} & 16.68\% &  \textbf{1.75} & 4.14\% & \textbf{699} & 0.55\% \\
\midrule
\multirow{12}{*}{80}
 &  \multirow{2}{*}{\cplex$^*$} & \UO{} & 49,535.81 & 1.46\% & 32.23 & 2.06\% & 11,083 & 0.06\% \\
 &  & \CS{} & 1201.99 & 12.13\% & 25.20 & 10.51\% & 28,817 & 0.43\% \\
 &  & \HSB{} & \textbf{12.09} & 2.27\% & \textbf{1.89} & 2.91\% & \textbf{764} & 0.81\% \\
\cmidrule{2-9}
 &  \multirow{2}{*}{\cplex} & \UO{} & \OOM{} & \FAIL{} & \FAIL{} & \FAIL{} & \FAIL{} & \FAIL{} \\
 &  & \CS{} & 1561.22 & 37.43\% & 30.02 & 28.98\% & 28,375 & 1.31\% \\
 &  & \HSB{} & \textbf{17.41} & 5.50\% & \textbf{2.27} & 12.67\% & \textbf{1939} & 0.19\% \\
 \cmidrule{2-9}
 & \multirow{2}{*}{\scip} & \UO{} & \OOM{} & \FAIL{} & \FAIL{} & \FAIL{} & \FAIL{} & \FAIL{} \\
 & & \HSB{} & \textbf{65.59} & 13.91\% &  \textbf{15.79} & 8.54\% & \textbf{3591} & 0.12\% \\
\cmidrule{2-9}
 & \multirow{3}{*}{\gurobi} & \UO{} & \OOM{} & \FAIL{} & \FAIL{} & \FAIL{} & \FAIL{} & \FAIL{} \\
 &  & \CS{} & 646.59 & 0.51\% & 57.16 & 1.00\% & 16,562 & 0.02\% \\
 &  & \HSB{} & \textbf{22.38} & 15.22\% &  \textbf{5.54} & 5.65\% & \textbf{1991} & 0.27\% \\
 \midrule
\multirow{12}{*}{160}
 &  \multirow{2}{*}{\cplex$^*$} & \UO{} & \OOM{} & \FAIL{} & \FAIL{} & \FAIL{} & \FAIL{} & \FAIL{} \\
 &  & \CS{} & 7009.58 & 24.54\% & 14.91 & 53.28\% & 83010 & 7.86\% \\
 &  & \HSB{} & \textbf{48.70} & 1.97\% & \textbf{8.75} & 1.57\% & \textbf{2379} & 0.20\% \\
\cmidrule{2-9}
 & \multirow{2}{*}{\cplex} & \CS{} & \OOM{} & \FAIL{} & \FAIL{} & \FAIL{} & \FAIL{} & \FAIL{} \\
 &  & \HSB{} & \textbf{68.62} & 1.88\% & \textbf{10.02} & 8.74\% & \textbf{6533} & 1.79\% \\
\cmidrule{2-9}
 & \scip{} & \HSB{} & \textbf{156.53} & 7.06\% &  \textbf{56.61} & 7.85\% & \textbf{11884} & 0.04\% \\
\cmidrule{2-9}
 & \multirow{2}{*}{\gurobi} & \CS{} & \OOM{} & \FAIL{} & \FAIL{} & \FAIL{} & \FAIL{} & \FAIL{} \\
 &  & \HSB{} & \textbf{141.45} & 3.68\% & \textbf{20.01} & 2.62\% & \textbf{7574} & 3.22\% \\
\bottomrule
\end{tabular}}
\end{table}

\begin{table}[hbtp]
\centering
\caption{Run times and memory usage for the \MM{} and \HMM{} benchmarks, with inputs encoded through the objective function. 
For \MM{} we compare \PO{}, \CS{}, \HSBCS{}, and \HSBCS{}$^\dagger$.  
For \HMM{}, we report \HSB{}.
The $^\dagger$ denotes the \HSBCS{} variant in which \CS{} is applied within manually flattened blocks. 
Each result is based on 3~runs with a 23-hour time limit, 372~GB memory, and 8~CPUs, except that \HSBCS{}$^\dagger$ uses a 126~GB memory limit. 
The \IQR{} is reported as a percentage of the median. 
A superscript $^*$ indicates dual simplex algorithm.
\OOM{} indicates out-of-memory, \OOT{} indicates timeout, and \FAIL{} denotes solver failure.} 
\label{tab:time-mm-notforced-complete}
\scalebox{0.6}{
\begin{tabular}{cccccccccc}
\toprule
\multicolumn{4}{c}{Info} & \multicolumn{4}{c}{Time (\Hour{})} & \multicolumn{2}{c}{Mem (\GB{})}  \\
\cmidrule(lr){1-4} \cmidrule(lr){5-8} \cmidrule(lr){9-10}
$n$ & solver & problem & \OPT{} &  \multicolumn{2}{c}{Elapsed}  & \multicolumn{2}{c}{Presolve} & \multicolumn{2}{c}{RSS} \\
 &  &  &  &  \Med{}& \IQR{}& \Med{}& \IQR{}& \Med{}& \IQR{}\\
\midrule
\multirow{24}{*}{8}
& \multirow{6}{*}{\cplex$^*$} &  
\multirow{4}{*}{\MM} & 
\PO{} & \OOT{} & \FAIL{} & \FAIL{} & \FAIL{} & \FAIL{} & \FAIL{} \\
 &      &              & \CS{}  &  \OOT{} & \FAIL{} & \FAIL{} & \FAIL{} & \FAIL{} & \FAIL{} \\
\cmidrule(lr){4-10}
  &  &   & \HSBCS{} &  \textbf{0.55} & 0.85\% & 0.00 & 0.57\% & 5.57 & 0.13\%  \\ 
  &   &  & \HSBCS{}$^\dagger$ &  1.57 & 0.95\% & 0.00 & 0.58\% & \textbf{3.14} & 0.37\%  \\ 
\cmidrule(lr){3-10}
  &  &  \multirow{1}{*}{\HMM}   
  & \HSB{}  & 1.81 & 1.74\% & 0.00 & 0.96\% & 2.17 & 0.24\% \\ 
\cmidrule{2-10}
 & \multirow{6}{*}{\cplex} &  
 \multirow{4}{*}{\MM} & 
 \PO{} &  \OOT{}  & \FAIL{} & \FAIL{} & \FAIL{} & \FAIL{} & \FAIL{} \\
 &         &                & \CS{}       & 0.82 & 0.26\% & 0.01 & 0.62\% & 43.86 & 0.22\% \\ 
 \cmidrule(lr){4-10}
  &  &   & \HSBCS{} & 0.43 & 0.47\% & 0.00 & 0.48\% & 17.35 & 0.33\% \\ 
  &  &   & \HSBCS{}$^\dagger$ & \textbf{0.06} & 0.26\% & 0.00 & 0.32\% & \textbf{11.60} & 1.34\%  \\ 
 \cmidrule(lr){3-10}
 &        &  \multirow{1}{*}{\HMM} 
 & \HSB{}  & 0.02 & 0.44\% & 0.00 & 0.64\% & 6.89 & 0.03\% \\ 
\cmidrule{2-10}
  & \multirow{6}{*}{\scip} &  
  \multirow{4}{*}{\MM} & 
  \PO{}  & \OOT{} & \FAIL{} & \FAIL{} & \FAIL{} & \FAIL{} & \FAIL{} \\
   &          &               & \CS{}  &  \OOT{} & \FAIL{} & \FAIL{} & \FAIL{} & \FAIL{} & \FAIL{} \\ 
  \cmidrule(lr){4-10}
  &   &  & \HSBCS{}  & \OOT{} & \FAIL{} & \FAIL{} & \FAIL{}& \FAIL{} & \FAIL{} \\ 
  &   &  & \HSBCS{}$^\dagger$ & \textbf{0.38} & 0.88\% & \textbf{0.02} & 0.06\% & \textbf{16.11} & 0.01\% \\ 
  \cmidrule(lr){3-10}
  &     &  \multirow{1}{*}{\HMM}   
  & \HSB{}  & 0.56 & 1.17\% & 0.01 & 0.93\% & 11.14 & 0.06\% \\ 
\cmidrule{2-10}
  & \multirow{6}{*}{\gurobi} &  
  \multirow{4}{*}{\MM} & 
  \PO{}  & \OOT{} & \FAIL{} & \FAIL{} & \FAIL{} & \FAIL{} & \FAIL{} \\
   &    &       & \CS{}  & 16.20 & 0.58\% & 0.02 & 0.66\% & 33.31 & 0.01\% \\  
  \cmidrule(lr){4-10}
    &  &   & \HSBCS{} &  0.50 & 0.61\% & 0.01 & 0.42\% & 13.81 & 0.04\% \\ 
  &   &  & \HSBCS{}$^\dagger$ & \textbf{0.38} & 0.43\% & 0.01 & 0.77\% & \textbf{9.27} & 3.06\% \\ 
  \cmidrule(lr){3-10}
    &     &  \multirow{1}{*}{\HMM}   
    & \HSB{}  & 0.05 & 2.43\% & 0.00 & 1.19\% & 8.35 & 22.31\% \\ 
\midrule
\multirow{16}{*}{10}
  & \multirow{3}{*}{\cplex$^*$}  & 
  \multirow{2}{*}{\MM} & 
  \HSBCS{} &  \OOT{} & \FAIL{} & \FAIL{} & \FAIL{} & \FAIL{} & \FAIL{} \\
  &  &   & \HSBCS{}$^\dagger$ & \OOT{} & \FAIL{} & \FAIL{} & \FAIL{} & \FAIL{} & \FAIL{}  \\ 
  \cmidrule(lr){3-10}
  &       &  \multirow{1}{*}{\HMM}   
  & \HSB{}  & 0.92 & 1.36\% & 0.00 & 0.17\% & 5.20 & 0.08\% \\ 
\cmidrule{2-10}
& \multirow{5}{*}{\cplex} & 
\multirow{3}{*}{\MM} & 
\CS{} &  3.69 & 0.29\% & 0.02 & 0.76\% & 88.13 & 0.40\% \\ 
  \cmidrule(lr){4-10}
  &  &   & \HSBCS{} &  0.93 & 0.30\% & 0.01 & 0.19\% & 45.80 & 0.31\%  \\ 
  &  &   & \HSBCS{}$^\dagger$ &  \textbf{0.14} & 0.57\% & 0.01 & 0.08\% & \textbf{29.22} & 0.42\%\\ 
\cmidrule(lr){3-10}
& &  \multirow{1}{*}{\HMM}   
& \HSB{}   & 0.08 & 0.60\% & 0.00 & 0.38\% & 16.53 & 0.42\% \\ 
\cmidrule{2-10}
& \multirow{2}{*}{\scip} &  
\multirow{1}{*}{\MM} & 
\HSBCS{}$^\dagger$ &  \OOT{} & \FAIL{} & \FAIL{} & \FAIL{} & \FAIL{} & \FAIL{}  \\ 
  \cmidrule(lr){3-10}
  &                 &  \multirow{1}{*}{\HMM}   
  & \HSB{}   & 2.07 & 1.31\% & 0.03 & 1.04\% & 26.37 & 0.04\% \\
\cmidrule{2-10}
  & \multirow{5}{*}{\gurobi} & 
  \multirow{3}{*}{\MM} & 
  \CS{}  & \OOT{} & \FAIL{} & \FAIL{} & \FAIL{} & \FAIL{} & \FAIL{} \\ 
    \cmidrule(lr){4-10}
  &  &   & \HSBCS{} & \textbf{1.43} & 0.19\% & 0.03 & 0.86\% & 36.48 & 0.03\% \\ 
  &  &   & \HSBCS{}$^\dagger$ &  3.34 & 0.27\% & \textbf{0.02} & 0.34\% & \textbf{21.99} & 0.16\% \\ 
  \cmidrule(lr){3-10}
  &       &  \multirow{1}{*}{\HMM}   
  & \HSB{} & 0.21 & 2.43\% & 0.01 & 0.72\% & 31.44 & 0.68\% \\ 
\midrule
\multirow{12}{*}{12}
 & \multirow{3}{*}{\cplex$^*$} &  
   \multirow{2}{*}{\MM} & 
   \HSBCS{} &\OOM{}& \FAIL{} & \FAIL{} & \FAIL{} & \FAIL{} & \FAIL{} \\ 
  &  &   & \HSBCS{}$^\dagger$ &\OOM{}& \FAIL{} & \FAIL{} & \FAIL{} & \FAIL{} & \FAIL{} \\ 
  \cmidrule(lr){3-10}
  &    &  \multirow{1}{*}{\HMM}   
  & \HSB{}  & 3.82 & 6.67\% & 0.01 & 5.80\% & 10.09 & 0.04\% \\ 
\cmidrule{2-10}
& \multirow{4}{*}{\cplex} & 
\multirow{3}{*}{\MM} & 
\CS{} &  3.51 & 0.14\% & 0.03 & 1.14\% & 146.73 & 1.24\% \\ 
    \cmidrule(lr){4-10}
  &  &   & \HSBCS{} & 1.43 & 0.19\% & 0.04 & 0.54\% & 138.58 & 2.22\%  \\ 
  &  &   & \HSBCS{}$^\dagger$ & \textbf{0.45} & 0.85\% & \textbf{0.02} & 0.90\% & \textbf{85.32} & 0.03\%  \\ 
\cmidrule(lr){3-10}
& &  \multirow{1}{*}{\HMM}   
& \HSB{}  & 0.30 & 2.24\% & 0.01 & 0.28\% & 32.31 & 0.02\% \\
\cmidrule{2-10}
& \multirow{1}{*}{\scip} &  
\multirow{1}{*}{\HMM}    
&  \HSB{}  & 4.50 & 3.41\% & 0.08 & 6.23\% & 53.36 & 0.01\% \\ 
\cmidrule{2-10}
  & \multirow{3}{*}{\gurobi} &  
  \multirow{2}{*}{\MM} & 
  \HSBCS{}  & \OOM{} & \FAIL{} & \FAIL{} & \FAIL{} & \FAIL{} & \FAIL{} \\ 
  &  &   & \HSBCS{}$^\dagger$ &  \textbf{11.81} & 0.78\% & 0.05 & 0.74\% & \textbf{66.89} & 0.01\% \\ 
  \cmidrule(lr){3-10}
  &    &  \multirow{1}{*}{\HMM}   
  & \HSB{} & \OOM{}  &\FAIL{} & \FAIL{}& \FAIL{}& \FAIL{}& \FAIL{}\\
\midrule
\multirow{10}{*}{16}
 & \multirow{1}{*}{\cplex$^*$} &  
\multirow{1}{*}{\HMM}      
 &  \HSB{}  & \OOT{} & \FAIL{}& \FAIL{}&\FAIL{} &\FAIL{} & \\
\cmidrule{2-10}
& \multirow{5}{*}{\cplex} & 
\multirow{3}{*}{\MM} & 
\CS{} &  18.68 & 1.06\% & 0.07 & 3.93\% & 320.95 & 0.02\% \\  
    \cmidrule(lr){4-10}
  &  &   & \HSBCS{} & \OOM{} & \FAIL{} & \FAIL{} & \FAIL{} & \FAIL{} & \FAIL{} \\ 
  &  &   & \HSBCS{}$^\dagger$ &  \textbf{4.42} & 0.06\% & 0.07 & 0.62\% & \textbf{282.48} & 0.99\% \\ 
\cmidrule(lr){3-10}
& &  \multirow{1}{*}{\HMM} 
& \HSB{}   & 2.75 & 1.07\% & 0.03 & 2.40\% & 100.33 & 0.05\% \\
\cmidrule{2-10}
& \multirow{1}{*}{\scip} &  
 \multirow{1}{*}{\HMM} 
& \HSB{}  & \OOM{} & \FAIL{} & \FAIL{} & \FAIL{} & \FAIL{} & \FAIL{} \\
\cmidrule{2-10}
  & \multirow{2}{*}{\gurobi} &  
  \multirow{1}{*}{\MM} & 
  \HSBCS{}$^\dagger$ & \textbf{2.96} & 1.22\% & \textbf{0.16} & 1.65\% & \textbf{217.83} & 0.00\% \\ 
  \cmidrule(lr){3-10}
  &    &  \multirow{1}{*}{\HMM}
  & \HSB{}  & 2.18 & 0.17\% & 0.10 & 0.11\% & 64.37 & 0.01\% \\ 
\bottomrule
\end{tabular}
}
\end{table}

\clearpage
\section{Additional Solution Accuracy Results}
\label{sec:appendix-solution_accuracy}
This appendix provides detailed solution accuracy results for all benchmarks in Table~\ref{tab:benchmark-config}, to complement the overall comparison presented in Section~\ref{sec:resultSolutionAccuracy}.
Tables~\ref{tab:sol-accuracy-spdj} to~\ref{tab:sol-accuracy-ms} report solution accuracy of \CS{} and \HSB{} relative to the ground truth solution. In these tables, \OPT{} denotes the optimization method compared with \UO{}, and \Med{} denotes the median across runs. \% Reduction vs \UO{} is computed as $(\UO{}-\OPT{})/\UO{} \times 100$.
A mismatch is a variable whose value differs from its ground-truth value. Mismatch percentage is the percentage of mismatched variables in \OPT{}, and \Diff{} denotes its difference from \UO{} in percentage points. 
Positive values indicate improvement relative to \UO{}.

\begin{table}[hbtp]
\centering
\caption{Solution accuracy for the \SPDJlong{} benchmark. Positive values indicate improvement relative to \UO{}.}
\label{tab:sol-accuracy-spdj}
\resizebox{\textwidth}{!}{
\begin{tabular}{c c c c c c c c c c c }
\toprule
\multicolumn{3}{c}{Info} & \multicolumn{2}{c}{MAE} & \multicolumn{2}{c}{RMSE} & \multicolumn{2}{c}{Mismatch Count} & \multicolumn{2}{c}{Mismatch Percentage} \\
\cmidrule(lr){1-3} \cmidrule(lr){4-5} \cmidrule(lr){6-7} \cmidrule(lr){8-9} \cmidrule(lr){10-11}
\makecell{Size} &
\makecell{Solver} &
\makecell{\OPT{}} &
\makecell{\Med{}} &
\makecell{\% Reduction\\vs\ \UO{}} &
\makecell{\Med{}} &
\makecell{\% Reduction\\vs\ \UO{}} &
\makecell{\Med} &
\makecell{\% Reduction\\vs\ \UO{}} &
\makecell{\Med} &
\makecell{\Diff{}\\vs\ \UO{}}\\
\midrule
\multirow{4}{*}{10} & \multirow{2}{*}{\cplex{}} & \CS{} & 0.226 & +6.4\% & 0.467 & +3.3\% & 146,274 & +27.3\% & 32.3\% & +5.1 \\
   &  & \HSB{} & 0.182 & +24.4\% & 0.426 & +11.8\% & 48,159 & +76.1\% & 20.1\% & +17.3 \\
\cmidrule(lr){2-11}
   & \multirow{2}{*}{\gurobi{}} & \CS{} & 0.216 & +7.3\% & 0.460 & +3.7\% & 115,420 & +23.8\% & 25.5\% & +2.7 \\
   &  & \HSB{} & 0.182 & +21.7\% & 0.425 & +10.9\% & 49,243 & +67.5\% & 20.6\% & +7.6 \\
\bottomrule
\end{tabular}
}
\end{table}

\begin{table}[hbtp]
\centering
\caption{Solution accuracy for the \KSlong{} benchmark. Positive values indicate improvement relative to \UO{}.}
\label{tab:sol-accuracy-kd}
\resizebox{\textwidth}{!}{
\begin{tabular}{c c c c c c c c c c c}
\toprule
\multicolumn{3}{c}{Info} & \multicolumn{2}{c}{MAE} & \multicolumn{2}{c}{RMSE} & \multicolumn{2}{c}{Mismatch Count} & \multicolumn{2}{c}{Mismatch Percentage} \\
\cmidrule(lr){1-3} \cmidrule(lr){4-5} \cmidrule(lr){6-7} \cmidrule(lr){8-9} \cmidrule(lr){10-11}
\makecell{Size} &
\makecell{Solver} &
\makecell{\OPT{}} &
\makecell{\Med{}} &
\makecell{\% Reduction\\vs\ \UO{}} &
\makecell{\Med{}} &
\makecell{\% Reduction\\vs\ \UO{}} &
\makecell{\Med} &
\makecell{\% Reduction\\vs\ \UO{}} &
\makecell{\Med} &
\makecell{\Diff{}\\vs\ \UO{}}\\
\midrule
\multirow{4}{*}{10} & \multirow{2}{*}{\cplex{}} & \CS{} & 0.182 & +3.7\% & 0.413 & +1.0\% & 31,497 & +10.8\% & 34.2\% & -2.9 \\
   &  & \HSB{} & 0.168 & +11.3\% & 0.407 & +2.3\% & 15,253 & +56.8\% & 20.4\% & +10.9 \\
\cmidrule(lr){2-11}
   & \multirow{2}{*}{\gurobi{}} & \CS{} & 0.194 & -2.3\% & 0.428 & -2.8\% & 23,855 & +58.2\% & 25.9\% & +24.8 \\
   &  & \HSB{} & 0.168 & +11.4\% & 0.404 & +3.2\% & 14,253 & +75.0\% & 19.1\% & +31.6 \\
\cmidrule(lr){1-11}
  \multirow{4}{*}{40} & \multirow{2}{*}{\cplex{}} & \CS{} & 0.161 & +1.5\% & 0.393 & +0.7\% & 115,681 & +30.5\% & 20.4\% & +3.8 \\
   &  & \HSB{} & 0.150 & +8.7\% & 0.375 & +5.1\% & 64,899 & +61.0\% & 18.2\% & +6.1 \\
\cmidrule(lr){2-11}
   & \multirow{2}{*}{\gurobi{}} & \CS{} & 0.171 & -2.8\% & 0.404 & -0.3\% & 114,833 & +8.5\% & 20.3\% & -2.0 \\
   &  & \HSB{} & 0.146 & +12.1\% & 0.369 & +8.3\% & 73,957 & +41.0\% & 20.7\% & -2.5 \\
\bottomrule
\end{tabular}
}
\end{table}

\begin{table}[hbtp]
\centering
\caption{Solution accuracy for the \MSTlong{} benchmark. Positive values indicate improvement relative to \UO{}.}
\label{tab:sol-accuracy-mst}
\resizebox{\textwidth}{!}{
\begin{tabular}{c c c c  c c c c c c c}
\toprule
\multicolumn{3}{c}{Info} & \multicolumn{2}{c}{MAE} & \multicolumn{2}{c}{RMSE} & \multicolumn{2}{c}{Mismatch Count} & \multicolumn{2}{c}{Mismatch Percentage} \\
\cmidrule(lr){1-3} \cmidrule(lr){4-5} \cmidrule(lr){6-7} \cmidrule(lr){8-9} \cmidrule(lr){10-11}
\makecell{Size} &
\makecell{Solver} &
\makecell{\OPT{}} &
\makecell{\Med{}} &
\makecell{\% Reduction\\vs\ \UO{}} &
\makecell{\Med{}} &
\makecell{\% Reduction\\vs\ \UO{}} &
\makecell{\Med} &
\makecell{\% Reduction\\vs\ \UO{}} &
\makecell{\Med} &
\makecell{\Diff{}\\vs\ \UO{}}\\
\midrule
\multirow{4}{*}{3} & \multirow{2}{*}{\cplex{}} & \CS{} & 0.114 & +14.6\% & 0.328 & +8.2\% & 25,532 & +43.4\% & 19.2\% & +10.1 \\
 &    & \HSB{} & 0.050 & +62.5\% & 0.221 & +38.1\% & 5,363 & +88.1\% & 7.5\% & +21.7 \\
\cmidrule(lr){2-11}
 &   \multirow{2}{*}{\gurobi{}} & \CS{} & 0.111 & +44.1\% & 0.327 & +6.6\% & 22,717 & +76.3\% & 17.1\% & +45.0 \\
 &    & \HSB{} & 0.051 & +74.4\% & 0.224 & +35.8\% & 5,199 & +94.6\% & 7.3\% & +54.8 \\
\cmidrule(lr){1-11}
  \multirow{4}{*}{5} & \multirow{2}{*}{\cplex{}} & \CS{} & 0.136 & +11.5\% & 0.362 & +5.8\% & 100,988 & +31.9\% & 18.6\% & +5.7 \\
   &  & \HSB{} & 0.066 & +57.0\% & 0.230 & +40.2\% & 25,261 & +83.0\% & 10.7\% & +13.5 \\
\cmidrule(lr){2-11}
   & \multirow{2}{*}{\gurobi{}} & \CS{} & 0.131 & +12.3\% & 0.358 & +7.2\% & 81,769 & +26.9\% & 15.0\% & +3.2 \\
   &  & \HSB{} & 0.080 & +46.3\% & 0.281 & +27.1\% & 21,764 & +80.5\% & 9.2\% & +9.1 \\
\bottomrule
\end{tabular}
}
\end{table}

\begin{table}[hbtp]
\centering
\caption{Solution accuracy for the \BSlong{} benchmark. Positive values indicate improvement relative to \UO{}.}
\label{tab:sol-accuracy-bs}
\resizebox{\textwidth}{!}{
\begin{tabular}{c c c c c c c c c c c}
\toprule
\multicolumn{3}{c}{Info} & \multicolumn{2}{c}{MAE} & \multicolumn{2}{c}{RMSE} & \multicolumn{2}{c}{Mismatch Count} & \multicolumn{2}{c}{Mismatch Percentage} \\
\cmidrule(lr){1-3} \cmidrule(lr){4-5} \cmidrule(lr){6-7} \cmidrule(lr){8-9} \cmidrule(lr){10-11}
\makecell{Size} &
\makecell{Solver} &
\makecell{\OPT{}} &
\makecell{\Med{}} &
\makecell{\% Reduction\\vs\ \UO{}} &
\makecell{\Med{}} &
\makecell{\% Reduction\\vs\ \UO{}} &
\makecell{\Med} &
\makecell{\% Reduction\\vs\ \UO{}} &
\makecell{\Med} &
\makecell{\Diff{}\\vs\ \UO{}}\\
\midrule
\multirow{8}{*}{160} & \multirow{2}{*}{\cplex{}} & \CS{} & 0.134 & +15.6\% & 0.276 & +26.4\% & 10,653 & -51.3\% & 46.0\% & -15.0 \\
   &  & \HSB{} & 0.099 & +37.3\% & 0.255 & +32.0\% & 4,108 & +41.6\% & 29.3\% & +1.7 \\
\cmidrule(lr){2-11}
   & \multirow{2}{*}{\cplex{}$^*$} & \CS{} & 0.111 & +30.0\% & 0.318 & +15.3\% & 7,244 & -2.9\% & 31.3\% & -0.3 \\
   &  & \HSB{} & 0.104 & +34.3\% & 0.286 & +23.8\% & 3,374 & +52.1\% & 24.0\% & +6.9 \\
\cmidrule(lr){2-11}
   & \multirow{2}{*}{\gurobi{}} & \CS{} & 0.111 & +41.6\% & 0.318 & +19.0\% & 6,077 & +55.7\% & 26.2\% & +34.1 \\
   &  & \HSB{} & 0.102 & +46.3\% & 0.261 & +33.6\% & 3,206 & +76.6\% & 22.8\% & +37.4 \\
\cmidrule(lr){2-11}
   & \multirow{2}{*}{\scip{}} & \CS{} & 0.121 & +16.9\% & 0.309 & +10.9\% & 7,633 & +27.6\% & 32.9\% & +13.5 \\
   &  & \HSB{} & 0.081 & +44.2\% & 0.262 & +24.5\% & 3,031 & +71.3\% & 21.6\% & +24.8 \\
\bottomrule
\end{tabular}
}
\end{table}

\begin{table}[hbtp]
\centering
\caption{Solution accuracy for the \LISLlong{} benchmark. Positive values indicate improvement relative to \UO{}.}
\label{tab:sol-accuracy-lisl}
\resizebox{\textwidth}{!}{
\begin{tabular}{c c c c c c c c c c c}
\toprule
\multicolumn{3}{c}{Info} & \multicolumn{2}{c}{MAE} & \multicolumn{2}{c}{RMSE} & \multicolumn{2}{c}{Mismatch Count} & \multicolumn{2}{c}{Mismatch Percentage} \\
\cmidrule(lr){1-3} \cmidrule(lr){4-5} \cmidrule(lr){6-7} \cmidrule(lr){8-9} \cmidrule(lr){10-11}
\makecell{Size} &
\makecell{Solver} &
\makecell{\OPT{}} &
\makecell{\Med{}} &
\makecell{\% Reduction\\vs\ \UO{}} &
\makecell{\Med{}} &
\makecell{\% Reduction\\vs\ \UO{}} &
\makecell{\Med} &
\makecell{\% Reduction\\vs\ \UO{}} &
\makecell{\Med} &
\makecell{\Diff{}\\vs\ \UO{}}\\
\midrule
\multirow{6}{*}{10} & \multirow{2}{*}{\cplex{}} & \CS{} & 0.118 & +20.3\% & 0.334 & +10.8\% & 19,568 & +43.2\% & 15.0\% & +8.4 \\
   &  & \HSB{} & 0.092 & +37.5\% & 0.225 & +39.9\% & 17,244 & +49.9\% & 27.7\% & -4.3 \\
\cmidrule(lr){2-11}
   & \multirow{2}{*}{\gurobi{}} & \CS{} & 0.123 & +17.5\% & 0.342 & +9.5\% & 20,904 & +32.9\% & 16.1\% & +5.2 \\
   &  & \HSB{} & 0.189 & -26.4\% & 0.370 & +2.1\% & 20,291 & +34.9\% & 32.6\% & -11.4 \\
\cmidrule(lr){2-11}
   & \multirow{2}{*}{\scip{}} & \CS{} & 0.111 & +16.4\% & 0.313 & +11.6\% & 29,213 & +27.7\% & 22.5\% & +5.1 \\
   &  & \HSB{} & 0.068 & +48.5\% & 0.254 & +28.2\% & 7,846 & +80.6\% & 12.6\% & +14.9 \\
\cmidrule(lr){1-11}
 \multirow{4}{*}{20} & \multirow{2}{*}{\cplex{}} & \CS{} & 0.118 & +17.1\% & 0.338 & +8.9\% & 64,119 & +35.7\% & 14.1\% & +5.9 \\
   &  & \HSB{} & 0.139 & +3.1\% & 0.280 & +24.7\% & 65,368 & +34.4\% & 34.2\% & -14.2 \\
\cmidrule(lr){2-11}
   & \multirow{2}{*}{\gurobi{}} & \CS{} & 0.122 & +14.7\% & 0.342 & +8.1\% & 66,877 & +23.6\% & 14.7\% & +2.9 \\
   &  & \HSB{} & 0.239 & -67.3\% & 0.420 & -12.8\% & 74,767 & +14.6\% & 39.1\% & -21.5 \\
\bottomrule
\end{tabular}
}
\end{table}

\begin{table}[hbtp]
\centering
\caption{Solution accuracy for the \SAMlong{} benchmark. Positive values indicate improvement relative to \UO{}.}
\label{tab:sol-accuracy-sam}
\resizebox{\textwidth}{!}{
\begin{tabular}{c c c c c c c c c c c}
\toprule
\multicolumn{3}{c}{Info} & \multicolumn{2}{c}{MAE} & \multicolumn{2}{c}{RMSE} & \multicolumn{2}{c}{Mismatch Count} & \multicolumn{2}{c}{Mismatch Percentage} \\
\cmidrule(lr){1-3} \cmidrule(lr){4-5} \cmidrule(lr){6-7} \cmidrule(lr){8-9} \cmidrule(lr){10-11}
\makecell{Size} &
\makecell{Solver} &
\makecell{\OPT{}} &
\makecell{\Med{}} &
\makecell{\% Reduction\\vs\ \UO{}} &
\makecell{\Med{}} &
\makecell{\% Reduction\\vs\ \UO{}} &
\makecell{\Med} &
\makecell{\% Reduction\\vs\ \UO{}} &
\makecell{\Med} &
\makecell{\Diff{}\\vs\ \UO{}}\\
\midrule
\multirow{6}{*}{10} & \multirow{2}{*}{\cplex{}} & \CS{} & 0.139 & +12.8\% & 0.370 & +2.5\% & 6,578 & +69.7\% & 16.5\% & +28.4 \\
   &  & \HSB{} & 0.091 & +43.0\% & 0.240 & +36.8\% & 6,030 & +72.2\% & 22.1\% & +22.8 \\
\cmidrule(lr){2-11}
   & \multirow{2}{*}{\gurobi{}} & \CS{} & 0.165 & -8.4\% & 0.386 & -1.8\% & 12,059 & +24.7\% & 30.2\% & +2.9 \\
  &  & \HSB{} & 0.097 & +36.6\% & 0.301 & +20.6\% & 3,519 & +78.0\% & 12.9\% & +20.2 \\
\cmidrule(lr){2-11}
   & \multirow{2}{*}{\scip{}} & \CS{} & 0.151 & +21.1\% & 0.359 & +9.8\% & 11,687 & +54.6\% & 29.2\% & +24.0 \\
   &  & \HSB{} & 0.086 & +55.1\% & 0.270 & +32.2\% & 3,961 & +84.6\% & 14.5\% & +38.7 \\
\cmidrule(lr){1-11}
  \multirow{4}{*}{20} & \multirow{2}{*}{\cplex{}} & \CS{} & 0.159 & +31.0\% & 0.383 & -12.4\% & 22,220 & +69.5\% & 25.1\% & +43.9 \\
   &  & \HSB{} & 0.116 & +49.6\% & 0.328 & +3.8\% & 9,830 & +86.5\% & 17.1\% & +51.9 \\
\cmidrule(lr){2-11}
   & \multirow{2}{*}{\gurobi{}} & \CS{} & 0.160 & +38.2\% & 0.387 & +1.6\% & 18,508 & +73.9\% & 20.9\% & +46.3 \\
   &  & \HSB{} & 0.105 & +59.4\% & 0.306 & +22.3\% & 9,083 & +87.2\% & 15.8\% & +51.4 \\
\cmidrule(lr){1-11}
  \multirow{4}{*}{40} & \multirow{2}{*}{\cplex{}} & \CS{} & 0.159 & +17.2\% & 0.383 & +8.8\% & 60,424 & +13.5\% & 30.9\% & -0.6 \\
   &  & \HSB{} & 0.117 & +38.9\% & 0.334 & +20.4\% & 19,186 & +72.5\% & 15.9\% & +14.4 \\
\cmidrule(lr){2-11}
   & \multirow{2}{*}{\gurobi{}} & \CS{} & 0.171 & +5.5\% & 0.403 & +3.2\% & 40,544 & +27.6\% & 20.7\% & +3.6 \\
   &  & \HSB{} & 0.113 & +37.6\% & 0.329 & +21.1\% & 16,952 & +69.7\% & 14.0\% & +10.3 \\
\bottomrule
\end{tabular}
}
\end{table}

\begin{table}[hbtp]
\centering
\caption{Solution accuracy for the \TSlong{} benchmark. Positive values indicate improvement relative to \UO{}.}
\label{tab:sol-accuracy-ts}
\resizebox{\textwidth}{!}{
\begin{tabular}{c c c c c c c c c c c}
\toprule
\multicolumn{3}{c}{Info} & \multicolumn{2}{c}{MAE} & \multicolumn{2}{c}{RMSE} & \multicolumn{2}{c}{Mismatch Count} & \multicolumn{2}{c}{Mismatch Percentage} \\
\cmidrule(lr){1-3} \cmidrule(lr){4-5} \cmidrule(lr){6-7} \cmidrule(lr){8-9} \cmidrule(lr){10-11}
\makecell{Size} &
\makecell{Solver} &
\makecell{\OPT{}} &
\makecell{\Med{}} &
\makecell{\% Reduction\\vs\ \UO{}} &
\makecell{\Med{}} &
\makecell{\% Reduction\\vs\ \UO{}} &
\makecell{\Med} &
\makecell{\% Reduction\\vs\ \UO{}} &
\makecell{\Med} &
\makecell{\Diff{}\\vs\ \UO{}}\\
\midrule
\multirow{4}{*}{10} & \multirow{2}{*}{\cplex{}} & \CS{} & 0.136 & +12.8\% & 0.358 & +7.2\% & 41,374 & +51.7\% & 18.1\% & +10.9 \\
   &  & \HSB{} & 0.097 & +37.6\% & 0.304 & +21.3\% & 13,974 & +83.7\% & 12.1\% & +16.9 \\
\cmidrule(lr){2-11}
   & \multirow{2}{*}{\gurobi{}} & \CS{} & 0.134 & +11.4\% & 0.362 & +5.9\% & 36,194 & +45.7\% & 15.8\% & +6.7 \\
   &  & \HSB{} & 0.220 & -44.8\% & 0.419 & -9.1\% & 39,207 & +41.2\% & 32.7\% & -10.1 \\
\cmidrule(lr){1-11}
  \multirow{4}{*}{20} & \multirow{2}{*}{\cplex{}} & \CS{} & 0.137 & +13.4\% & 0.362 & +7.2\% & 117,887 & +47.2\% & 16.6\% & +9.2 \\
   &  & \HSB{} & 0.116 & +26.6\% & 0.335 & +14.2\% & 39,434 & +82.3\% & 12.8\% & +13.0 \\
\cmidrule(lr){2-11}
   & \multirow{2}{*}{\gurobi{}} & \CS{} & 0.136 & +12.0\% & 0.366 & +6.2\% & 105,718 & +32.3\% & 14.9\% & +3.2 \\
   &  & \HSB{} & 0.344 & -122.1\% & 0.534 & -36.9\% & 153,254 & +1.9\% & 48.0\% & -29.9 \\
\bottomrule
\end{tabular}
}
\end{table}

\begin{table}[hbtp]
\centering
\caption{Solution accuracy for the \MSlong{} benchmark. Positive values indicate improvement relative to \UO{}.}
\label{tab:sol-accuracy-ms}
\resizebox{\textwidth}{!}{
\begin{tabular}{c c c c c c c c c c c}
\toprule
\multicolumn{3}{c}{Info} & \multicolumn{2}{c}{MAE} & \multicolumn{2}{c}{RMSE} & \multicolumn{2}{c}{Mismatch Count} & \multicolumn{2}{c}{Mismatch Percentage} \\
\cmidrule(lr){1-3} \cmidrule(lr){4-5} \cmidrule(lr){6-7} \cmidrule(lr){8-9} \cmidrule(lr){10-11}
\makecell{Size} &
\makecell{Solver} &
\makecell{\OPT{}} &
\makecell{\Med{}} &
\makecell{\% Reduction\\vs\ \UO{}} &
\makecell{\Med{}} &
\makecell{\% Reduction\\vs\ \UO{}} &
\makecell{\Med} &
\makecell{\% Reduction\\vs\ \UO{}} &
\makecell{\Med} &
\makecell{\Diff{}\\vs\ \UO{}}\\
\midrule
\multirow{8}{*}{5} & \multirow{2}{*}{\cplex{}} & \CS{} & 0.097 & +7.8\% & 0.292 & +4.1\% & 5,934 & +57.8\% & 24.6\% & +28.0 \\
   &  & \HSB{} & 0.044 & +57.9\% & 0.196 & +35.7\% & 1,459 & +89.6\% & 10.4\% & +42.2 \\
\cmidrule(lr){2-11}
   & \multirow{2}{*}{\cplex{}$^*$} & \CS{} & 0.101 & +3.7\% & 0.301 & +1.1\% & 7,799 & +44.5\% & 32.3\% & +20.3 \\
   &  & \HSB{} & 0.044 & +57.9\% & 0.196 & +35.7\% & 1,459 & +89.6\% & 10.4\% & +42.2 \\
\cmidrule(lr){2-11}
   & \multirow{2}{*}{\gurobi{}} & \CS{} & 0.092 & +35.8\% & 0.284 & +2.8\% & 6,076 & +60.5\% & 25.2\% & +32.4 \\
   &  & \HSB{} & 0.039 & +72.9\% & 0.189 & +35.5\% & 1,090 & +92.9\% & 7.3\% & +50.3 \\
\cmidrule(lr){2-11}
   & \multirow{2}{*}{\scip{}} & \CS{} & 0.091 & +14.8\% & 0.277 & +9.0\% & 6,444 & +38.0\% & 26.7\% & +12.2 \\
   &  & \HSB{} & 0.038 & +64.0\% & 0.176 & +42.1\% & 1,407 & +86.5\% & 9.5\% & +29.4 \\
\cmidrule(lr){1-11}
  \multirow{7}{*}{40} & \multirow{2}{*}{\cplex{}} & \CS{} & 0.167 & +5.8\% & 0.402 & +3.0\% & 70,774 & +29.1\% & 21.9\% & +8.2 \\
   &  & \HSB{} & 0.085 & +52.3\% & 0.290 & +30.1\% & 13,386 & +86.6\% & 10.4\% & +19.8 \\
\cmidrule(lr){2-11}
   & \multirow{2}{*}{\cplex{}$^*$} & \CS{} & 0.167 & +5.9\% & 0.403 & +2.7\% & 67,588 & +32.3\% & 20.9\% & +9.2 \\
   &  & \HSB{} & 0.085 & +52.3\% & 0.290 & +30.1\% & 13,386 & +86.6\% & 10.4\% & +19.8 \\
\cmidrule(lr){2-11}
   & \multirow{2}{*}{\gurobi{}} & \CS{} & 0.149 & +16.1\% & 0.381 & +7.8\% & 58,381 & +48.0\% & 18.1\% & +15.8 \\
   &  & \HSB{} & 0.078 & +56.3\% & 0.257 & +37.9\% & 15,553 & +86.1\% & 12.0\% & +21.8 \\
\cmidrule(lr){2-11}
   & \multirow{1}{*}{\scip{}} & \HSB{} & 0.084 & +51.6\% & 0.271 & +32.5\% & 15,687 & +82.4\% & 12.1\% & +14.7 \\
\cmidrule(lr){1-11}
  \multirow{2}{*}{80} & \multirow{2}{*}{\cplex{}$^*$} & \CS{} & 0.184 & +4.1\% & 0.425 & +1.6\% & 214,586 & +27.2\% & 22.1\% & +7.9 \\
   &  & \HSB{} & 0.098 & +48.9\% & 0.311 & +28.1\% & 32,710 & +88.9\% & 10.9\% & +19.1 \\
\bottomrule
\end{tabular}
}
\end{table}

\clearpage
\section{Additional Condition Number Results}
\label{sec:appendix-solution_kappa}
This appendix provides detailed condition number results for all benchmarks in Table~\ref{tab:benchmark-config}, to complement the overall comparison presented in Section~\ref{sec:resultConditionNumberKappa}.
Tables~\ref{tab:kappaRes_spdj} to~\ref{tab:kappaRes_ms} report the median basis condition number (kappa) for the unscaled basis for each solved LP instance, together with the corresponding conditioning class specified by \cplex{} to interpret the reported kappa values (Table~\ref{tab:mip-kappa-classes}).

\begin{table}[hbtp]
\centering
\caption{Median of basis condition numbers for the \SPDJlong{} benchmark, where inputs are encoded through the objective function.}
\label{tab:kappaRes_spdj}
\scalebox{1}{
\begin{tabular}{c c c r c}
\toprule
\makecell{$n$} & \makecell{solver} & \makecell{\OPT{}} & \makecell{condition number\\(\Med{})} & \makecell{conditioning\\class} \\
\midrule
\multirow{3}{*}{10} & \multirow{3}{*}{\cplex} & \CS{} & $7.10 \times 10^{5}$ & Stable \\
 &  & \HSB{} & $3.80 \times 10^{4}$ & Stable \\
 &  & \UO{} & $2.10 \times 10^{6}$ & Stable \\
\midrule
\multirow{1}{*}{20} & \multirow{1}{*}{\cplex} & \HSB{} & $6.90 \times 10^{7}$ & Suspicious \\
\midrule
\multirow{1}{*}{30} & \multirow{1}{*}{\cplex} & \HSB{} & $4.30 \times 10^{5}$ & Stable \\
\midrule
\multirow{1}{*}{40} & \multirow{1}{*}{\cplex} & \HSB{} & $2.30 \times 10^{4}$ & Stable \\
\bottomrule
\end{tabular}}
\end{table}

\begin{table}[hbtp]
\centering
\caption{Median of basis condition numbers for the \KSlong{} benchmark, where inputs are encoded through the objective function.}
\label{tab:kappaRes_kd}
\scalebox{1}{
\begin{tabular}{c c c r c}
\toprule
\makecell{$n$} & \makecell{solver} & \makecell{\OPT{}} & \makecell{condition number\\(\Med{})} & \makecell{conditioning\\class} \\
\midrule
\multirow{3}{*}{10} & \multirow{3}{*}{\cplex} & \CS{} & $1.90 \times 10^{5}$ & Stable \\
 &  & \HSB{} & $3.50 \times 10^{3}$ & Stable \\
 &  & \UO{} & $4.00 \times 10^{5}$ & Stable \\
\midrule
\multirow{3}{*}{40} & \multirow{3}{*}{\cplex} & \CS{} & $5.80 \times 10^{5}$ & Stable \\
 &  & \HSB{} & $6.00 \times 10^{3}$ & Stable \\
 &  & \UO{} & $7.80 \times 10^{6}$ & Stable \\
\midrule
\multirow{1}{*}{80} & \multirow{1}{*}{\cplex} & \HSB{} & $9.90 \times 10^{3}$ & Stable \\
\midrule
\multirow{1}{*}{160} & \multirow{1}{*}{\cplex} & \HSB{} & $1.00 \times 10^{4}$ & Stable \\
\bottomrule
\end{tabular}}
\end{table}

\begin{table}[hbtp]
\centering
\caption{Median of basis condition numbers for the \MSTlong{} benchmark, where inputs are encoded through the objective function.}
\label{tab:kappaRes_mst}
\scalebox{1}{
\begin{tabular}{c c c r c}
\toprule
\makecell{$n$} & \makecell{solver} & \makecell{\OPT{}} & \makecell{condition number\\(\Med{})} & \makecell{conditioning\\class} \\
\midrule
\multirow{4}{*}{3} & \multirow{1}{*}{\cplex$^*$} & \CS{} & $2.90 \times 10^{9}$ & Suspicious \\
\cmidrule(lr){2-5}
 & \multirow{3}{*}{\cplex} & \CS{} & $4.40 \times 10^{10}$ & Unstable \\
 &  & \HSB{} & $1.60 \times 10^{6}$ & Stable \\
 &  & \UO{} & $1.30 \times 10^{11}$ & Unstable \\
\midrule
\multirow{3}{*}{5} & \multirow{3}{*}{\cplex} & \CS{} & $1.50 \times 10^{13}$ & Unstable \\
 &  & \HSB{} & $1.50 \times 10^{8}$ & Suspicious \\
 &  & \UO{} & $1.90 \times 10^{13}$ & Unstable \\
\midrule
\multirow{1}{*}{8} & \multirow{1}{*}{\cplex} & \HSB{} & $2.20 \times 10^{10}$ & Unstable \\
\midrule
\multirow{1}{*}{12} & \multirow{1}{*}{\cplex} & \HSB{} & $2.10 \times 10^{14}$ & Ill-posed \\
\bottomrule
\end{tabular}}
\end{table}

\begin{table}[hbtp]
\centering
\caption{Median of basis condition numbers for the \BSlong{} benchmark, where inputs are encoded through the objective function.}
\label{tab:kappaRes_bs}
\scalebox{1}{
\begin{tabular}{c c c r c}
\toprule
\makecell{$n$} & \makecell{solver} & \makecell{\OPT{}} & \makecell{condition number\\(\Med{})} & \makecell{conditioning\\class} \\
\midrule
\multirow{6}{*}{160} & \multirow{3}{*}{\cplex$^*$} & \CS{} & $2.80 \times 10^{6}$ & Stable \\
 &  & \HSB{} & $1.30 \times 10^{6}$ & Stable \\
 &  & \UO{} & $1.30 \times 10^{6}$ & Stable \\
\cmidrule(lr){2-5}
 & \multirow{3}{*}{\cplex} & \CS{} & $2.70 \times 10^{6}$ & Stable \\
 &  & \HSB{} & $1.20 \times 10^{6}$ & Stable \\
 &  & \UO{} & $1.60 \times 10^{6}$ & Stable \\
\midrule
\multirow{4}{*}{640} & \multirow{2}{*}{\cplex$^*$} & \CS{} & $5.00 \times 10^{6}$ & Stable \\
 &  & \HSB{} & $2.40 \times 10^{6}$ & Stable \\
\cmidrule(lr){2-5}
 & \multirow{2}{*}{\cplex} & \CS{} & $5.00 \times 10^{6}$ & Stable \\
 &  & \HSB{} & $2.60 \times 10^{6}$ & Stable \\
\midrule
\multirow{3}{*}{1,280} & \multirow{2}{*}{\cplex$^*$} & \CS{} & $6.50 \times 10^{6}$ & Stable \\
 &  & \HSB{} & $3.10 \times 10^{6}$ & Stable \\
\cmidrule(lr){2-5}
 & \multirow{1}{*}{\cplex} & \HSB{} & $3.40 \times 10^{6}$ & Stable \\
\midrule
\multirow{2}{*}{2,560} & \multirow{1}{*}{\cplex$^*$} & \HSB{} & $4.10 \times 10^{6}$ & Stable \\
\cmidrule(lr){2-5}
 & \multirow{1}{*}{\cplex} & \HSB{} & $5.40 \times 10^{6}$ & Stable \\
\midrule
\multirow{2}{*}{5,120} & \multirow{1}{*}{\cplex$^*$} & \HSB{} & $5.90 \times 10^{6}$ & Stable \\
\cmidrule(lr){2-5}
 & \multirow{1}{*}{\cplex} & \HSB{} & $6.00 \times 10^{6}$ & Stable \\
\bottomrule
\end{tabular}}
\end{table}

\begin{table}[hbtp]
\centering
\caption{Median of basis condition numbers for the \LISLlong{} benchmark, where inputs are encoded through the objective function.}
\label{tab:kappaRes_lisl}
\scalebox{1}{
\begin{tabular}{c c c r c}
\toprule
\makecell{$n$} & \makecell{solver} & \makecell{\OPT{}} & \makecell{condition number\\(\Med{})} & \makecell{conditioning\\class} \\
\midrule
\multirow{3}{*}{10} & \multirow{3}{*}{\cplex} & \CS{} & $1.50 \times 10^{10}$ & Unstable \\
 &  & \HSB{} & $2.50 \times 10^{9}$ & Suspicious \\
 &  & \UO{} & $1.10 \times 10^{10}$ & Unstable \\
\midrule
\multirow{3}{*}{20} & \multirow{3}{*}{\cplex} & \CS{} & $7.30 \times 10^{10}$ & Unstable \\
 &  & \HSB{} & $2.10 \times 10^{10}$ & Unstable \\
 &  & \UO{} & $4.30 \times 10^{11}$ & Unstable \\
\midrule
\multirow{1}{*}{30} & \multirow{1}{*}{\cplex} & \HSB{} & $3.10 \times 10^{9}$ & Suspicious \\
\midrule
\multirow{2}{*}{40} & \multirow{2}{*}{\cplex} & \CS{} & $1.20 \times 10^{11}$ & Unstable \\
 &  & \HSB{} & $1.90 \times 10^{10}$ & Unstable \\
\midrule
\multirow{1}{*}{80} & \multirow{1}{*}{\cplex} & \HSB{} & $4.00 \times 10^{10}$ & Unstable \\
\bottomrule
\end{tabular}}
\end{table}

\begin{table}[hbtp]
\centering
\caption{Median of basis condition numbers for the \SAMlong{} benchmark, where inputs are encoded through the objective function.}
\label{tab:kappaRes_sam}
\scalebox{1}{
\begin{tabular}{c c c r c}
\toprule
\makecell{$n$} & \makecell{solver} & \makecell{\OPT{}} & \makecell{condition number\\(\Med{})} & \makecell{conditioning\\class} \\
\midrule
\multirow{3}{*}{10} & \multirow{3}{*}{\cplex} & \CS{} & $4.90 \times 10^{5}$ & Stable \\
 &  & \HSB{} & $5.50 \times 10^{4}$ & Stable \\
 &  & \UO{} & $1.50 \times 10^{5}$ & Stable \\
\midrule
\multirow{3}{*}{20} & \multirow{3}{*}{\cplex} & \CS{} & $4.10 \times 10^{5}$ & Stable \\
 &  & \HSB{} & $6.50 \times 10^{4}$ & Stable \\
 &  & \UO{} & $7.30 \times 10^{6}$ & Stable \\
\midrule
\multirow{3}{*}{40} & \multirow{3}{*}{\cplex} & \CS{} & $1.50 \times 10^{6}$ & Stable \\
 &  & \HSB{} & $2.10 \times 10^{4}$ & Stable \\
 &  & \UO{} & $1.70 \times 10^{5}$ & Stable \\
\midrule
\multirow{2}{*}{80} & \multirow{2}{*}{\cplex} & \CS{} & $8.30 \times 10^{5}$ & Stable \\
 &  & \HSB{} & $2.30 \times 10^{4}$ & Stable \\
\midrule
\multirow{1}{*}{160} & \multirow{1}{*}{\cplex} & \HSB{} & $2.50 \times 10^{4}$ & Stable \\
\midrule
\multirow{1}{*}{320} & \multirow{1}{*}{\cplex} & \HSB{} & $1.20 \times 10^{5}$ & Stable \\
\bottomrule
\end{tabular}}
\end{table}

\begin{table}[hbtp]
\centering
\caption{Median of basis condition numbers for the \TSlong{} benchmark, where inputs are encoded through the objective function.}
\label{tab:kappaRes_ts}
\scalebox{1}{
\begin{tabular}{c c c r c}
\toprule
\makecell{$n$} & \makecell{solver} & \makecell{\OPT{}} & \makecell{condition number\\(\Med{})} & \makecell{conditioning\\class} \\
\midrule
\multirow{3}{*}{10} & \multirow{3}{*}{\cplex} & \CS{} & $1.30 \times 10^{6}$ & Stable \\
 &  & \HSB{} & $1.00 \times 10^{5}$ & Stable \\
 &  & \UO{} & $9.40 \times 10^{6}$ & Stable \\
\midrule
\multirow{3}{*}{20} & \multirow{3}{*}{\cplex} & \CS{} & $7.20 \times 10^{8}$ & Suspicious \\
 &  & \HSB{} & $7.80 \times 10^{4}$ & Stable \\
 &  & \UO{} & $1.30 \times 10^{7}$ & Suspicious \\
\midrule
\multirow{1}{*}{40} & \multirow{1}{*}{\cplex} & \HSB{} & $3.70 \times 10^{5}$ & Stable \\
\midrule
\multirow{1}{*}{80} & \multirow{1}{*}{\cplex} & \HSB{} & $1.30 \times 10^{6}$ & Stable \\
\bottomrule
\end{tabular}}
\end{table}

\begin{table}[hbtp]
\centering
\caption{Median of basis condition numbers for the \MSlong{} benchmark, where inputs are encoded through the objective function.}
\label{tab:kappaRes_ms}
\scalebox{1}{
\begin{tabular}{c c c r c}
\toprule
\makecell{$n$} & \makecell{solver} & \makecell{\OPT{}} & \makecell{condition number\\(\Med{})} & \makecell{conditioning\\class} \\
\midrule
\multirow{6}{*}{5} & \multirow{3}{*}{\cplex$^*$} & \CS{} & $3.00 \times 10^{4}$ & Stable \\
 &  & \HSB{} & $1.70 \times 10^{3}$ & Stable \\
 &  & \UO{} & $8.10 \times 10^{4}$ & Stable \\
\cmidrule(lr){2-5}
 & \multirow{3}{*}{\cplex} & \CS{} & $2.40 \times 10^{4}$ & Stable \\
 &  & \HSB{} & $1.70 \times 10^{3}$ & Stable \\
 &  & \UO{} & $3.90 \times 10^{6}$ & Stable \\
\midrule
\multirow{6}{*}{10} & \multirow{3}{*}{\cplex$^*$} & \CS{} & $1.50 \times 10^{5}$ & Stable \\
 &  & \HSB{} & $1.50 \times 10^{4}$ & Stable \\
 &  & \UO{} & $2.40 \times 10^{4}$ & Stable \\
\cmidrule(lr){2-5}
 & \multirow{3}{*}{\cplex} & \CS{} & $7.50 \times 10^{4}$ & Stable \\
 &  & \HSB{} & $2.60 \times 10^{3}$ & Stable \\
 &  & \UO{} & $2.40 \times 10^{6}$ & Stable \\
\midrule
\multirow{6}{*}{20} & \multirow{3}{*}{\cplex$^*$} & \CS{} & $3.20 \times 10^{4}$ & Stable \\
 &  & \HSB{} & $3.70 \times 10^{3}$ & Stable \\
 &  & \UO{} & $2.90 \times 10^{4}$ & Stable \\
\cmidrule(lr){2-5}
 & \multirow{3}{*}{\cplex} & \CS{} & $2.30 \times 10^{4}$ & Stable \\
 &  & \HSB{} & $1.50 \times 10^{4}$ & Stable \\
 &  & \UO{} & $9.20 \times 10^{6}$ & Stable \\
\midrule
\multirow{3}{*}{40} & \multirow{3}{*}{\cplex} & \CS{} & $1.10 \times 10^{8}$ & Suspicious \\
 &  & \HSB{} & $3.40 \times 10^{4}$ & Stable \\
 &  & \UO{} & $3.00 \times 10^{7}$ & Suspicious \\
\midrule
\multirow{2}{*}{80} & \multirow{2}{*}{\cplex} & \CS{} & $1.40 \times 10^{5}$ & Stable \\
 &  & \HSB{} & $8.60 \times 10^{3}$ & Stable \\
\midrule
\multirow{1}{*}{160} & \multirow{1}{*}{\cplex} & \HSB{} & $2.60 \times 10^{7}$ & Suspicious \\
\bottomrule
\end{tabular}}
\end{table}

\end{document}